\RequirePackage{rotating}

\documentclass[fleqn,usenatbib]{mnras}
\usepackage[dvipsnames]{xcolor}
\usepackage{newtxtext,newtxmath}
\usepackage[normalem]{ulem}
\usepackage[T1]{fontenc}

\DeclareRobustCommand{\VAN}[3]{#2}
\let\VANthebibliography\thebibliography
\def\thebibliography{\DeclareRobustCommand{\VAN}[3]{##3}\VANthebibliography}

\usepackage{booktabs}
\usepackage{pdflscape}
\usepackage{numprint}
\usepackage{siunitx}

\npdecimalsign{.}
\nprounddigits{4}

\usepackage{graphicx}	 % Including figure files
\usepackage{amsmath}	% Advanced maths commands

\newcommand{\foreground}{CL J021741.7-034546}

\title[Outskirts of XLSSC 122]{Initial Investigations of the Outskirts of XLSSC 122}

\author[Todd et. al]{Eleanore. B. Todd,$^{1}$
Jon. P. Willis,$^{1}$
Rebecca. E. A. Canning,$^{2}$
Oph\'elie. K. Leste, $^{1}$
Rahma. Alfarsy, $^{2}$
\newauthor
Steven W. Allen, $^{3,6,8}$
Gabriel Brammer, $^{4}$
Joseph. N. Burchett, $^{5}$
Adam. B. Mantz, $^{6}$
Spencer. A. Stanford $^{7}$
\\
$^{1}$Department of Physics and Astronomy, University of Victoria, 3800 Finnerty Road, Victoria, BC, Canada\\
$^{2}$Institute of Cosmology \& Gravitation, University of Portsmouth, Dennis Sciama Building, Portsmouth, PO1 3FX, UK\\
$^{3}$Department of Physics, Stanford University, 382 Via Pueblo Mall, Stanford, CA 94305, USA\\
$^{4}$Cosmic Dawn centre (DAWN), Niels Bohr Institute, University of Copenhagen, Jagtvej 128, København N, DK-2200, Denmark\\
$^{5}$Department of Astronomy, New Mexico State University, Las Cruces, NM 88003, USA\\
$^{6}$Kavli Institute for Particle Astrophysics and Cosmology, Stanford University, 452 Lomita Mall, Stanford, CA 93405, USA\\
$^{7}$Department of Physics, University of California, One Shields Avenue, Davis, CA 95616, USA\\
$^{8}$SLAC National Accelerator Laboratory, 2575 Sand Hill Road, Menlo Park, CA 94025, USA\\
}

\date{Accepted XXX. Received YYY; in original form ZZZ}

\pubyear{\the\year{}}

\begin{document}
\label{firstpage}
\pagerange{\pageref{firstpage}--\pageref{lastpage}}
\maketitle

\begin{abstract}
We investigate the redshift $1.98$ galaxy cluster XLSSC 122 using the Hubble Space Telescope (HST) from the core of the cluster out to $3\,\mathrm{Mpc}$, a scale equivalent to 10 times the $R_{500} = 295 \, \mathrm{kpc}$ radius. We present an expanded photometric and spectroscopic catalogue of the cluster, bringing the total number of spectroscopically classified member galaxies to $74$, with $35$ new member galaxies added in the outer regions of the cluster. We compute the radial galaxy number density profile in the cluster, and observe no clear evidence of infalling groups or cosmic filaments. We observe a clear bimodal colour relation in member galaxies, with red fraction increasing towards the cluster centre. This rapid increase of red fraction upon infall is indicative of a fast quenching mechanism, such as ram pressure stripping, as galaxies enter the cluster centre. We fit a luminosity function to the cluster members, finding a similar low mass slope but fainter scale magnitude than z = 1 clusters of similar temperature, implying a similar galaxy evolution rate to clusters at lower redshift.
\end{abstract}
\begin{figure*}
	\includegraphics[width=\textwidth]{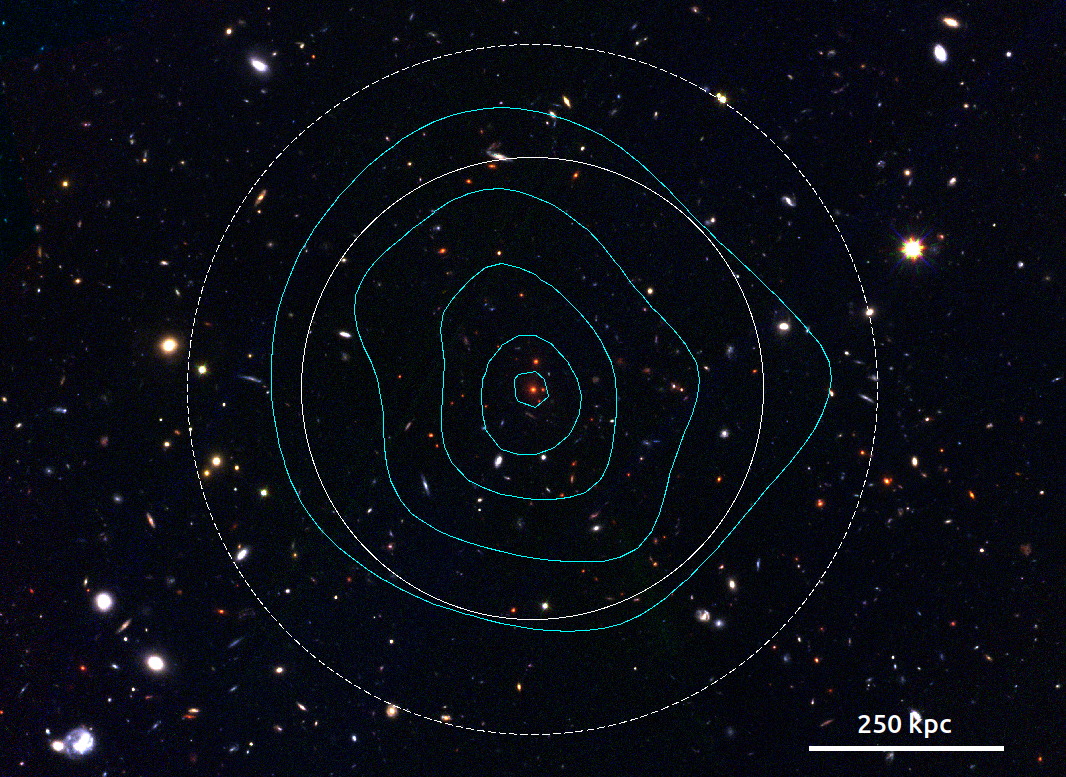}
    \caption{A 3 colour image of XLSSC
    122, r,g,b = F140W, F105W, F814W. Central red large galaxy is the BCG, id = 313. Cyan contours are X-ray contours from \citep{mantzXXLSurveyXVII2018}, The solid line is $R_{500} = 295\,\mathrm{kpc}$, the dashed line is $R_{200} = 440\,\mathrm{kpc}$.}
    \label{fig:XLSSC122_falsecolor}
\end{figure*}


%%%%%%%%%%%%%%%%%%%%%%%%%%%%%%%%%%%%%%%%%%%%%%%%%%

%%%%%%%%%%%%%%%%% BODY OF PAPER %%%%%%%%%%%%%%%%%%

\section{Introduction}
% \subsection{Galaxy Clusters}
Galaxy clusters frame the formation and evolution of galaxies across cosmic time \citep{dresslerGalaxyMorphologyRich1980,pengMASSENVIRONMENTDRIVERS2012,wetzelGalaxyEvolutionGroups2013}. Cluster cores host a greater number of red galaxies than the surrounding field \citep{baloghBimodalGalaxyColor2004,gladdersRedSequenceClusterSurvey2005,rykoffRedMaPPerAlgorithmSDSS2014,strazzulloRedSequenceBirth2016,baloghEvidenceChangeDominant2016} while their outskirts host evolving galaxies. \citep{mcgeeAccretionGalaxiesGroups2009,webbGOGREENSurveyPostinfall2020,wernerSatelliteQuenchingWas2022}. Galaxies in cluster outskirts experience a cessation of star formation (quenching) in part due to the interplay of member galaxies with a hot intracluster medium (ICM) \citep{gunnInfallMatterClusters1972,abadiRamPressureStripping1999,mcgeeAccretionGalaxiesGroups2009}, where gas is stripped from a galaxy upon collision with the ICM. \citep{paccagnellaOmegaWINGSFirstComplete2017,brownColdGasStripping2017, wattsVERTICOEnvironmentallyDriven2023}. The bimodal colour distributions of cluster galaxies from the local universe to at least $z = 2$ \citep{baloghBimodalGalaxyColor2004,strazzulloGalaxyPopulationsMost2019, willisSpectroscopicConfirmationMature2020a} imply that rapid quenching upon cluster entry is present from early cosmic times. 
 
The cluster galaxy population colour bimodality \citep{butcherEvolutionGalaxiesClusters1984,urquhartEnvironmentalButcherOemlerEffect2010} can be described as a double Gaussian of red and blue galaxies, with the proportion of red galaxies to blue increasing with cluster ICM density \citep{baloghBimodalGalaxyColor2004}. Due to this quiescent galaxy prevalence and hot ICM, clusters are often observed as overdensities of massive red galaxies \citep{dresslerGalaxyMorphologyRich1980,hoggOverdensitiesGalaxyEnvironments2003,gladdersNewMethodGalaxy2000,rykoffRedMaPPerAlgorithmSDSS2014,strazzulloPassiveGalaxiesTracers2015} with extended X-ray emission \citep{sarazinXrayEmissionClusters1986}. 

The proportion of red galaxies in a cluster (red fraction) can be used to compare the effect of environmental density on the star formation history of galaxies \citep{baloghEvidenceChangeDominant2016}, and has been used as strong indication of quenching activity \citep{willisDistantGalaxyClusters2013,strazzulloRedSequenceBirth2016,rykoffRedMaPPerAlgorithmSDSS2014}. The change in red fraction over regions of varying density probes the prominence of the cluster environment on the quenching of their member galaxies \citep{raichoorStarFormationEnvironment2012}.

% \subsection{Quenching in cluster environments}
Quenching mechanisms can be broadly divided between mass quenching and environmental quenching \citep{pengMassEnvironmentDrivers2010,kawinwanichakijEffectLocalEnvironment2017}. Quenching mechanisms due to a galaxy's mass include rapid consumption of the gas content due to star formation \citep{mcgeeAccretionGalaxiesGroups2009}, or the heating and removal of gas in the galaxy due to the feedback from active galactic nuclei (AGN \cite{circostaSUPERUnbiasedStudy2018}). Environmental quenching occurs when galaxies interact with other objects, which may result in an increase in star formation, or a removal of gas due to the turbulent interaction \citep{mooreGalaxyHarassmentEvolution1996}. The galaxy may quench rapidly after this interaction as gas viable for star formation is removed or consumed \citep{mcgeeOverconsumptionOutflowsQuenching2014}. 

The quenching mechanism of interest to us in the study of dense cluster environments is the interaction between the galaxy and the hot plasma ICM, known as ram pressure stripping \citep{gunnInfallMatterClusters1972}. When a galaxy enters a cluster environment and passes through its ICM we often observe a compression and expulsion of gas from the galaxy \citep{werlePoststarburstGalaxiesCenters2022}. This pressure can remove the gas reservoir necessary for star formation \citep{brownColdGasStripping2017}. Ram pressure stripped, as well as other fast quenched galaxies, may appear as post-starburst galaxies. A post-starburst galaxy has quenched, but retains a remnant population of intermediate age stars offering a unique spectral signature\citep{poggiantiStarFormationHistories1999,paccagnellaOmegaWINGSFirstComplete2017}.

Galaxies in hot, dense, cluster environments have a higher rate of quenching than low density regions \citep{baldryGalaxyBimodalityStellar2006,pengMASSENVIRONMENTDRIVERS2012}. At higher cluster masses, ram pressure stripping is more effective, scaling with both ICM density and infall velocity \citep{gunnInfallMatterClusters1972,mccarthyRamPressureStripping2008} (both higher in massive cluster environments). A massive cluster is also often located at the intersection of a greater number of filaments of the cosmic web \citep{bondHowFilamentsGalaxies1996a}. Just as a galaxy in a cluster is quenched, locally overdense regions such as filaments can quench or partially quench galaxies crossing them, reducing the galaxy's gas supply upon reaching the cluster \citep{liEVOLUTIONGROUPGALAXIES2012}. Intriguingly, the dominant quenching mechanism has shifted over time. In the local universe, environmental quenching \citep{gunnInfallMatterClusters1972,baloghOriginStarFormation2000} has been observed as the dominant quenching mechanism in galaxy clusters \citep{darvishEFFECTSLOCALENVIRONMENT2016, pengMassEnvironmentDrivers2010}, where galaxies quench at a rate different from that expected from their mass. This picture is not as clear in the early universe.

Pushing our observations of quenching in cluster environments into the early universe probes a region past where the dominant quenching mechanism appears to shift \citep{baloghEvidenceChangeDominant2016, trudeauMassiveDistantClusters2024}. Recently, the GOGREEN survey has conducted an extensive study on galaxies from $z = 1$ to $z = 1.5$ \citep{baloghGeminiObservationsGalaxies2017}, finding that the quenching mechanism in galaxies at $z > 1$ is more difficult to explain with environment dominated quenching, implying an early mass based quenching \citep{webbGOGREENSurveyPostinfall2020}. Notably, massive quenched galaxies appear to quench earlier, at $z > 2$, possibly during a pre-infall stage. This is counteracted by a larger sample of lower mass galaxies rapidly quenching due to the dense environment during infall  \citep{mcnabGOGREENSurveyTransition2021}. This time scale roughly aligns to other observational evidence of a change in quenching mechanism at $z = 1.5$ \citep{nantaisEvidenceStrongEvolution2017}, showing an increase in the strength of environmental quenching they attribute to an increase in ICM density at this time.

Effectively, the local universe has a fairly consistent environmental quenching which is independent of galaxy mass \citep{baldryGalaxyBimodalityStellar2006,pengMassEnvironmentDrivers2010, vanderburgStellarMassFunction2018} and this consistent quenching is not observed in the early universe. \citep{reevesGOGREENSurveyDependence2021} find that in distant clusters, higher mass galaxies quench with greater efficiency. This could be explained by a dominance of a mass quenching mechanism over environmental quenching. In fact, when observing a greater sample of galaxies from $z = 1.2$ to $z = 0.7$, it appears that the environmental quenching efficiency is relatively constant for galaxies of  $log(\frac{<M_*>}{M_\odot}) > 10.26$ \citep{lemauxPersistenceColourdensityRelation2019}. This implies that environmental quenching of low mass galaxies drives the change in dominant quenching mechanism, not an increased quenching efficiency of high mass galaxies, as seen within $z=0.5$ \citep{houGroupDynamicsPlay2013}. In a more recent study with a larger sample size, it seems that the impact of environmental quenching may only start to dominate after $z \approx 1.5$ \citep{trudeauMassiveDistantClusters2024}. In short, we need data on z > 1.5 clusters to further constrain the dominant quenching method\citep{strazzulloClusterGalaxiesXMMU2010}.

Galaxy clusters in the local universe have had more time to relax, accrete more mass, and heat their ICMs, providing an environment readily able to quench through ram pressure stripping or galaxy galaxy interaction, as seen in the presence of post-starburst and stripped galaxies in local clusters \citep{brownColdGasStripping2017,wattsVERTICOEnvironmentallyDriven2023,brownVERTICOVIIEnvironmental2023}. By observing galaxy clusters in the early universe of a similar ICM temperature and density, we can test to see if the observed variation in dominant quenching mechanisms is due to the natural evolution of a galaxy cluster, or an effect external to the cluster's composition. For example, the observed mass quenching could be caused by pre-infall processing \citep{fujitaPreProcessingGalaxiesEntering2004,liEVOLUTIONGROUPGALAXIES2012}.

Complicating this matter further, galaxies in some of the few distant ($z\approx 2$) clusters we have observed still possess old stellar populations \citep{willisSpectroscopicConfirmationMature2020a, trudeauMassiveDistantClusters2024,strazzulloRedSequenceBirth2016,webbGOGREENSurveyPostinfall2020}, despite existing in an era with high average star formation \citep{madauCosmicStarFormationHistory2014}, implying early quenching in cluster environments. However, acquiring large samples of galaxy clusters at high redshift has proven difficult \citep{vanderwelFundamentalPlaneField2004,gobatStarFormationHistories2008}. The concept of a cluster is also less clearly defined at earlier times. The progenitors of today's massive clusters such as Coma are predicted to exist at $z = 2$ as clusters which are a small fraction of that mass \citep{chiangAncientLightYoung2013}. To control for these effects, we want to observe and connect galaxies directly to the environment, tracing their red fraction across environmental density.

XLSSC 122 is a $z = 1.98$ galaxy cluster discovered in the XMM-Newton Large Scale Structure Survey (XMM-LSS) with X-ray emission \citep{willisDistantGalaxyClusters2013} and further supported by a Sunyaev Zeldovich decrement radio detection \citep{mantzXXLSurveyDetection2014,mantzXXLSurveyXVII2018}. XLSSC 122 also provides a large population of quenched galaxies \citep{noordehQuiescentGalaxiesVirialized2021}, Fig. \ref{fig:XLSSC122_falsecolor} shows a three colour image of the cluster core with X-ray contours from \citep{mantzXXLSurveyXVII2018}. The observed temperature of XLSSC 122's ICM is $kT = 5\pm0.7$ keV \citep{mantzXXLSurveyXVII2018}, and it may grow to be a high ($5-8\times10^{14}M_\odot$) mass cluster by $z = 0$ \citep{chiangAncientLightYoung2013}. This cluster exhibits an evolved population of red galaxies near its centre, indicating a significant evolution before the time of observation. The morphology, dark matter halo history and star formation history \citep{lesteMorphologicalAnalysisGalaxy2024} have been studied for the cluster core using previous HST cycle 25 observations. The star formation histories of these galaxies appear to extend into the very first Gyrs of the universe, as seen in \citep{trudeauXXLSurveyXLIX2022}. 

This paper presents an updated photometric and spectroscopic catalogue of cluster members including new Hubble Space Telescope cycle 30 observations of the cluster outskirts out to 10 virial radii. With this data, we can observe the properties of member galaxies across a range of environmental densities, allowing us to determine how these galaxies are affected by their infall into the cluster environment. The inclusion of fields past 10 virial radii of the cluster allow us to include a comparison to a region which is closer to the $z = 1.98$ field than a cluster environment. With a single, well characterized cluster along with a wealth of radio, x-ray, and optical observations we can investigate the influence of a cluster environment on its member galaxies, from cluster centre to outskirts. 

Section 2 describes the creation of the photometric and spectroscopic catalogues from the combined cycle 25 and cycle 30 HST data sets. Section 3 presents our findings from the dataset, highlighting the unique features of the cluster. Section 4 presents our discussion of the results and their implication for galaxy evolution in cluster environments in the early universe. Finally we summarize our findings in Section 5. 

% \subsection{cosmology}
We assume a flat $\Lambda$CDM cosmology with $H_0 = 69.32 \, \mathrm{km \, s^{-1} \, Mpc^{-1}}$, $\Omega_{\mathrm{m}} = 0.286$, and $\Omega_{\Lambda} = 0.714$ \citep{bennettNineyearWilkinsonMicrowave2013}. Magnitudes are expressed using the AB magnitude system \citep{okeAbsoluteSpectralEnergy1974}.

\section{Photometric and Spectroscopic Catalog}
The updated catalogue of cluster members for XLSSC 122 is found in Table. \ref{photocat1}. The spectroscopic cluster membership classification and photometric pipeline is described below.
\subsection{HST data} \label{HST data}
\begin{figure}
\centering
	\includegraphics[width = 0.5\textwidth]{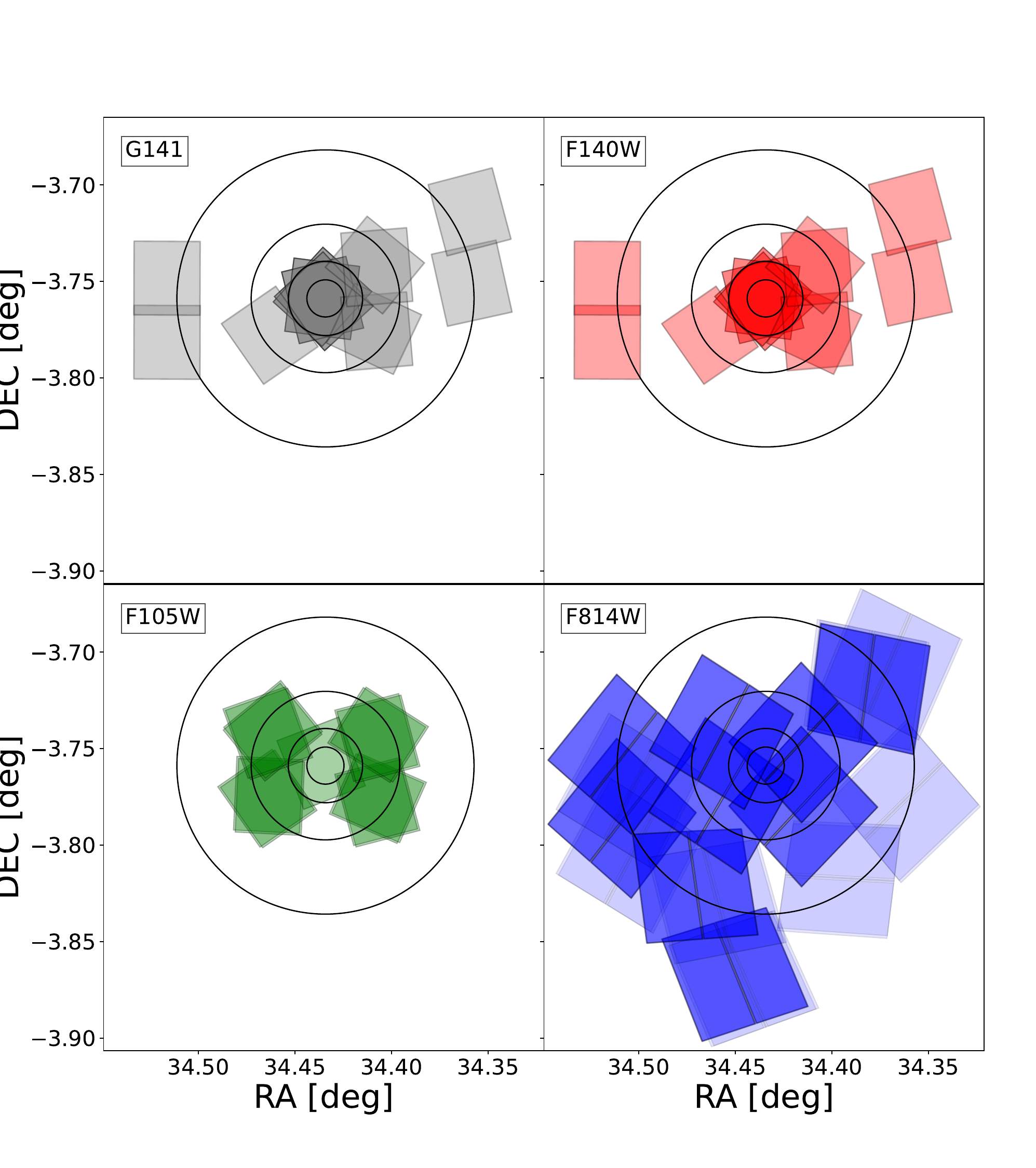}
    \caption{A map of the exposure area of the HST data used in this paper from cycles 25 and 30. Higher opacity indicates a greater number of exposures. G141 and F140W are missing exposures due to a failure to acquire guide stars in some targets. The concentric circles are centred on XLSSC 122, with radii of $R_{500}$, $2\,R_{500}$, $4\,R_{500}$, and $8\,R_{500}$}
    \label{fig:Exposure Map}
\end{figure}

We utilize observations of XLSSC 122 made using the Hubble Space Telescope's (HST) Wide Field Camera 3 (WFC3) and Advanced Camera for Surveys (ACS). The imaging used the WFC3 IR F140W and F105W and ACS F814W photometric bands. We further employ the G141 grism to extract slitless spectrography over the same area as the F140W photometry. Observations of the cluster core were taken in 2017-2018 as part of HST cycle 25 and 8 fields in the cluster outskirts were taken in 2023-2024 as part of HST cycle 30. In total, these observations cover 9 fields, with the F814W band UV exposures taken in parallel imaging mode to overlap with the F140W and F105W IR exposures. These regions cover the centre of the cluster, 4 side regions, and 4 farther regions along the same axis as seen in Fig. \ref{fig:Exposure Map}. These observations provide spectroscopic and photometric data in at least 1 band for all fields as seen in Table. \ref{tab:Exposure Table}.

\begin{table*}
    \centering
    \begin{tabular}{c|ccccccc}
        \toprule
        Target & RA [deg] & DEC [deg] & Cycle & G141EXP (s) & F105WEXP (s) & F140WEXP (s) & F814EXP$^a$ (s)\\
        \midrule
        Core & 34.4341& -3.7587& 25 & 26540 & 2611 & 5170 & 0 \\
        NE & 34.4619& -3.7432& 30 & 0 & 4235 & 0 & 3856 \\
        SE & 34.4619& -3.7764& 30 & 4011 & 4235 & 1011 & 7998 \\
        NW & 34.4094& -3.7432& 30 & 8023 & 4235 & 2023 & 12140 \\
        SW & 34.4094& -3.7764& 30 & 8023 & 4235 & 2023 & 12140 \\
        NE-Far & 34.4535& -3.7438& 30 & 3808 & 0 & 1011 & 4264 \\
        SE-Far & 34.4535& -3.7770& 30 & 3911 & 0 & 1011 & 4264 \\
        NW-Far & 34.4178& -3.7438& 30 & 4011 & 0 & 1011 & 4264 \\
        SW-Far & 34.4178& -3.7770& 30 & 4011 & 0 & 1011 & 4264 \\
        \bottomrule
 \multicolumn{8}{l}{$^a$F814W exposures are offset from their target RA and DEC}\\
    \end{tabular}
    \caption{Details of the location, cycle, and filter exposure times from the HST cycle 25 and 30 observations of XLSSC 122. Target names are shortened and paraphrased from their original full target names.}
    \label{tab:Exposure Table}
\end{table*}

The F105W, F140W, and F814W images were processed using the calwf3 reduction pipeline through the Grizli \citep{brammerGrizliGrismRedshift2019} software wrapper. The reduction included bias subtraction, dark current subtraction, flat field and gain correction, cosmic ray rejection, and a non-linearity correction. The images were inspected between each reduction step, and default parameters were used to create the calibrated images and weight maps. These calibrated images are then combined into tiled mosaics.

\subsection{Mosaics and Photometry}
Contiguous mosaics for XLSSC 122 were created using the Astrodrizzle and Tweakreg packages from DrizzlePac \citep{gonzagaDrizzlePacHandbook2012}. These drizzled images combine overlapping images on a sub-pixel scale, increasing the signal to noise ratio (SNR) and resolution. This initial drizzling did not require a thorough alignment, as images were taken as part of the same observation sequence using the same guide stars. These images were subsequently aligned to the source catalog of the drizzled core region of XLSSC 122 from \citep{willisSpectroscopicConfirmationMature2020a} created using SExtractor \citep{bertinSExtractorSoftwareSource1996a}. As each image was added to the mosaic, its astrometric points were added to a running catalogue and used to align the next image. Final mosaics and inverse variance weight maps (IVM) for the F140W filter were created for three contiguous overlapping exposure regions as seen in the F140W filter exposure map footprint area (Fig. \ref{fig:Exposure Map}), with relative astrometry within each contiguous region. A similar mosaic in F105W was created for the central region. The final mosaics and IVMs for the F814W band were created for overlap with the F140W contiguous regions, despite their span over the entire field. A false colour RGB image of the central region of XLSSC 122 combining these contiguous mosaics is shown in Fig. \ref{fig:XLSSC122_falsecolor}

Initial catalogs using single image mode from SExtractor were extracted for each contiguous mosaic for each filter. Overlapping mosaics in different filters were matched with a tolerance of $0.3$ arcseconds, or 5 pixels on the F140W mosaics to find the mean and standard deviation of the shift between objects in the mosaics. Mosaics were then shifted by the mean difference in sky positions between filters to match to the F140W mosaic after an iterative mean cut. This changed the reference point, the distance scale, and rotation of the image. Finally, aligned mosaics in F105W and F814W were resampled using SWARP \citep{bertinSWarpResamplingCoadding2010} to match the pixel scale and location of the F140W mosaics. This ensured that pixel positions and sizes match between mosaics.

Photometric information was extracted from each region by applying SExtractor to the mosaics and corresponding IVM image. Fluxes were computed from two apertures, a 13 pixel ($0.78\, \mathrm{arcsecond}$) fixed aperture and a variable elliptical aperture with a Kron factor of $0.8$ as used in \citep{willisSpectroscopicConfirmationMature2020a}. Extractions for F105W and F814W were made using SExtractor double image mode, using the F140W image as reference for astrometric position. In this way the aperture size and shape are consistent between mosaics. 

\subsection{Spectral Extraction \& Classification}  \label{Spectral Extraction}
We extracted spectroscopic information from the grism images using Grizli. The G141 spectroscopic images were calibrated using a similar process to the photometric images using Grizli and the calwf3 software as described in Sec. \ref{HST data}. Once calibrated, the G141 images images were then aligned to F140W images taken over the same fields. To do this, the intermediate F140W mosaics for each pointing were matched with each G141 image primarily overlapping that field. Then, a polynomial based contamination model was fit to each G141 image. By treating the spectroscopic extraction on a field by field basis, we sought to reduce the uncertainty added to the object position introduced by creating the mosaics between HST cycles. Our interest here is the matching of the position of the object in the F140W field to the G141 fields, and the SNR of the photometric images in the overlap regions was not our primary concern. We acknowledge that this prevents the combination of beams from overlapping fields due to the shift in their relative positions. Beams of width 20 pixels were extracted along the dispersion angle of each object and were averaged into a single one-dimensional spectral trace.

We compute spectroscopic redshifts, focusing on $z = 1.98$, using the G141 filter. The range of the G141 filter captures multiple rest frame optical features of $z = 2$ objects, including the the 4000 \AA \ break and the [OIII]5007 \AA  \ emission line.  Three template sets were fit to each object's beams, a continuum based set containing galaxy templates and absorption features in young and old stellar populations, the same continuum based set with three additional line complex templates (OII+Ne, OIII+Hb+Hg+Hd, and Ha+NII+SII+SIII+He+PaB), and the continuum based set with several additional individual line features including individual versions of each of the complex lines. The line features used include the Balmer emission and absorption lines, along with metal lines including oxygen, silicon, and neon. Template fitting was performed using a grid between $z =0-4.0$, with an initial course grid of $\delta z = 0.004$ and a finer grid of $\delta z = 0.0005$ around local minima. Based on the result of these 3 template fits hereafter referred to as guess 1, guess 2, and guess 3, objects were visually inspected for best template fit and initial redshift guess. Objects were then placed into categories based on a number of selection cuts. These categories divided the objects based on reliability of the guess. Objects of category A and B are bright, uncontaminated, high SNR objects with a redshift guess within $0.1$ of $z = 1.98$, only differing in whether the continuum guess and line complex guess agreed. Category C lacks a redshift guess within $0.1$ of $z = 1.98$, but has one its second or third lowest local minima in the $\chi^2$ distribution in that range. Categories D and E were carefully inspected, containing either high contamination from nearby sources, or low SNRs. Category F, G, H, and I objects were not  inspected as they contained none of their top three local minima in the $\chi^2$ distribution near $z = 1.98$, were dimmer than 25th magnitude in F140W, lie in the stellar locus in FWHM/magnitude space, or have a SNR  less than one respectively.

Objects which did not appear to be contaminated and had fits roughly approximating the data were then processed through a final template fit using the third template set containing individual lines with a Gaussian prior about the best initial guess. This was to identify galaxies which fit well at $z = 1.98$. If these galaxies could not be fit to $z = 1.98$ with a boost to the likelihood, we attempted a similar fit at different local minima in the initial $\chi^2$ distribution. The objects which could be fit by a template combination at $z = 1.98$ were then passed through the same fitting process with a tophat prior between $z = 0.5$ and $z = 3.5$ to obtain a fit unbiased from the Gaussian prior. These final two template fits were performed using a Markov Chain Monte Carlo (MCMC) method using the emcee python package \citep{foreman-mackeyEmceeMCMCHammer2013} with 256-512 steps and 128-196 walkers.

Objects were then visually inspected for goodness of fit, and rerun at varying initial positions, beam sizes, and beam selections to reduce contamination from nearby sources. A spectral SNR was computed between $1.3$ and $1.55\,\mathrm{\mu m}$ as in \citep{willisSpectroscopicConfirmationMature2020a}.  Then, objects with an SNR < 5 were inspected for the presence or absence of emission lines. Objects without emission features at this level of noise were discarded. Then, objects were divided into Gold, Silver, and Bronze categories based on the integral of the redshift probability density function ($p_{memb}$) over the range $1.96 < z < 2.00$. With Gold defined as $p_{memb} > 0.5$, and Silver defined as $p_{memb} > 0.1$. The Bronze category for XLSSC 122 includes objects which would lie in the Gold or Silver categories except for their low SNR. There is a structure of objects that fit to $z = 1.93$ in the vicinity of XLSSC 122. We conduct a similar fitting process for these objects, changing only the central redshift from $z = 1.98$ to $z = 1.93$, and incorporate the probability of membership in this second structure into the classification of XLSSC 122 objects to avoid contamination. The details of the final categories for both structures are described in Table. \ref{Specmemb}, we do not include Bronze membership for the $1.93$ structure. We identify $63$ gold members, $11$ silver members, and $8$ bronze members, adding $34$ members in the gold and silver categories in addition to those found in \citep{willisSpectroscopicConfirmationMature2020a}.

\begin{table}
    \centering
    \caption{Spectroscopic membership criteria for clusters in XLSSC 122 and \foreground. Note, no cut is made on clustercentric distance as we are including infalling galaxies as far away as possible.}
    \label{Specmemb}
    \begin{tabular}{lccc}
        \toprule
        Category & $p_{\text{memb},1}$ & $p_{\text{memb},2}$ & S/N\\ 
        \midrule
        Gold & $\geq0.5$ & $<0.5$ & $>3$\\ 
        Silver & $\geq0.1,<0.5$ & $<0.5$ & $>3$\\
        Bronze & $\geq0.1$ & $<0.5$ & $>1,\leq3$\\
        Foreground Gold & $<0.5$ & $>0.5$ & $>3$ \\
        Foreground Silver & $<0.1$ & $>0.1$ & $>3$ \\
        NonMember & $<0.1$ & $<0.1$ & $>3$ \\
        \bottomrule
    \end{tabular}
\end{table}

\subsection{Spectroscopic Incompleteness}
\begin{figure}
	\includegraphics[width = 0.48\textwidth]{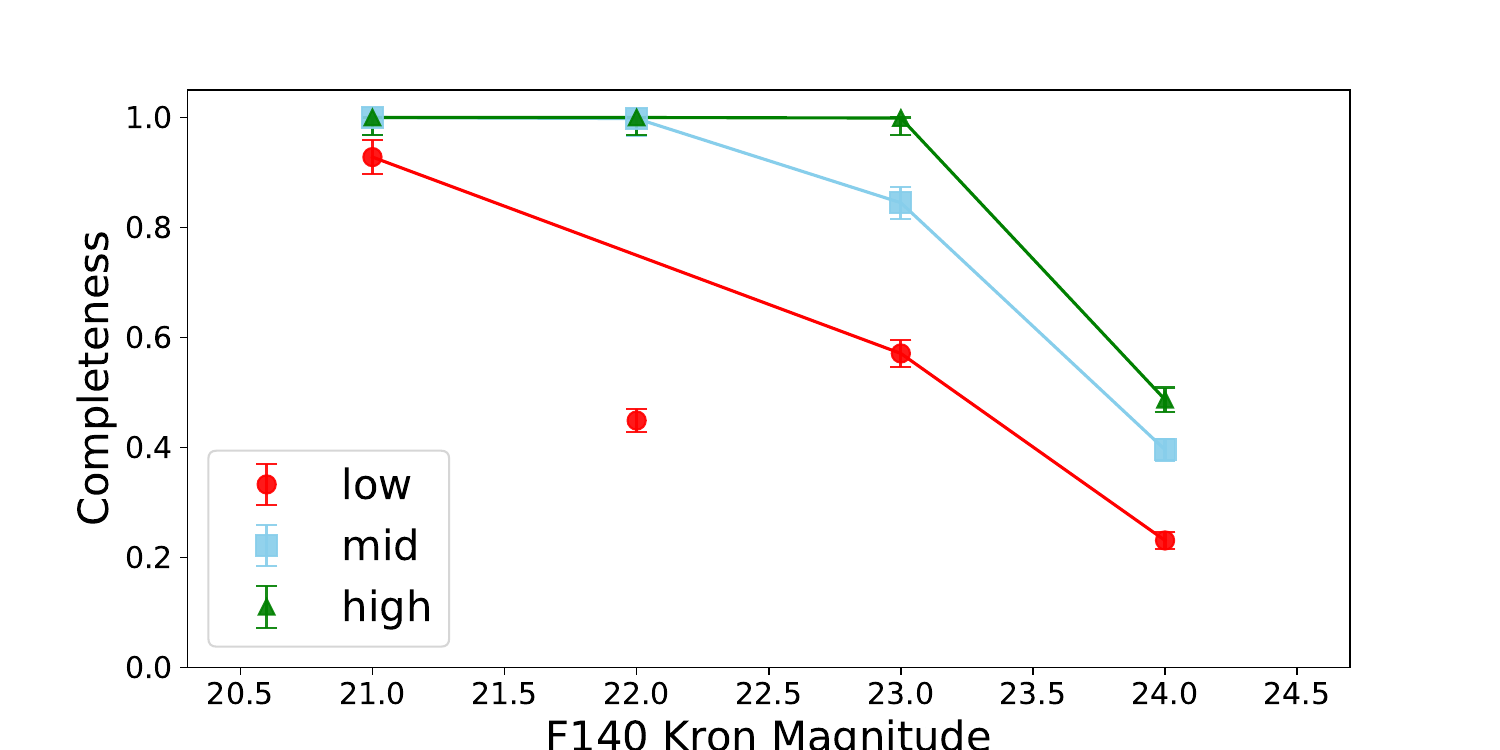}
    \caption{Simulated spectroscopic incompleteness across magnitudes and exposure time regions. Low exposure time regions are those with $\mathrm{exptime} < 4000\,\mathrm{s}$ from Table. \ref{tab:Exposure Table}, mid exposure time regions are those with $\approx 8000\,\mathrm{s}$, the only high exposure time region is the cycle 25 G141 exposure stack with a total of $\approx 26,500\,\mathrm{s}$.}
    \label{fig:spectroincomp}
\end{figure}

We estimate the spectroscopic incompleteness on the identification of cluster galaxies by creating mock G141 spectra and passing them through the spectral classification pipeline. We use a single template of an old stellar population included in the set of templates we use for our fits \citep{brammerEAZYFastPublic2008a}. We note that for areas with large numbers of blue galaxies, such as the outskirts, this does underestimate the completeness, as star forming galaxies with emission lines have a higher detection rate. We estimate the background noise, object size, amplitude, and extent for the two cycles and three exposure regions. The G141 data taken in cycle 25 was collected over 48 exposures of t $\approx 550\,\mathrm{s}$ each. The data taken in cycle 25 was collected over 4 or 8 exposures of t $\approx 1000\,\mathrm{s}$ each, leading to a different contribution of noise per beam. To estimate the amplitude of the signal for each beam, we extract the 1 dimensional spectra for 4 sample objects of F140W magnitudes 21 through 24 for both cycle 25 and cycle 30. We scale the amplitude of each simulated beam such that its integrated 1 dimensional flux is equal to a real object of that magnitude. We then pass these simulated beams through our classification pipeline, stopping at the initial $\chi^2$ grid search, using only the continuum templates. We classify any galaxy passing the local minimum test as being successfully identified. A plot of the spectroscopic completeness for the 3 exposure time regions across magnitude is shown in Fig. \ref{fig:spectroincomp}. During this process, low exposure time objects at 22nd magnitude had an unexpectedly low rate of recovery. We were unable to find a suitable reference spectra for 22nd magnitude that was uncontaminated. As such we interpolate over 22nd magnitude in the low exposure time region. Inclusion of the point does not significantly alter the findings presented in the rest of this work.

\section{Results}

\begin{figure*}
	\includegraphics[width=\textwidth]{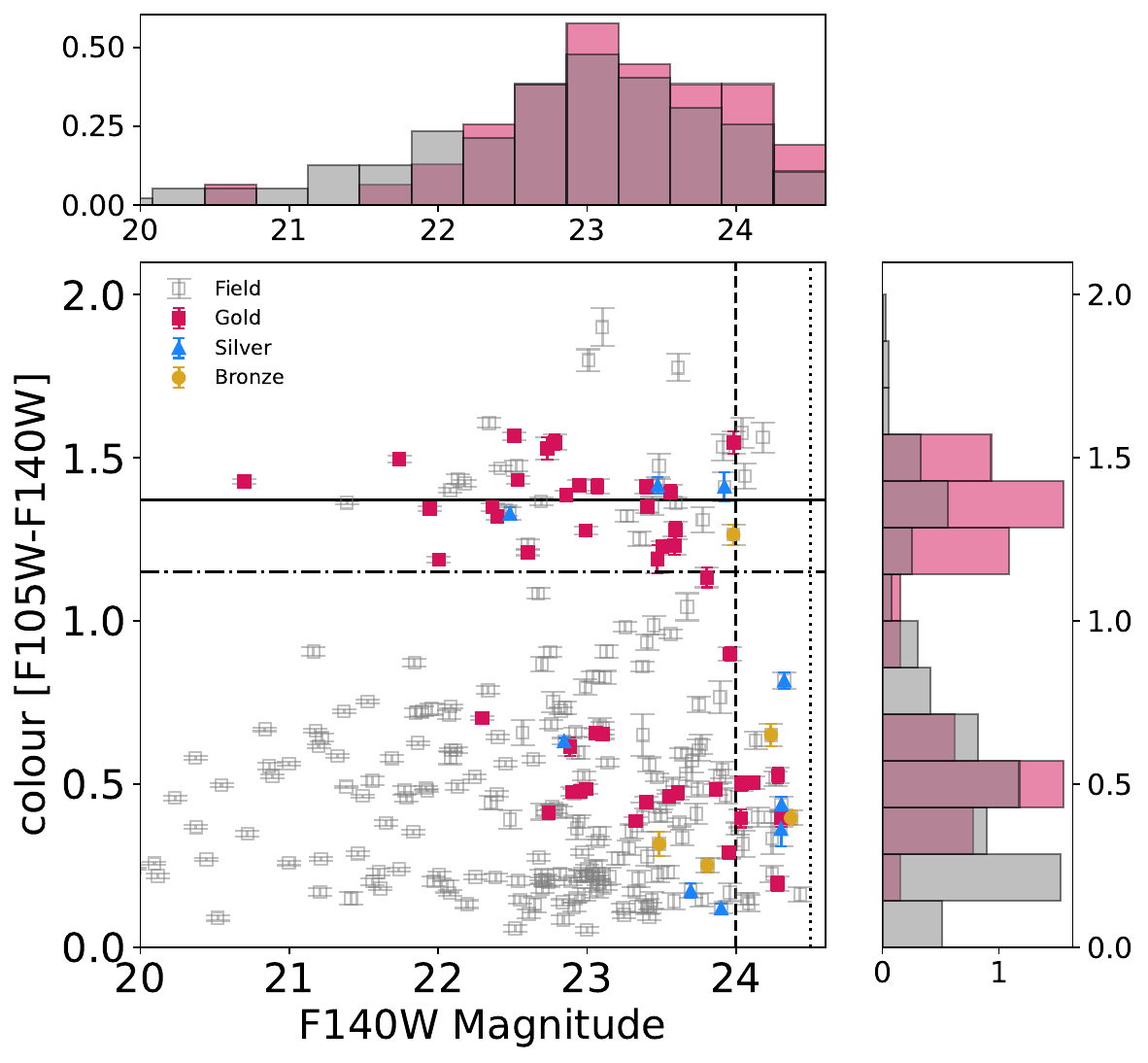}
    \caption{A Colour-Magnitude Diagram of the full XLSSC 122 photometric field with colour SNR > 5. Red, blue, and yellow points are gold, silver, and bronze members respectively. Membership is spectroscopically estimated within $z = (1.96,2.00)$. Grey points represent all photometrically extracted objects in the area which have not been spectroscopically classified as cluster members. Vertical lines designate the $24$ and $24.5$ magnitude limits on cluster membership from prior core observations. The dash-dot horizontal line at $1.15$ in colour indicates the red-sequence blue-cloud cut. The solid horizontal black line indicates the mean colour of the gold red sequence members. The histograms indicate the normalized number density of cluster members (red) vs non-cluster-members (grey) in intervals of $0.25$ over colour and magnitude space corresponding to the CMD. Past F140W = 23.5 spectroscopic completeness drops quickly, especially in low exposure time regions, which is shown in the scarcity of spectroscopically observed objects around F140W = 24.}
    \label{fig:XLSSC122_CMD_F105-F140}
\end{figure*}

\begin{figure*}
	\includegraphics[width=\textwidth]{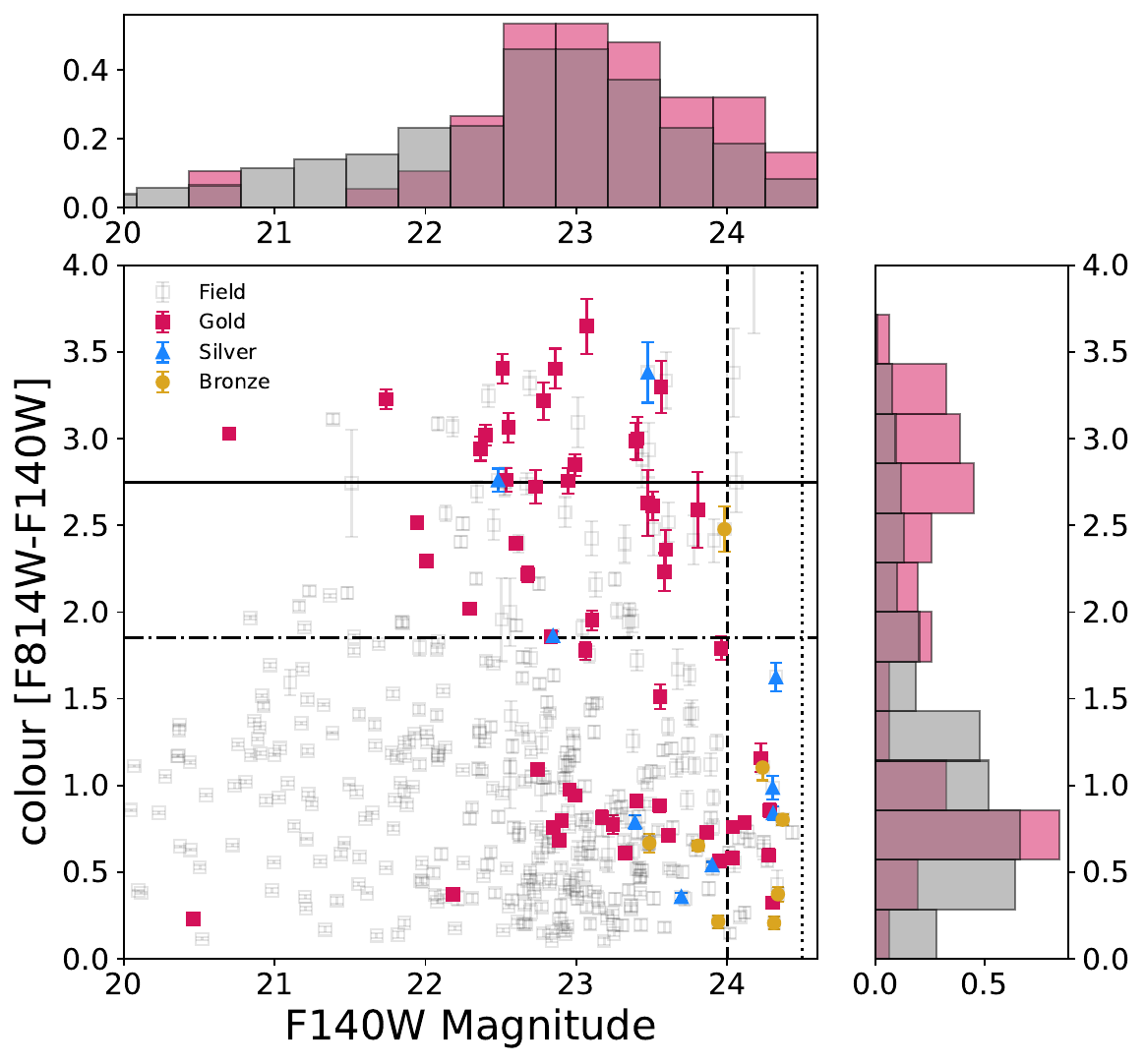}
    \caption{A Colour-Magnitude Diagram of the full XLSSC 122 photometric field with colour SNR > 5. Red, blue, and yellow points are gold, silver, and bronze members respectively. Membership is spectroscopically estimated within $z = (1.96,2.00)$. Grey points represent all photometrically extracted objects in the area which have not been spectroscopically classified as cluster members. Vertical lines designate the $24$ and $24.5$ magnitude limits on cluster membership from prior core observations. The dash-dot horizontal line at $1.85$ in colour indicates the red-sequence blue-cloud cut from \protect\cite{noordehQuiescentGalaxiesVirialized2021}. The solid horizontal black line indicates the mean colour of the gold red sequence members. The histograms indicate the normalized number density of cluster members (red) vs non-cluster-members (grey) in intervals of $0.25$ over colour and magnitude space corresponding to the CMD. Past F140W = 23.5 spectroscopic completeness drops quickly, especially in low exposure time regions, which is shown in the scarcity of spectroscopically observed objects around F140W = 24.}
    \label{fig:XLSSC122_CMD_F814-F140}
\end{figure*}

XLSSC 122 shows a strong bimodality in the colour of member galaxies. We characterize the red fraction and density estimates out to several virial radii. 

\subsection{Colour Bimodality and Red Fraction} \label{Red Fraction}
\begin{figure*}
	\includegraphics[width = \textwidth]{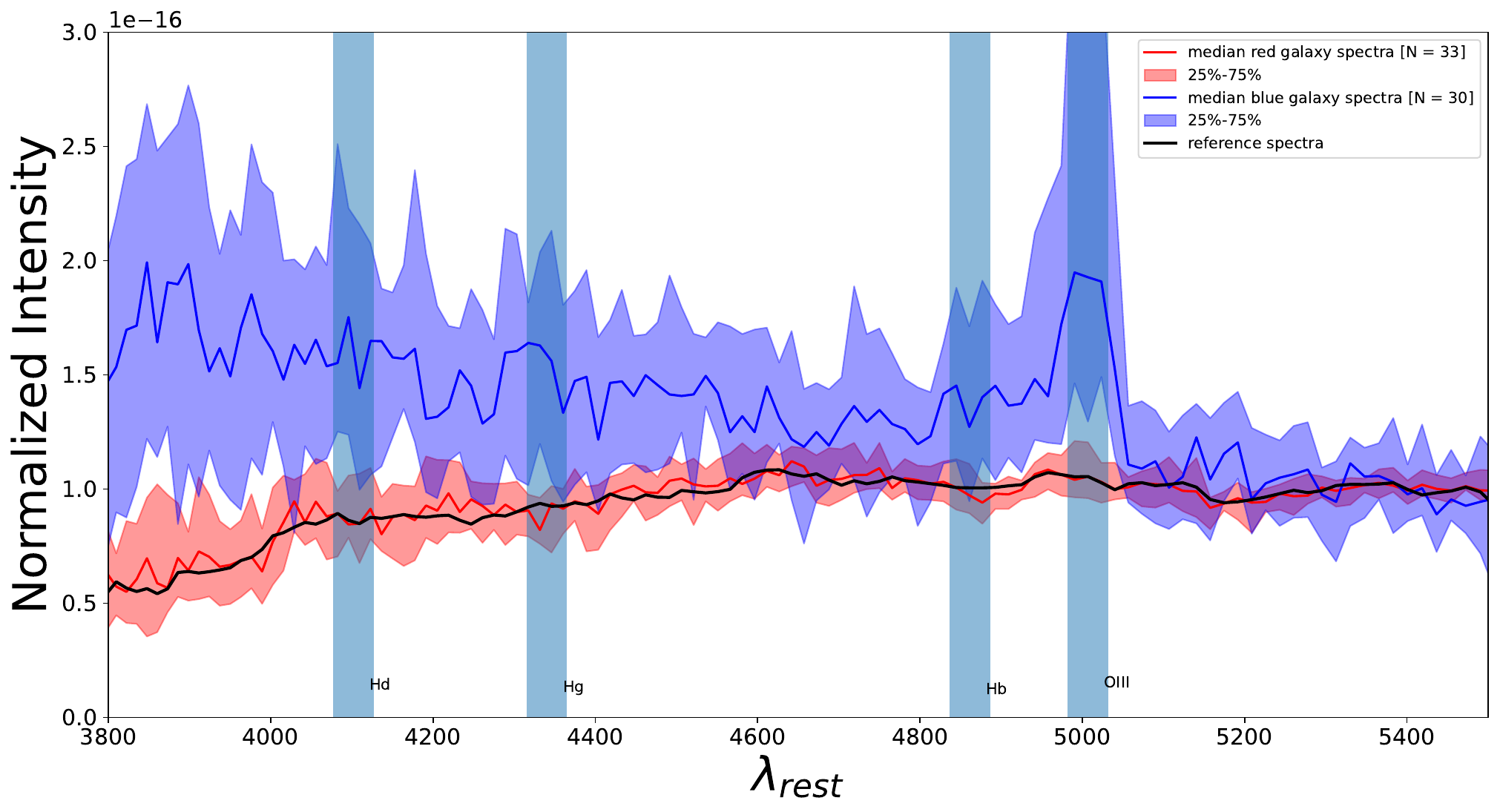}
    \caption{The sensitivity corrected and normalized spectra. The median red and blue galaxy spectra as divided by a cut of $\mathrm{F814W-F140W} = 1.85$ for all $63$ gold cluster members of XLSSC 122 are shown in red and blue respectively. All spectra are normalized to the BCG of XLSSC 122 (in black) at 5300\,\AA. Shaded regions indicate the first and third quartiles.}
    \label{fig:median rvb main}
\end{figure*}

The colour as a function of magnitude is shown in Fig. \ref{fig:XLSSC122_CMD_F105-F140} and Fig. \ref{fig:XLSSC122_CMD_F814-F140}, where we can identify the red and blue populations, and very few intermediate colour galaxies, consistent with other studies of high redshift galaxies \citep{strazzulloRedSequenceBirth2016}. These two populations are distinct and identifiable with a simple colour cut of $\mathrm{F105W-F140W} = 1.15$ and less distinct but separable with a colour cut of $\mathrm{F814W-F140W} = 1.85$ as used in \citep{noordehQuiescentGalaxiesVirialized2021} for the CANDELS field, where the arbitrary colour cut of $1.85$ is selected with their analysis relatively unchanged for a cut $0.2$ redder or bluer. Going forward we will be using both colours, F105W-F140W has a more distinct red and blue separation, while F814W-F140W has coverage over all fields out to larger radii. The clear separation between colours, especially in F105W-F140W, which separates the red and blue starlight at $z = 1.98$, shows us two distinct stages in the star formation histories of the member galaxies. We see this in Fig. \ref{fig:median rvb main}, where we show the median spectra between the red and blue populations. These spectra have been normalized to a common mean intensity in the rest frame 5200 Angstrom (\AA) range. We find a median rest frame [3800-4000\,\AA] region $\approx 3$ times higher in the blue population than the red. The region where we expect to find young blue stars. We also observe the clear presence of O[III]5007\,\AA  \ emission features twice as large as the underlying continuum of the red and blue galaxies. It seems fair to say that these populations are drawn from two distinct stellar populations, the blue population containing younger stars, and the red containing older stars. The O[III] emission features indicate star formation.

\begin{figure}
	\includegraphics[width = 0.48\textwidth]{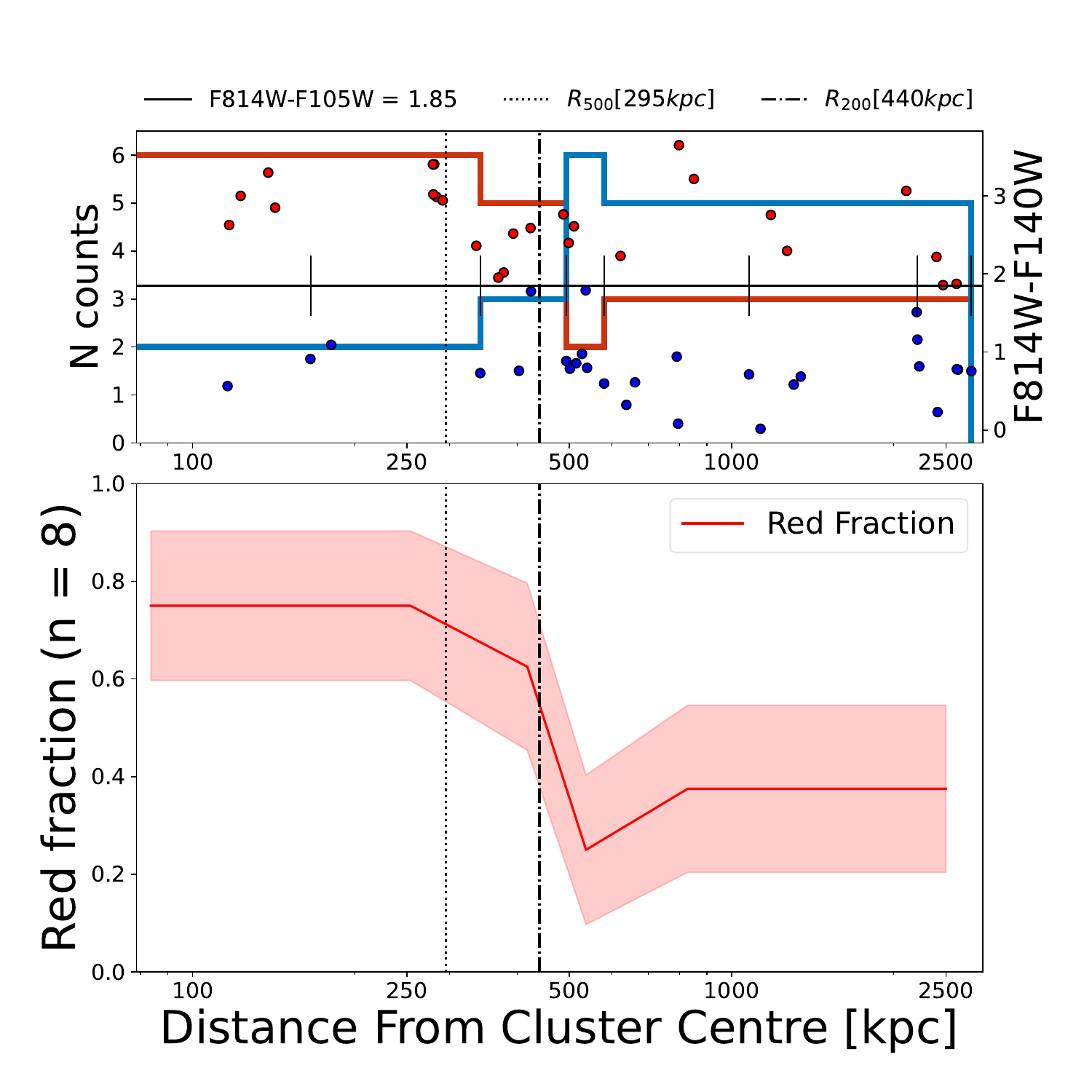}
    \caption{\textit{Top}: Galaxy colour and distance from cluster centre, with red and blue points indicating red and blue galaxies as defined by a $1.85$ colour cut on F814W-F140W. Galaxies are binned by colour, with bins chosen such that each bin contains 8 galaxies. Bin edges indicated by small black vertical lines along the solid colour cut line. \textit{Bottom}: The red fraction of galaxies in XLSSC 122 as a function of distance from the cluster centre. The shaded region indicates the 1 $\sigma$ binomial uncertainty.}
    \label{fig:red_frac_over_radius}
\end{figure}

We have observed a change in $f_R$ from $0.44$ to $0.78$ in the relative number distribution of red and blue galaxies in XLSSC 122 across clustercentric radii. The red fraction as a function of clustercentric radius is shown in Fig. \ref{fig:red_frac_over_radius} with the two colour cut populations, along with a scattering of the points in colour-radius space. We use a  constant number bin method to estimate red fraction, which bridges the different scales in number density over radius. We note that this red fraction varies with radius as seen in \citep{noordehQuiescentGalaxiesVirialized2021} for galaxies in XLSSC 122. The red fraction increases within the virial radius from $0.44\pm0.17$ in the outskirts to $0.78\pm0.14$ in the core. Red fraction here is calculated using only galaxies with $F140W \leq 24$. As an outside comparison with a non-cluster field, \citep{noordehQuiescentGalaxiesVirialized2021} uses the CANDELS field \citep{koekemoerCANDELSCosmicAssembly2011,brammer3DHSTWidefieldGrism2012,skelton3DHSTWFC3selectedPhotometric2014,momcheva3DHSTSurveyHubble2016}, an HST field with galaxies in the $z = 1.9-2.1$ range taken with the F814W and F140W filters. The CANDELS field quenched fraction is reported as $0.20\pm0.02$ in \citep{noordehQuiescentGalaxiesVirialized2021} when split along $\mathrm{F814W-F140W} = 1.85$, which is lower than the quenched fraction observed at the very outskirts of XLSSC 122. However, the outer edges of galaxy clusters at $z > 1.4$ appear to experience quenching enhancements relative to galaxies in lower density environments \citep{wernerSatelliteQuenchingWas2022}. The CANDELS field red fraction also uses a magnitude cut of $F140W \leq 24$.

\subsection{Cluster Membership and Spatial Distribution} \label{Clustermemb}
\begin{figure*}
	\includegraphics[width = \textwidth]{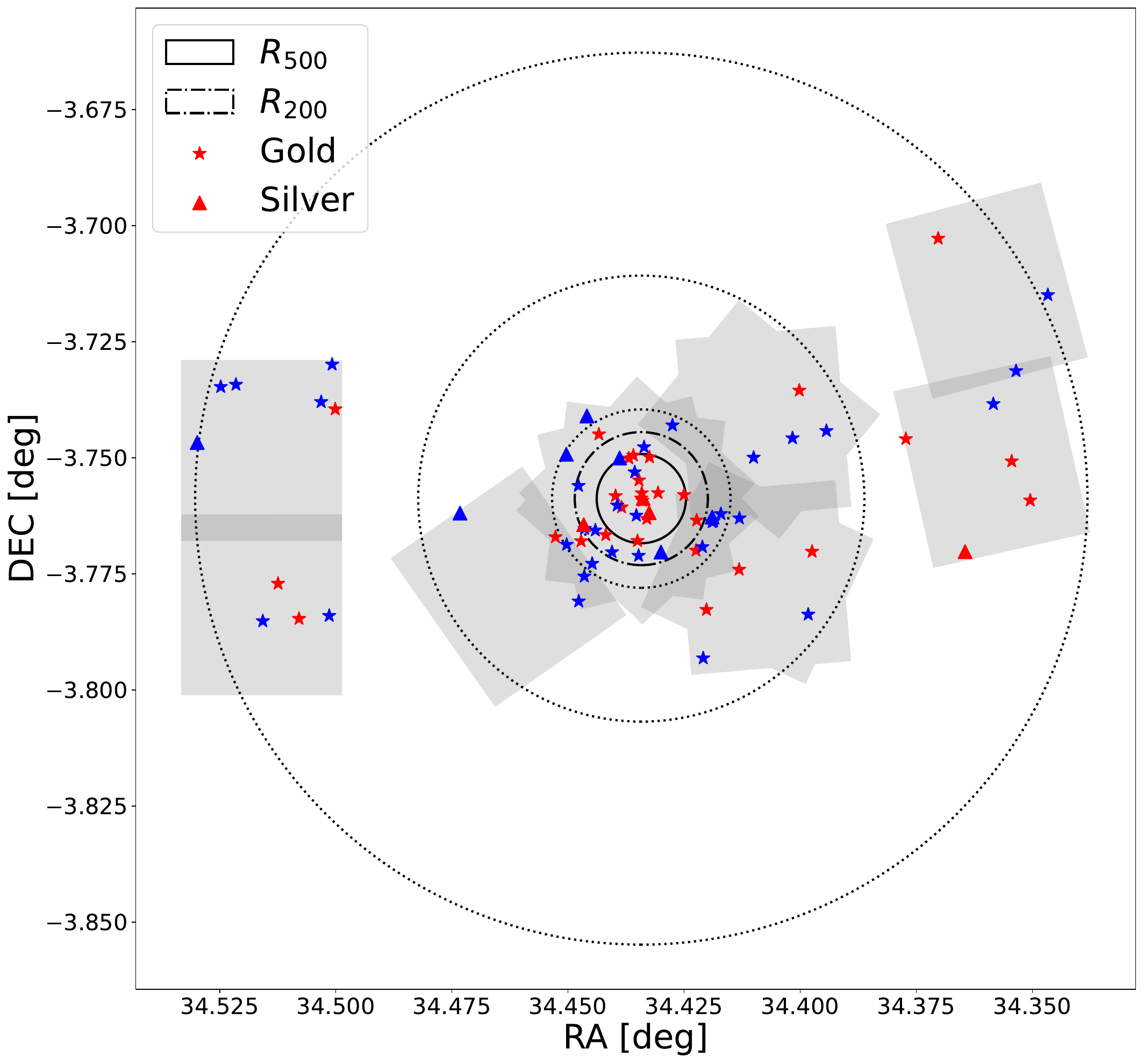}
    \caption{A map of XLSSC 122's cluster members. The grey outlines indicate regions with G141 coverage, where spectroscopic cluster membership can be determined. Gold and silver members are shown as stars and triangles respectively. Cluster members are coloured red or blue based on the colour cuts in $\mathrm{F105W-F140W} = 1.15$ in the central region, and $\mathrm{F814W-F140W} = 1.85$ in the far regions. The two innermost solid and dash-dot circles are $R_{500}$ and $R_{200}$ respectively. The dotted circles indicate, from inside to out, 2, 5, and 10 times $R_{500}$.}
    \label{fig:clustermap}
\end{figure*}

The new galaxies are primarily in the regions observed by the most recent cycle 30 observations, but include a few galaxies in the cycle 25 region. This is because we have re-reduced the cycle 25 observations for consistency between objects. The approximate doubling of cluster members was about what we expected given that the galaxy number density falls at larger clustercentric distances. This density falloff  can be observed in our map of the objects divided by category found in Fig. \ref{fig:clustermap}, which shows the distribution of cluster galaxies across the F140W observing fields. The highest concentration of cluster members still lies in the centre, corresponding to the cycle 25 data taken in 2017/2018. There is a gradual drop-off of object density into the cycle 30 data, with fewer objects in the farther fields. The falloff, though due in part to the decreasing density of cluster members with larger distance from cluster centre, is compounded with the reduced depth of the exposures in those regions. We estimate the total number of objects we would expect to miss with a magnitude cut of $F140W < 24.5$. We expect to be missing $12\%$ of cycle 25 objects, $23\%$ in the next nearest cycle 30 exposure regions, and $55\%$ in the farthest and lowest cycle 30 exposure time regions, including the south east region.

An NFW profile describes a system of collisionless particles, which we use to model the density of the cluster galaxy population.
% \begin{equation}\label{double power law}
%     \sigma_{DPL}(r) = \frac{f_0}{(\frac{1}{x^a})+(\frac{s}{x^b})}+c, x = \frac{r}{R_s}
% \end{equation}

\begin{equation}\label{3dNFW}
\scriptsize
\rho_{NFW}(x) = \frac{1}{x(1+x)^2}+c
\end{equation}

% \begin{equation}\label{NFW profile}
%     \sigma_{NFW}(r) = f_0\int_R^\infty
%     \rho_{NFW}(r)\cdot  \frac{r}{(r^2-R^2)}dr+c
% \end{equation}
\begin{equation}\label{PiecewiseNFW}
\scriptsize
\sigma_{NFW}(x) =
\begin{cases}

\dfrac{f_0}{x^{2} - 1}
\left[
1 - \dfrac{2}{\sqrt{1 - x^{2}}}
\,\operatorname{arctanh}
\!\left(
\sqrt{\dfrac{1 - x}{1 + x}}
\right)
\right]+c,
& 0 \le x < 1 \\[1.25em]
\dfrac{f_0}{3} + c,
& x = 1 \\[1.25em]
\dfrac{f_0}{x^{2} - 1}
\left[
1 - \dfrac{2}{\sqrt{x^{2} - 1}}
\,\arctan
\!\left(
\sqrt{\dfrac{x - 1}{1 + x}}
\right)
\right]+c,
& x > 1
\end{cases}
\end{equation}

The NFW surface density at radius r expressed by $\sigma_{NFW}(r)$ is the integrated NFW 3 dimensional profile \citep{navarroUniversalDensityProfile1997}  computed analytically using equation 8 from \cite{lazarAnalyticSurfaceDensity2024a}. Here $\rho_{NFW}(r)$ is given in Eq. \ref{3dNFW} with scale radius $R_s$ and normalizing constant $f_0$. For both equations, c is an additive density constant and x = $\frac{r}{R_s}$.

\begin{figure}
	\includegraphics[width = 0.48\textwidth]{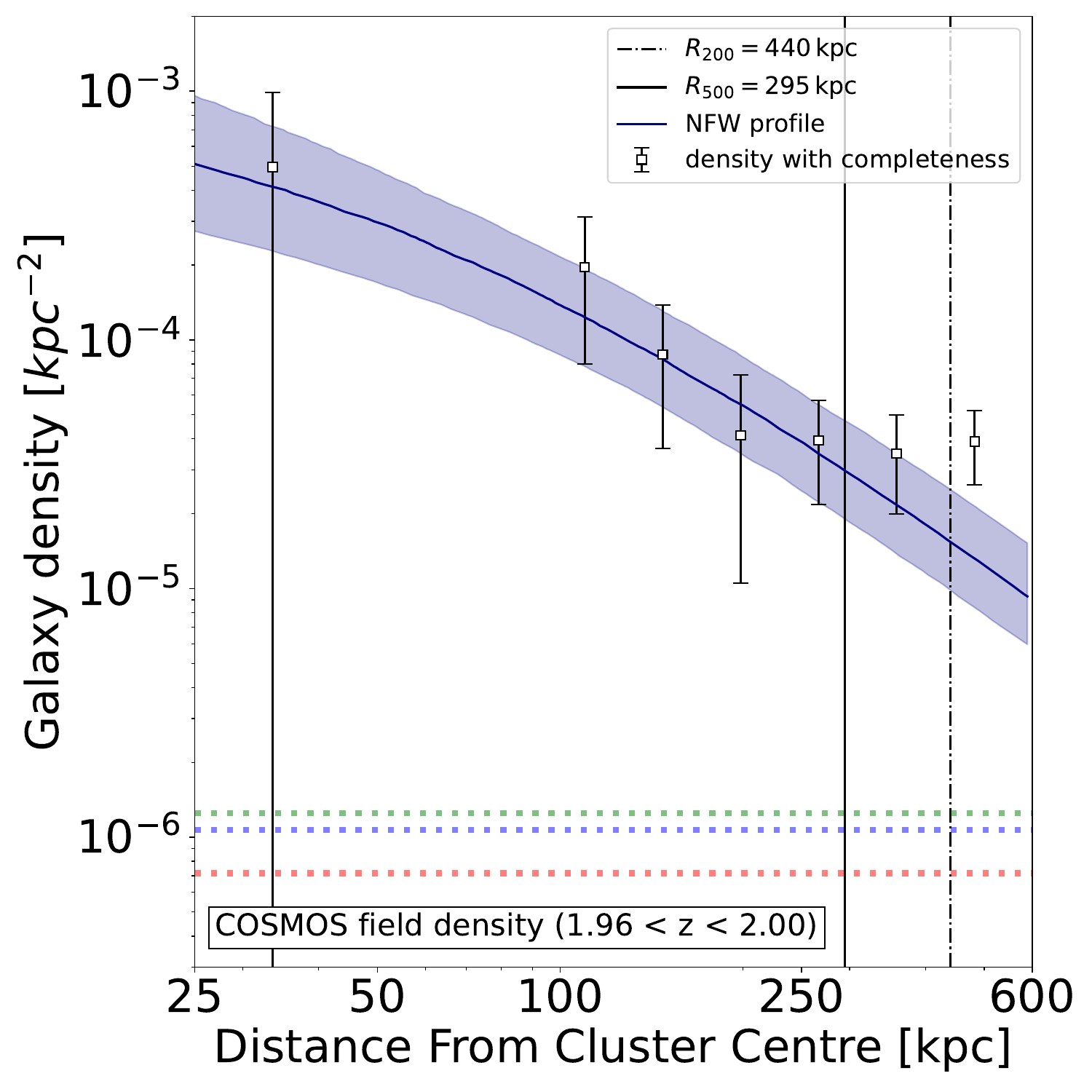}
    \caption{The number density of objects in XLSSC 122 over the distance from the cluster centre, taking into account both missing exposure region and incompleteness. A projected NFW profile \citep{navarroUniversalDensityProfile1997} (navy) is fit using a $\chi^2$ goodness-of-fit grid-search. The shaded regions indicate 16th and 84th model percentiles of the full probability space, while the solid line indicates the minimum of the $\chi^2$ distribution. Overlayed are $R_{500} = 295\,\mathrm{kpc}$, $2R_{500} = 590\,\mathrm{kpc}$  and $R_{200} = 440 \,\mathrm{kpc}$. Errors are based on propagated $\sqrt{N}$ counting uncertainty.  Horizontal dotted lines indicate the completeness corrected background galaxy densities expected from the COSMOS field \citep{brammer3DHSTWidefieldGrism2012,skelton3DHSTWFC3selectedPhotometric2014}. Green, blue, and red indicate high, medium, and low exposure time incompleteness curves shown in Fig. \ref{fig:spectroincomp}.}
    \label{fig:densityvRad_profilefits}
\end{figure}

We compute the overall density of objects per unit area over fixed radius interval bins in Fig. \ref{fig:densityvRad_profilefits} and fit an NFW profile to $2R_{500} = 590\, kpc$. This radius was chosen to restrict our profile fit to the area with highest spectroscopic completeness. The fit was done using a $\chi^2$ goodness of fit method over a grid of input parameters. We find an NFW profile fit with no significant local overdensity. The  mean scale radius from the marginalized probability distribution (given by $P = \exp(-\chi^2/2)/\sum{exp(-\chi^2/2)}$) with 16th and 84th percentile uncertainties is $R_s = 110^{+63}_{-65}\, kpc$.

\subsection{Red Sequence Luminosity Function} \label{LF}
\begin{figure*}
    \centering
	\includegraphics[width = \textwidth]{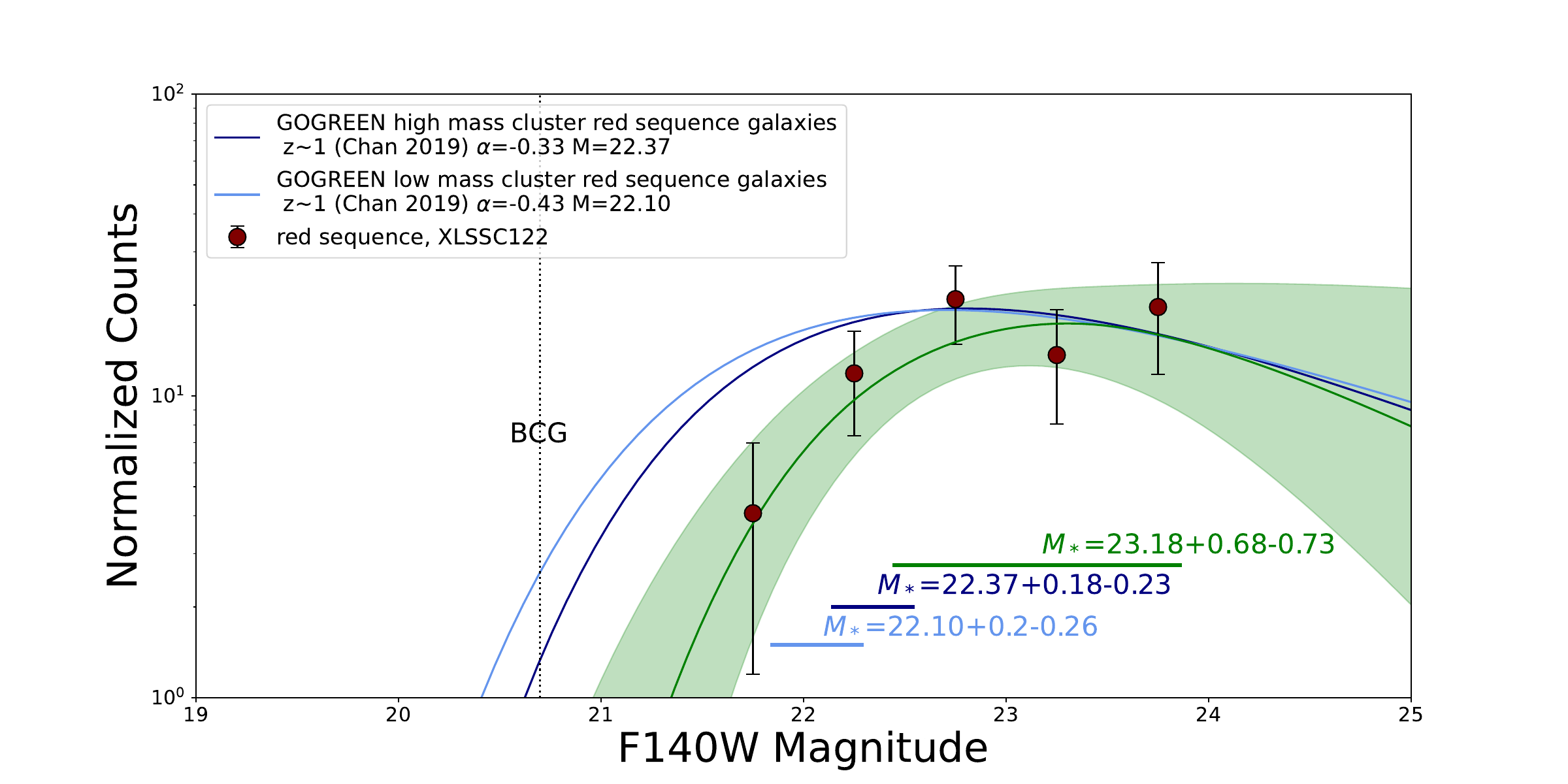}
    \caption{The F140W band completeness and area corrected rest magnitude luminosity (maroon) fit with the  \protect\cite{schechterAnalyticExpressionLuminosity1976} function as given in Eq. \ref{schecter} (green). The solid green line is the minimum of the $\chi^2$ distribution. The green shaded region spans the 16th to 84th percentiles of the Schecter $\chi^2$ goodness-of-fit full probability space. Curves are normalized for comparison to low and high mass luminosity functions for redshift 1 clusters from \protect\cite{chanRestframe$H$bandLuminosity2019}.}
    \label{fig:LF}
\end{figure*}

We measure the luminosity function in XLSSC 122's red cluster members and provide a comparison to the luminosity function of other clusters at $z \approx 1$ from \citep{chanRestframe$H$bandLuminosity2019} in Fig. \ref{fig:LF}. We perform a correction between the F140W [1.4 $\mu$] filter and the rest frame F160W [1.5 $\mu$] HST filter to estimate the H band used in \citep{chanRestframe$H$bandLuminosity2019}. We find a correction factor including both distance modulus and k correction of -45.767 from $m_{\text{F140W},{z=2}}$ to $M_{\text{F160W},{z=0}}$ using an old stellar population template from \citep{brammerEAZYFastPublic2008a}. The number of counts at each magnitude is the sum of magnitudes in the three different exposure time regions. We normalize by area to account for the varying counting area of the three regions. Each magnitude count is then scaled by a factor of $\frac{1}{c}$, where c is the interpolated completeness at that magnitude and exposure time. We then fit a Schecter function in magnitude space shown in Eq. \ref{schecter} \citep{schechterAnalyticExpressionLuminosity1976}. 

\begin{equation}\label{schecter}
\scriptsize
    \Phi(M_i) = \Phi_0\cdot(0.4\cdot ln(10))\cdot (10^{0.4(M-M_i)}))^{\alpha+1}\cdot \exp(-10^{0.4\cdot(M-M_i)})
\end{equation}

Where M is the scale magnitude, $\Phi_0$ is a normalizing factor, and $\alpha$ is the dim slope. We use a $\chi^2$ goodness of fit test sampled over a grid and approximate uncertainties on each parameter by marginalizing the normalized posterior probabilities given by $P = \exp(-\chi^2/2)/\sum{exp(-\chi^2/2)}$. Uncertainties are then the 16th to 84th percentiles of the resulting probability distribution. The resulting mean posterior fits with uncertainties are $\alpha = 0.03^{+0.98}_{-0.95}$ and $M_* =23.18^{+0.68}_{-0.73}$.

Comparing this luminosity function to the stellar mass function of galaxies in the GOGREEN survey \citep{vanderburgGOGREENSurveyDeep2020}, we find a low mass slope, $\alpha$, within uncertainty for quiescent galaxies in cluster environments in their sample. This downward slope does extend into a region of high incompleteness (<50\% for the medium and low exposure time regions past $\mathrm{F140W} = 23.5$) and is sensitive to small variations in the count or completion correction. The scale magnitude for XLSSC 122's Schecter fit is fainter by about 1 magnitude. Passive evolution alone would predict a brighter magnitude in the older XLSSC 122 sample \citep{chanRestframe$H$bandLuminosity2019}. The dominant effect here may be the difference in mass between the two cluster samples and the k correction. The clusters in \cite{chanRestframe$H$bandLuminosity2019} have mass $M_{200} = 1\times10^{14}M_\odot-7.2\times 10^{14}M_\odot$. If we convert these to $M_{500}$ using the mass concentration relation from \cite{duffyDarkMatterHalo2008} assuming a relatively unchanging NFW profile with concentration c $\approx$ 3, $M_{500}\approx0.6M_{200}$. Thus, the clusters in \cite{chanRestframe$H$bandLuminosity2019} are about 1-7 times the mass of XLSSC 122 with $M_{500} = 6.3\times10^{13}M_{\odot}$ \citep{mantzXXLSurveyXVII2018}. The k correction also assumed a single red galaxy template. The k correction varies between red and blue galaxies from $-45.3$ and $-46$ when using all available templates used to fit redshifts for cluster members \citep{brammerEAZYFastPublic2008a}. However, the 0.25 difference between the maximum and chosen k correction is not enough to entirely explain the difference in scale magnitude.

\subsection{A second galaxy cluster at z = 1.93}
A significant, physically unrelated, overdensity of galaxies was found in \citep{willisSpectroscopicConfirmationMature2020a} located at $z = 1.93$ near the edge of the cycle 25 data. This object was not specifically identified in the X-ray or radio observations made of XLSSC 122 \citep{mantzXXLSurveyXVII2018}, however a review of the X-ray contours does show a small excess of x-ray emission in the direction of this second cluster. In this structure we identify a total of $36$ gold members and $6$ silver members across the cycle 25 and cycle 30 HST data. We decided on a barycentric approach for the cluster centre, taking the mean position of the $12$ nearest galaxies to the brightest galaxy in the central region. There is, similar to XLSSC 122, a red fraction enhancement in the cluster core. 
With a larger picture of this sample, we are confident in naming this a cluster with designation \foreground. We follow a similar analysis of \foreground \ as for XLSSC 122, analyzing the colour bimodality and red fraction of member galaxies. We also present a catalogue of the \foreground \ cluster members alongside the updated catalogue of XLSSC 122 in Table. \ref{Foregroundcat}. 

We find that the colour bimodality using the F105W and F140W filters is preserved between XLSSC 122 and \foreground. What differs is the difference in the fraction of red to blue galaxies. XLSSC 122 has a red fraction of $0.5\pm 0.13$, compared to \foreground's $0.32\pm 0.13$ over the same exposure area and magnitude cut (F140W = 24.5) as XLSSC 122.

\begin{landscape}
    \begin{table}
    \centering
    \caption{The reduced catalogue of photometric and spectroscopic information for the galaxy cluster XLSSC122. F140W kron magnitude, colours from 13 px (0.78 arcsecond) circular apertures, and colour errors are in AB magnitude.}
    \label{photocat1}
    \begin{tabular}{c %id
    c %RA
    c %DEC
    c %z
    cccccccc}
        \toprule
        \hline
         id & {RA [deg]} & {Dec [deg]} & {Redshift} & CM & M [F140W] & $M_{\sigma} [F140W]$ & colour [F105-F140] & $\textrm{colour}_{\sigma} [F105-F140]$ & colour [F814-F105] & $\textrm{colour}_{\sigma} [F814-F105]$ \\
        \midrule
        \hline
        181 & 34.5157 & -3.7852 & 1.967 & G & 23.24 & 0.03 &...&...&...&...\\
186 & 34.5079 & -3.7847 & 1.994 & G & 22.68 & 0.02 &...&...&...&...\\
202 & 34.5014 & -3.784 & 1.974 & G & 23.56 & 0.04 &...&...&...&...\\
291 & 34.5124 & -3.7771 & 2.006 & G & 22.83 & 0.02 &...&...&...&...\\
603 & 34.5298 & -3.7467 & 2.0 & S & 23.39 & 0.03 &...&...&...&...\\
644 & 34.5215 & -3.7342 & 1.976 & G & 22.85 & 0.02 &...&...&...&...\\
776 & 34.5008 & -3.7299 & 1.958 & G & 23.17 & 0.03 &...&...&...&...\\
808 & 34.5248 & -3.7347 & 1.999 & G & 22.18 & 0.01 &...&...&...&...\\
811 & 34.5001 & -3.7395 & 1.968 & G & 22.55 & 0.02 &...&...&...&...\\
823 & 34.5032 & -3.738 & 1.968 & G & 24.23 & 0.05 &...&...&...&...\\
848 & 34.3645 & -3.7702 & 2.003 & S & 22.37 & 0.02 &...&...&...&...\\
934 & 34.3505 & -3.7592 & 1.974 & G & 22.51 & 0.02 &...&...&...&...\\
1012 & 34.3545 & -3.7507 & 1.976 & G & 22.02 & 0.02 &...&...&...&...\\
1066 & 34.3773 & -3.7459 & 1.992 & G & 22.28 & 0.03 &...&...&...&...\\
1141 & 34.3584 & -3.7384 & 1.993 & G & 20.46 & 0.01 &...&...&...&...\\
1230 & 34.3535 & -3.7313 & 1.987 & G & 22.57 & 0.02 &...&...&...&...\\
1396 & 34.3467 & -3.7149 & 1.974 & G & 23.17 & 0.03 &...&...&...&...\\
1499 & 34.3703 & -3.7028 & 1.981 & G & 21.07 & 0.01 &...&...&...&...\\
1654 & 34.4209 & -3.7932 & 1.986 & G & 23.3 & 0.02 & 0.08 & 0.03 & -0.06 & 0.02 \\
1735 & 34.4118 & -3.7894 & 1.989 & B & 23.94 & 0.03 & 0.08 & 0.03 & 0.14 & 0.03 \\
1894 & 34.3983 & -3.7838 & 1.995 & G & 22.89 & 0.01 & 0.61 & 0.03 & 0.07 & 0.03 \\
1910 & 34.4202 & -3.7827 & 1.99 & G & 22.78 & 0.01 & 1.55 & 0.02 & 1.67 & 0.11 \\
1988 & 34.4477 & -3.7809 & 1.975 & G & 23.69 & 0.03 & 0.06 & 0.03 & 0.02 & 0.01 \\
2082 & 34.4791 & -3.7775 & 1.988 & B & 23.48 & 0.03 & 0.32 & 0.04 & 0.35 & 0.04 \\
2146 & 34.4465 & -3.7755 & 1.98 & G & 24.3 & 0.03 & 0.4 & 0.03 & -0.07 & 0.02 \\
2176 & 34.4132 & -3.7741 & 1.98 & G & 23.07 & 0.01 & 1.41 & 0.02 & 2.24 & 0.16 \\
2235 & 34.4448 & -3.7728 & 1.981 & G & 22.9 & 0.01 & 0.48 & 0.01 & 0.32 & 0.02 \\
2276 & 34.4348 & -3.7711 & 1.98 & G & 22.29 & 0.01 & 0.7 & 0.01 & 1.32 & 0.02 \\
2322 & 34.4405 & -3.7703 & 1.978 & G & 24.04 & 0.02 & 0.5 & 0.02 & 0.26 & 0.03 \\
2333 & 34.4224 & -3.77 & 1.997 & G & 22.6 & 0.01 & 1.21 & 0.01 & 1.19 & 0.03 \\
2334 & 34.3974 & -3.7702 & 1.997 & G & 22.95 & 0.01 & 1.42 & 0.02 & 1.34 & 0.08 \\
2336 & 34.43 & -3.7703 & 2.017 & S & 23.9 & 0.01 & 0.12 & 0.01 & 0.42 & 0.02 \\
2359 & 34.4211 & -3.7692 & 1.982 & G & 24.28 & 0.02 & 0.53 & 0.02 & 0.33 & 0.04 \\
2397 & 34.4503 & -3.7687 & 1.967 & G & 24.28 & 0.02 & 0.2 & 0.02 & 0.4 & 0.03 \\
2411 & 34.435 & -3.7679 & 1.997 & G & 22.51 & 0.01 & 1.57 & 0.01 & 1.84 & 0.09 \\
2414 & 34.4472 & -3.768 & 1.986 & G & 22.53 & 0.01 & 1.43 & 0.01 & 1.33 & 0.07 \\
2442 & 34.4419 & -3.7666 & 1.98 & G & 23.59 & 0.01 & 1.28 & 0.03 & 1.08 & 0.11 \\
2453 & 34.4276 & -3.7674 & 1.974 & B & 24.37 & 0.02 & 0.4 & 0.02 & 0.41 & 0.03 \\
2456 & 34.4527 & -3.7671 & 1.99 & G & 23.59 & 0.02 & 1.23 & 0.03 & 1.0 & 0.11 \\
2479 & 34.4434 & -3.7666 & 2.098 & B & 24.24 & 0.02 & 0.65 & 0.03 & 0.45 & 0.07 \\
2514 & 34.4441 & -3.7656 & 1.971 & G & 23.1 & 0.01 & 0.65 & 0.01 & 1.3 & 0.06 \\
2515 & 34.4233 & -3.7658 & 2.013 & B & 23.98 & 0.02 & 1.26 & 0.03 & 1.21 & 0.13 \\
2540 & 34.4464 & -3.7653 & 1.993 & G & 23.06 & 0.01 & 0.66 & 0.01 & 1.12 & 0.05 \\
2568 & 34.4466 & -3.7644 & 2.028 & S & 23.92 & 0.02 & 1.41 & 0.04 & 4.3 & 4.58 \\
2573 & 34.419 & -3.7638 & 1.978 & G & 23.55 & 0.02 & 0.46 & 0.02 & 0.42 & 0.03 \\
2590 & 34.4223 & -3.7635 & 1.981 & G & 21.94 & 0.0 & 1.35 & 0.01 & 1.17 & 0.03 \\
2605 & 34.4187 & -3.7637 & 1.982 & G & 24.12 & 0.02 & 0.5 & 0.02 & 0.28 & 0.03 \\
2633 & 34.433 & -3.7631 & 1.977 & G & 23.56 & 0.01 & 1.39 & 0.02 & 1.9 & 0.15 \\
                                                   
        \bottomrule
        \hline
    \end{tabular}
    \end{table}
\end{landscape}
\begin{landscape}
    \begin{table}
    \centering
    \label{photocat2}
    \begin{tabular}{c %id
    c %RA
    c %DEC
    c %z
    cccccccc}
        \toprule
        \hline
         id & {RA [deg]} & {Dec [deg]} & {Redshift} & CM & M [F140W] & $M_{\sigma} [F140W]$ & colour [F105-F140] & $\textrm{colour}_{\sigma} [F105-F140]$ & colour [F814-F105] & $\textrm{colour}_{\sigma} [F814-F105]$ \\
    \midrule
    \hline
    2648 & 34.4131 & -3.763 & 1.979 & G & 23.33 & 0.01 & 0.39 & 0.01 & 0.22 & 0.01 \\
2652 & 34.419 & -3.7628 & 1.968 & S & 24.32 & 0.02 & 0.82 & 0.02 & 0.81 & 0.08 \\
2659 & 34.4342 & -3.7587 & 1.981 & G & 20.7 & 0.0 & 1.43 & 0.01 & 1.6 & 0.03 \\
2678 & 34.4353 & -3.7624 & 1.968 & G & 23.95 & 0.02 & 0.29 & 0.02 & 0.27 & 0.02 \\
2683 & 34.4171 & -3.7621 & 1.982 & G & 23.96 & 0.02 & 0.9 & 0.02 & 0.89 & 0.07 \\
2687 & 34.4325 & -3.7618 & 2.032 & S & 23.47 & 0.01 & 1.42 & 0.03 & 1.97 & 0.18 \\
2699 & 34.4733 & -3.7619 & 2.001 & S & 24.3 & 0.05 & 0.36 & 0.05 & 0.62 & 0.05 \\
2721 & 34.4394 & -3.7603 & 1.995 & G & 23.4 & 0.01 & 0.44 & 0.02 & 0.47 & 0.03 \\
2722 & 34.4385 & -3.7607 & 1.973 & G & 22.99 & 0.01 & 1.28 & 0.01 & 1.57 & 0.06 \\
2771 & 34.434 & -3.7592 & 1.954 & G & 22.73 & 0.01 & 1.53 & 0.03 & 1.19 & 0.1 \\
2775 & 34.4341 & -3.7576 & 1.969 & G & 21.74 & 0.01 & 1.5 & 0.01 & 1.73 & 0.06 \\
2781 & 34.4338 & -3.7588 & 2.027 & S & 22.48 & 0.01 & 1.33 & 0.02 & 1.43 & 0.07 \\
2826 & 34.4398 & -3.7582 & 1.985 & G & 23.98 & 0.01 & 1.55 & 0.03 & 73.43 & 99.0 \\
2827 & 34.4306 & -3.7576 & 2.0 & G & 23.47 & 0.02 & 1.19 & 0.04 & 1.44 & 0.19 \\
2830 & 34.425 & -3.758 & 1.974 & G & 23.4 & 0.01 & 1.41 & 0.02 & 1.57 & 0.11 \\
2887 & 34.4477 & -3.756 & 1.965 & G & 23.8 & 0.02 & 1.13 & 0.03 & 1.45 & 0.22 \\
2912 & 34.4347 & -3.7548 & 1.988 & G & 23.41 & 0.01 & 1.35 & 0.02 & 1.65 & 0.12 \\
2958 & 34.4356 & -3.7531 & 1.967 & G & 22.74 & 0.01 & 0.41 & 0.01 & 0.68 & 0.01 \\
3037 & 34.4325 & -3.7499 & 1.96 & G & 22.4 & 0.01 & 1.32 & 0.01 & 1.7 & 0.06 \\
3049 & 34.4369 & -3.7501 & 1.979 & G & 22.86 & 0.01 & 1.39 & 0.01 & 2.02 & 0.12 \\
3050 & 34.41 & -3.75 & 1.993 & G & 22.99 & 0.01 & 0.49 & 0.01 & 0.46 & 0.01 \\
3053 & 34.4389 & -3.75 & 2.009 & S & 24.31 & 0.02 & 0.44 & 0.02 & 0.4 & 0.04 \\
3058 & 34.4359 & -3.7495 & 1.974 & G & 22.36 & 0.01 & 1.35 & 0.01 & 1.59 & 0.07 \\
3064 & 34.4504 & -3.7492 & 1.96 & S & 22.85 & 0.01 & 0.63 & 0.01 & 1.23 & 0.02 \\
3114 & 34.4336 & -3.7477 & 1.969 & G & 23.87 & 0.01 & 0.48 & 0.02 & 0.25 & 0.02 \\
3161 & 34.4017 & -3.7458 & 1.986 & G & 23.61 & 0.02 & 0.47 & 0.02 & 0.24 & 0.03 \\
3188 & 34.4434 & -3.7449 & 1.973 & G & 23.51 & 0.01 & 1.23 & 0.02 & 1.38 & 0.08 \\
3203 & 34.3944 & -3.7442 & 1.994 & G & 24.03 & 0.03 & 0.4 & 0.03 & 0.19 & 0.03 \\
3222 & 34.4275 & -3.743 & 1.966 & G & 22.96 & 0.01 & 0.48 & 0.01 & 0.5 & 0.02 \\
3273 & 34.4089 & -3.7412 & 1.957 & B & 24.31 & 0.03 & 0.1 & 0.03 & 0.11 & 0.03 \\
3283 & 34.4459 & -3.741 & 2.012 & S & 23.7 & 0.02 & 0.17 & 0.02 & 0.18 & 0.01 \\
3383 & 34.4002 & -3.7355 & 1.976 & G & 22.01 & 0.01 & 1.19 & 0.01 & 1.11 & 0.02 \\
3407 & 34.4016 & -3.7347 & 1.967 & B & 24.34 & 0.03 & 0.07 & 0.03 & 0.3 & 0.03 \\
3457 & 34.417 & -3.7307 & 1.994 & B & 23.81 & 0.02 & 0.25 & 0.02 & 0.4 & 0.02 \\
    \bottomrule
    \end{tabular}
    \end{table}
\end{landscape}
\begin{landscape}
    \begin{table}
    \centering
    \caption{The reduced catalogue of photometric and spectroscopic information for the galaxy cluster \foreground. F140 kron magnitude, colours from 13 px (0.78 arcsecond) circular apertures, and colour errors are in AB magnitude.}
    \label{Foregroundcat}
    \begin{tabular}{c %id
    c %RA
    c %DEC
    c %z
    ccccccc}
        \toprule
        \hline
         id & {RA [deg]} & {Dec [deg]} & {Redshift} & CM & M [F140W] & $M_{\sigma} [F140W]$ & colour [F105-F140] & $\textrm{colour}_{\sigma} [F105-F140]$ & colour [F814-F105] & $\textrm{colour}_{\sigma} [F814-F105]$ \\
\midrule
345 & 34.502 & -3.7708 & 1.932 & FG & 23.09 & 0.03 &...&...&...&...\\
481 & 34.5085 & -3.7567 & 1.925 & FG & 22.37 & 0.02 &...&...&...&...\\
536 & 34.511 & -3.7521 & 1.935 & FG & 23.7 & 0.03 &...&...&...&...\\
687 & 34.5179 & -3.7386 & 1.911 & FG & 22.92 & 0.02 &...&...&...&...\\
786 & 34.5031 & -3.73 & 1.93 & FG & 22.46 & 0.02 &...&...&...&...\\
789 & 34.5093 & -3.731 & 1.918 & FG & 21.81 & 0.01 &...&...&...&...\\
892 & 34.3593 & -3.7624 & 1.945 & FG & 20.98 & 0.01 &...&...&...&...\\
946 & 34.3697 & -3.7578 & 1.918 & FG & 21.11 & 0.01 &...&...&...&...\\
985 & 34.343 & -3.7537 & 1.92 & FG & 22.0 & 0.04 &...&...&...&...\\
1038 & 34.3514 & -3.7487 & 1.925 & FG & 22.75 & 0.02 &...&...&...&...\\
1173 & 34.37 & -3.7364 & 1.936 & FG & 21.29 & 0.01 &...&...&...&...\\
1432 & 34.3664 & -3.7117 & 1.936 & FG & 21.92 & 0.01 &...&...&...&...\\
1457 & 34.3615 & -3.7089 & 1.904 & FG & 21.1 & 0.01 &...&...&...&...\\
1891 & 34.3985 & -3.7837 & 1.941 & FG & 22.79 & 0.01 & 0.42 & 0.01 & 0.62 & 0.02 \\
1947 & 34.4213 & -3.7822 & 1.95 & FS & 24.08 & 0.03 & 0.13 & 0.03 & 0.11 & 0.02 \\
2015 & 34.4271 & -3.7797 & 1.92 & FG & 23.66 & 0.03 & 0.53 & 0.04 & 0.57 & 0.05 \\
2097 & 34.4533 & -3.7729 & 1.942 & FG & 19.11 & 0.0 & 0.2 & 0.01 & 0.15 & 0.0 \\
2148 & 34.4597 & -3.7746 & 1.941 & FS & 22.35 & 0.02 & 0.44 & 0.02 & 0.65 & 0.03 \\
2178 & 34.3951 & -3.7747 & 1.928 & FG & 24.13 & 0.02 & 0.63 & 0.03 & 0.1 & 0.04 \\
2194 & 34.3946 & -3.7725 & 1.927 & FG & 22.06 & 0.01 & 0.19 & 0.02 & 0.18 & 0.02 \\
2253 & 34.4569 & -3.7727 & 1.968 & FS & 22.94 & 0.02 & 0.02 & 0.02 & 0.17 & 0.01 \\
2256 & 34.4438 & -3.7719 & 1.922 & FG & 22.25 & 0.01 & 0.52 & 0.01 & 0.57 & 0.01 \\
2275 & 34.4226 & -3.7717 & 1.941 & FG & 23.62 & 0.01 & 0.38 & 0.01 & 0.14 & 0.01 \\
2518 & 34.4232 & -3.7661 & 1.937 & FG & 24.28 & 0.02 & 0.4 & 0.02 & 0.19 & 0.02 \\
2554 & 34.4316 & -3.7645 & 1.913 & FG & 22.25 & 0.01 & 0.22 & 0.01 & 0.34 & 0.01 \\
2598 & 34.4195 & -3.7626 & 1.946 & FG & 21.39 & 0.0 & 1.36 & 0.0 & 1.75 & 0.03 \\
2612 & 34.416 & -3.7631 & 1.933 & FG & 22.08 & 0.01 & 1.4 & 0.01 & 1.68 & 0.04 \\
2630 & 34.4743 & -3.7636 & 1.926 & FS & 23.73 & 0.03 & 0.07 & 0.03 & 0.05 & 0.02 \\
2655 & 34.4265 & -3.7625 & 1.942 & FG & 22.69 & 0.01 & 1.37 & 0.01 & 1.95 & 0.07 \\
2666 & 34.4602 & -3.7625 & 1.928 & FG & 23.09 & 0.02 & 0.35 & 0.02 & 0.36 & 0.01 \\
2675 & 34.4205 & -3.7614 & 1.916 & FG & 22.42 & 0.01 & 1.47 & 0.01 & 1.78 & 0.06 \\
2700 & 34.4297 & -3.7616 & 1.935 & FG & 23.89 & 0.02 & 0.52 & 0.02 & 0.29 & 0.03 \\
2715 & 34.4172 & -3.7613 & 1.933 & FG & 23.6 & 0.01 & 1.36 & 0.02 & 1.97 & 0.16 \\
2727 & 34.4178 & -3.76 & 1.944 & FG & 22.77 & 0.01 & 0.43 & 0.01 & 0.29 & 0.01 \\
2728 & 34.449 & -3.7605 & 1.947 & FG & 23.37 & 0.01 & 0.86 & 0.02 & 1.09 & 0.07 \\
2769 & 34.4218 & -3.7589 & 1.944 & FG & 22.08 & 0.0 & 0.74 & 0.01 & 0.45 & 0.01 \\
2785 & 34.4362 & -3.7594 & 1.963 & FG & 24.18 & 0.02 & 1.56 & 0.05 & 2.77 & 0.73 \\
2804 & 34.4538 & -3.7584 & 1.93 & FG & 23.77 & 0.02 & 0.62 & 0.03 & 0.79 & 0.07 \\
3015 & 34.4456 & -3.7514 & 1.933 & FG & 23.47 & 0.01 & 0.52 & 0.01 & 0.48 & 0.02 \\
3018 & 34.4519 & -3.751 & 1.936 & FG & 23.64 & 0.02 & 0.34 & 0.02 & 0.42 & 0.03 \\
3065 & 34.4002 & -3.7491 & 1.912 & FS & 22.94 & 0.02 & 0.6 & 0.02 & 0.59 & 0.03 \\
3350 & 34.408 & -3.7367 & 1.878 & FS & 22.69 & 0.01 & 0.22 & 0.02 & 0.29 & 0.02 \\
\bottomrule
\hline
\end{tabular}
\end{table}
\end{landscape}

\section{Discussion}
We discuss here the observed trend in galaxy properties as a function of radius and their implications for galaxy evolution in massive clusters in the early universe. Notably, the radius at which the red fraction increases rapidly aligns with an increase in expected ram pressure stripping. We observe no nearby local overdensities at $z = 1.98$. XLSSC 122 appears to be a mature galaxy cluster exhibiting quenching processes and a galaxy luminosity distribution similar to clusters at lower redshift shifted by about 1 magnitude fainter.

\subsection{Lack of Filamentary Structure or Infalling Groups in XLSSC 122}\label{structure}

Within a filamentary structure of dark matter, containing galaxies tracing its extent \citep{bondHowFilamentsGalaxies1996a}, one might expect to see streams or clumps of galaxies entering into a cluster environment over time, as observed in the local group \citep{kraljicGalaxyEvolutionMetric2018}, and out to $z = 1.1$ \citep{okabeHaloConcentrationGalaxy2019}. These groups would be local overdensities of galaxies offset from the cluster centre, as seen in \cite{dresslerIMACSClusterBuilding2013}. Before entering into a cluster environment, these galaxies would already exist in a range of overdense regions, experiencing a cluster-like pre-processing, or accelerated evolution \citep{fujitaPreProcessingGalaxiesEntering2004}.  In the local universe, environmental pre-processing in filaments and groups prior to infall into a cluster has been observed \citep{mcnabGOGREENSurveyTransition2021}.

We find only a small overdensity at 500 kpc which is only $1-2\sigma$ from the fit profiles. Considering the low number sample of galaxies and photometric and spectroscopic measurements taken in multiple exposure time ranges, we did not consider a $2\sigma$ overdensity sufficient to identify if there is a substructure falling in to XLSSC 122. A spherically symmetric galaxy number density profile in XLSSC 122 could indicate that XLSSC 122 is a mature cluster, with central member galaxies in a halo more than infalling filaments. However, we note that at larger radii, past $R_{200}$, an infalling group or filament could exist within a region with lower completeness.

\subsection{Rapid Quenching and Ram Pressure Stripping}\label{structure}
In Fig. \ref{fig:red_frac_over_radius}, we notice a rapid transition from blue to red galaxies around the virial radius of the cluster. This bears resemblance to the quenching of galaxies upon infall into clusters in the local universe\citep{wernerSatelliteQuenchingWas2022}. To investigate whether the observations in XLSSC 122 are consistent with this scenario, we follow the model of ram pressure stripping given in \cite{mccarthyRamPressureStripping2008}. Mechanisms such as galaxy-galaxy interaction, AGN and stellar heating, or the heating of the CGM may have a quenching effect, but occur over longer time scales \citep{dresslerGalaxyMorphologyRich1980,pengMassEnvironmentDrivers2010}.

We use the ram pressure stripping condition from \cite{mccarthyRamPressureStripping2008}, given in Eq. \ref{stripcond}.
\begin{equation}\label{stripcond}
\scriptsize
    \rho_{ICM}(r)\cdot v^2 > \frac{\pi}{2}\frac{GM_{gal}(R)\rho_{gas}(R)}{R}
\end{equation}
Where $\rho_{ICM}$ is the density of the ICM plasma at a distance r from the cluster centre, v is the three-dimensional velocity of the infalling galaxy and G is the gravitational constant. In this model, the infalling galaxy is assumed to be on a direct path to the centre of the cluster. We use the ICM density profile as found by \cite{mantzXXLSurveyXVII2018}, and $\rho_{gas}$ as modelled using an exponential gas profile.

We model the infalling galaxies using the masses found by \cite{trudeauXXLSurveyXLIX2022} for XLSSC 122 member galaxies. Using an NFW, Sersic, and exponential profile to model the dark matter, stars, and gas respectively. We then find the stripping radius using Eq. \ref{stripcond}. The details of the galaxy modelling and infall velocity estimation can be found in Appendix. \ref{RamPressDerivation}. Ram pressure can explain a weak stripping effect for galaxies in the mass range of XLSSC 122 at z = 2, particularly those of lower mass, near the virial radius. However, a spectroscopic confirmation of post starburst features using an instrument with higher spectral resolution than the HST G141 grism would be invaluable in constraining this population.

\section{Summary and Conclusion}
In this paper we present an updated and expanded photometric and spectroscopic catalogue of galaxies within the cluster XLSSC 122 and the cluster \foreground. This catalogue contains new information out past $10\,R_{vir}$ of XLSSC 122. We trace radial colour and density relations.
In summary:
\begin{enumerate}
    \item We have increased the total number of spectroscopically confirmed cluster members to $74$ in the Gold and Silver member categories for XLSSC 122.
    \item XLSSC 122 contains a high fraction of red galaxies overall at $49\% \pm12\%$, with upwards of $80\%$ red fraction within the closest 9 galaxies to the BCG. For comparison, the Virgo cluster in the local universe has a core red fraction of $90\%$ \citep{roedigerNextGenerationVirgo2017a}
    \item The luminosity function of XLSSC 122 member galaxies has a similar low mass slope but has a characteristic $M_*$ $\approx$ 1 magnitude fainter compared with clusters at $z \approx 1$ with similar ICM temperature.
    \item The clear colour bimodality indicates a rapid cessation of star formation upon entry into the cluster environment, with a strongly quenched core population.
    \item Given the apparent spherically symmetric arrangement of cluster galaxies around the core, we do not report a detection of any infalling groups or filaments. If there is no infalling structure, this would imply that XLSSC 122 is a mature galaxy cluster which has entered into a more relaxed state.
    \item   We use a toy model of ram pressure  stripping and find the observed rise in red fraction is consistent  with a weak ram pressure stripping scenario leading to lower gas reservoirs in middle to high mass galaxies, and more rapid quenching for low mass galaxies.
    \item The red fraction increase going from outskirts to cluster core shows that these galaxies quench fast and stay quenched. With the lack of observed $H\beta$ absorption features in the cluster core, it appears that these galaxies have been quenched for longer than the detection time of a recently quenched galaxy, or at least $0.5\,\mathrm{Gyr}$ \citep{werlePoststarburstGalaxiesCenters2022}.
\end{enumerate}

\section*{Acknowledgements}
The authors thank the anonymous referee for their detailed feedback. The paper benefited greatly from their comments. We are also grateful to the OUP for the fee waiver. This research is based on observations made with the NASA/ESA Hubble Space Telescope obtained from the Space Telescope Science Institute, which is operated by the Association of Universities for Research in Astronomy, Inc., under NASA contract NAS 5–26555. 

Support for program number HST-GO-17172 was provided by NASA through a grant from the Space Telescope Science Institute, which is operated by the Association of Universities for Research in Astronomy, Inc., under NASA contract NAS 5-26555
%%%%%%%%%%%%%%%%%%%%%%%%%%%%%%%%%%%%%%%%%%%%%%%%%%
\section*{Data Availability}

Full catalogues and reduced spectra for all galaxies in XLSSC 122 and \foreground \ are available at https://github.com/blonsbrough/Initial-Investigations-of-the-Outskirts-of-XLSSC-122. If you have questions about the catalogue, feel free to contact me at my email.
\\
\\
Email: blonsbrough@gmail.com

%%%%%%%%%%%%%%%%%%%% REFERENCES %%%%%%%%%%%%%%%%%%

\bibliographystyle{mnras}
\bibliography{Bibliography} 

@article{abadiRamPressureStripping1999,
  title = {Ram Pressure Stripping of Spiral Galaxies in Clusters},
  author = {Abadi, Mario G. and Moore, Ben and Bower, Richard G.},
  year = {1999},
  month = oct,
  journal = {Monthly Notices of the Royal Astronomical Society},
  volume = {308},
  pages = {947--954},
  issn = {0035-8711},
  doi = {10.1046/j.1365-8711.1999.02715.x},
  urldate = {2025-02-14}
}

@article{baldryGalaxyBimodalityStellar2006,
  title = {Galaxy Bimodality versus Stellar Mass and Environment},
  author = {Baldry, I. K. and Balogh, M. L. and Bower, R. G. and Glazebrook, K. and Nichol, R. C. and Bamford, S. P. and Budavari, T.},
  year = {2006},
  month = dec,
  journal = {Monthly Notices of the Royal Astronomical Society},
  volume = {373},
  pages = {469--483},
  issn = {0035-8711},
  doi = {10.1111/j.1365-2966.2006.11081.x},
  urldate = {2025-02-01}
}

@article{baloghBimodalGalaxyColor2004,
  title = {The {{Bimodal Galaxy Color Distribution}}: {{Dependence}} on {{Luminosity}} and {{Environment}}},
  shorttitle = {The {{Bimodal Galaxy Color Distribution}}},
  author = {Balogh, Michael L. and Baldry, Ivan K. and Nichol, Robert and Miller, Chris and Bower, Richard and Glazebrook, Karl},
  year = {2004},
  month = nov,
  journal = {The Astrophysical Journal},
  volume = {615},
  pages = {L101--L104},
  issn = {0004-637X},
  doi = {10.1086/426079},
  urldate = {2024-10-16}
}

@article{baloghEvidenceChangeDominant2016,
  title = {Evidence for a Change in the Dominant Satellite Galaxy Quenching Mechanism at z = 1},
  author = {Balogh, Michael L. and McGee, Sean L. and Mok, Angus and Muzzin, Adam and {van der Burg}, Remco F. J. and Bower, Richard G. and Finoguenov, Alexis and Hoekstra, Henk and Lidman, Chris and Mulchaey, John S. and Noble, Allison and Parker, Laura C. and Tanaka, Masayuki and Wilman, David J. and Webb, Tracy and Wilson, Gillian and Yee, Howard K. C.},
  year = {2016},
  month = mar,
  journal = {Monthly Notices of the Royal Astronomical Society},
  volume = {456},
  pages = {4364--4376},
  issn = {0035-8711},
  doi = {10.1093/mnras/stv2949},
  urldate = {2024-10-16}
}

@article{baloghGeminiObservationsGalaxies2017,
  title = {Gemini {{Observations}} of {{Galaxies}} in {{Rich Early Environments}} ({{GOGREEN}}) {{I}}: Survey Description},
  shorttitle = {Gemini {{Observations}} of {{Galaxies}} in {{Rich Early Environments}} ({{GOGREEN}}) {{I}}},
  author = {Balogh, Michael L. and Gilbank, David G. and Muzzin, Adam and Rudnick, Gregory and Cooper, Michael C. and Lidman, Chris and Biviano, Andrea and Demarco, Ricardo and McGee, Sean L. and Nantais, Julie B. and Noble, Allison and Old, Lyndsay and Wilson, Gillian and Yee, Howard K. C. and Bellhouse, Callum and Cerulo, Pierluigi and Chan, Jeffrey and {Pintos-Castro}, Irene and Simpson, Rane and {van der Burg}, Remco F. J. and Zaritsky, Dennis and Ziparo, Felicia and Alonso, Mar{\'i}a Victoria and Bower, Richard G. and De Lucia, Gabriella and Finoguenov, Alexis and Lambas, Diego Garcia and Muriel, Hernan and Parker, Laura C. and Rettura, Alessandro and Valotto, Carlos and Wetzel, Andrew},
  year = {2017},
  month = oct,
  journal = {Monthly Notices of the Royal Astronomical Society},
  volume = {470},
  pages = {4168--4185},
  issn = {0035-8711},
  doi = {10.1093/mnras/stx1370},
  urldate = {2025-02-06}
}

@article{baloghOriginStarFormation2000,
  title = {The {{Origin}} of {{Star Formation Gradients}} in {{Rich Galaxy Clusters}}},
  author = {Balogh, Michael L. and Navarro, Julio F. and Morris, Simon L.},
  year = {2000},
  month = sep,
  journal = {The Astrophysical Journal},
  volume = {540},
  pages = {113--121},
  issn = {0004-637X},
  doi = {10.1086/309323},
  urldate = {2025-02-06}
}

@article{behrooziAverageStarFormation2013,
  title = {The {{Average Star Formation Histories}} of {{Galaxies}} in {{Dark Matter Halos}} from Z=0-8},
  author = {Behroozi, Peter S. and Wechsler, Risa H. and Conroy, Charlie},
  year = {2013},
  month = may,
  journal = {The Astrophysical Journal},
  volume = {770},
  number = {1},
  eprint = {1207.6105},
  primaryclass = {astro-ph},
  pages = {57},
  issn = {0004-637X, 1538-4357},
  doi = {10.1088/0004-637X/770/1/57},
  urldate = {2025-05-02},
  archiveprefix = {arXiv}
}

@article{behrooziLackEvolutionGalaxy2013,
  title = {On the {{Lack}} of {{Evolution}} in {{Galaxy Star Formation Efficiency}}},
  author = {Behroozi, Peter S. and Wechsler, Risa H. and Conroy, Charlie},
  year = {2013},
  month = jan,
  journal = {The Astrophysical Journal},
  volume = {762},
  number = {2},
  eprint = {1209.3013},
  primaryclass = {astro-ph},
  pages = {L31},
  issn = {2041-8205, 2041-8213},
  doi = {10.1088/2041-8205/762/2/L31},
  urldate = {2025-05-02},
  archiveprefix = {arXiv}
}

@article{bennettNineyearWilkinsonMicrowave2013,
  title = {Nine-Year {{Wilkinson Microwave Anisotropy Probe}} ({{WMAP}}) {{Observations}}: {{Final Maps}} and {{Results}}},
  shorttitle = {Nine-Year {{Wilkinson Microwave Anisotropy Probe}} ({{WMAP}}) {{Observations}}},
  author = {Bennett, C. L. and Larson, D. and Weiland, J. L. and Jarosik, N. and Hinshaw, G. and Odegard, N. and Smith, K. M. and Hill, R. S. and Gold, B. and Halpern, M. and Komatsu, E. and Nolta, M. R. and Page, L. and Spergel, D. N. and Wollack, E. and Dunkley, J. and Kogut, A. and Limon, M. and Meyer, S. S. and Tucker, G. S. and Wright, E. L.},
  year = {2013},
  month = oct,
  journal = {The Astrophysical Journal Supplement Series},
  volume = {208},
  pages = {20},
  issn = {0067-0049},
  doi = {10.1088/0067-0049/208/2/20},
  urldate = {2025-02-14}
}

@article{bertinSExtractorSoftwareSource1996a,
  title = {{{SExtractor}}: {{Software}} for Source Extraction.},
  shorttitle = {{{SExtractor}}},
  author = {Bertin, E. and Arnouts, S.},
  year = {1996},
  month = jun,
  journal = {Astronomy and Astrophysics Supplement Series},
  volume = {117},
  pages = {393--404},
  issn = {0365-01380004-6361},
  doi = {10.1051/aas:1996164},
  urldate = {2025-04-30}
}

@article{bertinSWarpResamplingCoadding2010,
  title = {{{SWarp}}: {{Resampling}} and {{Co-adding FITS Images Together}}},
  shorttitle = {{{SWarp}}},
  author = {Bertin, Emmanuel},
  year = {2010},
  month = oct,
  journal = {Astrophysics Source Code Library},
  pages = {ascl:1010.068},
  urldate = {2025-04-30}
}

@book{binneyGalacticDynamics1987,
  title = {Galactic Dynamics},
  author = {Binney, James and Tremaine, Scott},
  year = {1987},
  month = jan,
  journal = {Princeton},
  urldate = {2025-05-26},
  publisher = {Princeton University Press}
}

@article{bondHowFilamentsGalaxies1996a,
  title = {How Filaments of Galaxies Are Woven into the Cosmic Web},
  author = {Bond, J. Richard and Kofman, Lev and Pogosyan, Dmitry},
  year = {1996},
  month = apr,
  journal = {Nature},
  volume = {380},
  pages = {603--606},
  issn = {0028-0836},
  doi = {10.1038/380603a0},
  urldate = {2024-12-16}
}

@article{brammer3DHSTWidefieldGrism2012,
  title = {{{3D-HST}}: {{A Wide-field Grism Spectroscopic Survey}} with the {{Hubble Space Telescope}}},
  shorttitle = {{{3D-HST}}},
  author = {Brammer, Gabriel B. and {van Dokkum}, Pieter G. and Franx, Marijn and Fumagalli, Mattia and Patel, Shannon and Rix, Hans-Walter and Skelton, Rosalind E. and Kriek, Mariska and Nelson, Erica and Schmidt, Kasper B. and Bezanson, Rachel and {da Cunha}, Elisabete and Erb, Dawn K. and Fan, Xiaohui and F{\"o}rster Schreiber, Natascha and Illingworth, Garth D. and Labb{\'e}, Ivo and Leja, Joel and Lundgren, Britt and Magee, Dan and Marchesini, Danilo and McCarthy, Patrick and Momcheva, Ivelina and Muzzin, Adam and Quadri, Ryan and Steidel, Charles C. and Tal, Tomer and Wake, David and Whitaker, Katherine E. and Williams, Anna},
  year = {2012},
  month = jun,
  journal = {The Astrophysical Journal Supplement Series},
  volume = {200},
  pages = {13},
  publisher = {IOP},
  issn = {0067-0049},
  doi = {10.1088/0067-0049/200/2/13},
  urldate = {2025-04-30}
}

@article{brammerEAZYFastPublic2008a,
  title = {{{EAZY}}: {{A Fast}}, {{Public Photometric Redshift Code}}},
  shorttitle = {{{EAZY}}},
  author = {Brammer, Gabriel B. and {van Dokkum}, Pieter G. and Coppi, Paolo},
  year = {2008},
  month = oct,
  journal = {The Astrophysical Journal},
  volume = {686},
  pages = {1503--1513},
  publisher = {IOP},
  issn = {0004-637X},
  doi = {10.1086/591786},
  urldate = {2025-05-30}
}

@article{brammerGrizliGrismRedshift2019,
  title = {Grizli: {{Grism}} Redshift and Line Analysis Software},
  shorttitle = {Grizli},
  author = {Brammer, Gabe},
  year = {2019},
  month = may,
  journal = {Astrophysics Source Code Library},
  pages = {ascl:1905.001},
  urldate = {2025-04-30}
}

@article{brownColdGasStripping2017,
  title = {Cold Gas Stripping in Satellite Galaxies: From Pairs to Clusters},
  shorttitle = {Cold Gas Stripping in Satellite Galaxies},
  author = {Brown, Toby and Catinella, Barbara and Cortese, Luca and Lagos, Claudia del P. and Dav{\'e}, Romeel and Kilborn, Virginia and Haynes, Martha P. and Giovanelli, Riccardo and Rafieferantsoa, Mika},
  year = {2017},
  month = apr,
  journal = {Monthly Notices of the Royal Astronomical Society},
  volume = {466},
  pages = {1275--1289},
  issn = {0035-8711},
  doi = {10.1093/mnras/stw2991},
  urldate = {2024-12-16}
}

@article{brownVERTICOVIIEnvironmental2023,
  title = {{{VERTICO}}. {{VII}}. {{Environmental Quenching Caused}} by the {{Suppression}} of {{Molecular Gas Content}} and {{Star Formation Efficiency}} in {{Virgo Cluster Galaxies}}},
  author = {Brown, Toby and Roberts, Ian D. and Thorp, Mallory and Ellison, Sara L. and Zabel, Nikki and Wilson, Christine D. and Bah{\'e}, Yannick M. and Bisaria, Dhruv and Bolatto, Alberto D. and Boselli, Alessandro and Chung, Aeree and Cortese, Luca and Catinella, Barbara and Davis, Timothy A. and {Jim{\'e}nez-Donaire}, Mar{\'i}a J. and Lagos, Claudia D. P. and Lee, Bumhyun and Parker, Laura C. and Smith, Rory and Spekkens, Kristine and Stevens, Adam R. H. and Villanueva, Vicente and Watts, Adam B.},
  year = {2023},
  month = oct,
  journal = {The Astrophysical Journal},
  volume = {956},
  pages = {37},
  publisher = {IOP},
  issn = {0004-637X},
  doi = {10.3847/1538-4357/acf195},
  urldate = {2025-04-25}
}

@article{butcherEvolutionGalaxiesClusters1984,
  title = {The Evolution of Galaxies in Clusters. {{V}}. {{A}} Study of Populations since {{Z}} 0.5.},
  author = {Butcher, H. and Oemler, Jr., A.},
  year = {1984},
  month = oct,
  journal = {The Astrophysical Journal},
  volume = {285},
  pages = {426--438},
  issn = {0004-637X},
  doi = {10.1086/162519},
  urldate = {2024-10-03}
}

@article{chanRestframe$H$bandLuminosity2019,
  title = {The {{Rest-frame}} \${{H}}\$-Band {{Luminosity Function}} of {{Red Sequence Galaxies}} in {{Clusters}} at \$1.0 {$<$} z {$<$} 1.3\$},
  author = {Chan, Jeffrey C. C. and Wilson, Gillian and Rudnick, Gregory and Muzzin, Adam and Balogh, Michael and Nantais, Julie and van der Burg, Remco F. J. and Cerulo, Pierluigi and Biviano, Andrea and Cooper, Michael C. and Demarco, Ricardo and Forrest, Ben and Lidman, Chris and Noble, Allison and Old, Lyndsay and {Pintos-Castro}, Irene and Reeves, Andrew M. M. and Webb, Kristi A. and Yee, Howard K. C. and Abdullah, Mohamed H. and Lucia, Gabriella De and Marchesini, Danilo and McGee, Sean L. and Stefanon, Mauro and Zaritsky, Dennis},
  year = {2019},
  month = aug,
  journal = {The Astrophysical Journal},
  volume = {880},
  number = {2},
  eprint = {1906.10707},
  primaryclass = {astro-ph},
  pages = {119},
  issn = {1538-4357},
  doi = {10.3847/1538-4357/ab2b3a},
  urldate = {2025-04-16},
  archiveprefix = {arXiv}
}

@misc{chiangAncientLightYoung2013,
  title = {Ancient {{Light}} from {{Young Cosmic Cities}}: {{Physical}} and {{Observational Signatures}} of {{Galaxy Proto-Clusters}}},
  shorttitle = {Ancient {{Light}} from {{Young Cosmic Cities}}},
  author = {Chiang, Yi-Kuan and Overzier, Roderik and Gebhardt, Karl},
  year = {2013},
  month = dec,
  publisher = {arXiv},
  doi = {10.48550/arXiv.1310.2938},
  urldate = {2024-11-27}
}

@article{circostaSUPERUnbiasedStudy2018,
  title = {{{SUPER}}. {{I}}. {{Toward}} an Unbiased Study of Ionized Outflows in z {$\sim$} 2 Active Galactic Nuclei: Survey Overview and Sample Characterization},
  shorttitle = {{{SUPER}}. {{I}}. {{Toward}} an Unbiased Study of Ionized Outflows in z {$\sim$} 2 Active Galactic Nuclei},
  author = {Circosta, C. and Mainieri, V. and Padovani, P. and Lanzuisi, G. and Salvato, M. and Harrison, C. M. and Kakkad, D. and Puglisi, A. and Vietri, G. and Zamorani, G. and Cicone, C. and Husemann, B. and Vignali, C. and Balmaverde, B. and Bischetti, M. and Bongiorno, A. and Brusa, M. and Carniani, S. and Civano, F. and Comastri, A. and Cresci, G. and Feruglio, C. and Fiore, F. and Fotopoulou, S. and Karim, A. and Lamastra, A. and Magnelli, B. and Mannucci, F. and Marconi, A. and Merloni, A. and Netzer, H. and Perna, M. and Piconcelli, E. and Rodighiero, G. and Schinnerer, E. and Schramm, M. and Schulze, A. and Silverman, J. and Zappacosta, L.},
  year = {2018},
  month = nov,
  journal = {Astronomy and Astrophysics},
  volume = {620},
  pages = {A82},
  publisher = {EDP},
  issn = {0004-6361},
  doi = {10.1051/0004-6361/201833520},
  urldate = {2025-04-25}
}

@article{darvishEFFECTSLOCALENVIRONMENT2016,
  title = {{{THE EFFECTS OF THE LOCAL ENVIRONMENT AND S}}TEL{{LAR MASS ON GALAXY QUENCHING TO}} z {$\sim$} 3},
  author = {Darvish, Behnam and Mobasher, Bahram and Sobral, David and Rettura, Alessandro and Scoville, Nick and Faisst, Andreas and Capak, Peter},
  year = {2016},
  month = jul,
  journal = {The Astrophysical Journal},
  volume = {825},
  number = {2},
  pages = {113},
  issn = {0004-637X},
  doi = {10.3847/0004-637X/825/2/113},
  urldate = {2025-02-10},
  langid = {english}
}

@article{dresslerGalaxyMorphologyRich1980,
  title = {Galaxy Morphology in Rich Clusters: Implications for the Formation and Evolution of Galaxies.},
  shorttitle = {Galaxy Morphology in Rich Clusters},
  author = {Dressler, A.},
  year = {1980},
  month = mar,
  journal = {The Astrophysical Journal},
  volume = {236},
  pages = {351--365},
  issn = {0004-637X},
  doi = {10.1086/157753},
  urldate = {2024-10-18}
}

@article{dresslerIMACSClusterBuilding2013,
  title = {The {{IMACS Cluster Building Survey}}. {{II}}. {{Spectral Evolution}} of {{Galaxies}} in the {{Epoch}} of {{Cluster Assembly}}},
  author = {Dressler, Alan and Oemler, Jr., Augustus and Poggianti, Bianca M. and Gladders, Michael D. and Abramson, Louis and Vulcani, Benedetta},
  year = {2013},
  month = jun,
  journal = {The Astrophysical Journal},
  volume = {770},
  pages = {62},
  issn = {0004-637X},
  doi = {10.1088/0004-637X/770/1/62},
  urldate = {2024-10-18}
}

@article{duffyDarkMatterHalo2008,
  title = {Dark Matter Halo Concentrations in the {{Wilkinson Microwave Anisotropy Probe}} Year 5 Cosmology},
  author = {Duffy, Alan R. and Schaye, Joop and Kay, Scott T. and Dalla Vecchia, Claudio},
  year = {2008},
  month = oct,
  journal = {Monthly Notices of the Royal Astronomical Society},
  volume = {390},
  pages = {L64-L68},
  publisher = {OUP},
  issn = {0035-8711},
  doi = {10.1111/j.1745-3933.2008.00537.x},
  urldate = {2025-05-13}
}

@article{duttonColdDarkMatter2014,
  title = {Cold Dark Matter Haloes in the {{Planck}} Era: Evolution of Structural Parameters for {{Einasto}} and {{NFW}} Profiles},
  shorttitle = {Cold Dark Matter Haloes in the {{Planck}} Era},
  author = {Dutton, Aaron A. and Macci{\`o}, Andrea V.},
  year = {2014},
  month = jul,
  journal = {Monthly Notices of the Royal Astronomical Society},
  volume = {441},
  pages = {3359--3374},
  publisher = {OUP},
  issn = {0035-8711},
  doi = {10.1093/mnras/stu742},
  urldate = {2025-06-06}
}

@article{foreman-mackeyEmceeMCMCHammer2013,
  title = {Emcee: {{The MCMC Hammer}}},
  shorttitle = {Emcee},
  author = {{Foreman-Mackey}, Daniel and Hogg, David W. and Lang, Dustin and Goodman, Jonathan},
  year = {2013},
  month = mar,
  journal = {Publications of the Astronomical Society of the Pacific},
  volume = {125},
  pages = {306},
  publisher = {IOP},
  issn = {0004-6280},
  doi = {10.1086/670067},
  urldate = {2025-04-30}
}

@article{freemanDisksSpiralS01970,
  title = {On the {{Disks}} of {{Spiral}} and {{S0 Galaxies}}},
  author = {Freeman, K. C.},
  year = {1970},
  month = jun,
  journal = {The Astrophysical Journal},
  volume = {160},
  pages = {811},
  publisher = {IOP},
  issn = {0004-637X},
  doi = {10.1086/150474},
  urldate = {2025-06-08}
}

@article{fujitaPreProcessingGalaxiesEntering2004,
  title = {Pre-{{Processing}} of {{Galaxies}} before {{Entering}} a {{Cluster}}},
  author = {Fujita, Yutaka},
  year = {2004},
  month = feb,
  journal = {Publications of the Astronomical Society of Japan},
  volume = {56},
  pages = {29--43},
  issn = {0004-6264},
  doi = {10.1093/pasj/56.1.29},
  urldate = {2025-02-06}
}

@article{gladdersNewMethodGalaxy2000,
  title = {A {{New Method For Galaxy Cluster Detection}}. {{I}}. {{The Algorithm}}},
  author = {Gladders, Michael D. and Yee, H. K. C.},
  year = {2000},
  month = oct,
  journal = {The Astronomical Journal},
  volume = {120},
  pages = {2148--2162},
  publisher = {IOP},
  issn = {0004-6256},
  doi = {10.1086/301557},
  urldate = {2025-04-24}
}

@article{gladdersRedSequenceClusterSurvey2005,
  title = {The {{Red-Sequence Cluster Survey}}. {{I}}. {{The Survey}} and {{Cluster Catalogs}} for {{Patches RCS}} 0926+37 and {{RCS}} 1327+29},
  author = {Gladders, Michael D. and Yee, H. K. C.},
  year = {2005},
  month = mar,
  journal = {The Astrophysical Journal Supplement Series},
  volume = {157},
  number = {1},
  pages = {1},
  publisher = {IOP Publishing},
  issn = {0067-0049},
  doi = {10.1086/427327},
  urldate = {2025-04-24},
  langid = {english}
}

@article{gobatStarFormationHistories2008,
  title = {Star Formation Histories of Early-Type Galaxies at z = 1.2 in Cluster and Field Environments},
  author = {Gobat, R. and Rosati, P. and Strazzullo, V. and Rettura, A. and Demarco, R. and Nonino, M.},
  year = {2008},
  month = sep,
  journal = {Astronomy and Astrophysics},
  volume = {488},
  pages = {853--860},
  issn = {0004-6361},
  doi = {10.1051/0004-6361:200809531},
  urldate = {2025-02-06}
}

@book{gonzagaDrizzlePacHandbook2012,
  title = {The {{DrizzlePac Handbook}}},
  author = {Gonzaga, S. and Hack, W. and Fruchter, A. and Mack, J.},
  year = {2012},
  month = jun,
  journal = {The DrizzlePac Handbook},
  urldate = {2025-04-30}
}

@article{gunnInfallMatterClusters1972,
  title = {On the {{Infall}} of {{Matter Into Clusters}} of {{Galaxies}} and {{Some Effects}} on {{Their Evolution}}},
  author = {Gunn, James E. and Gott, III, J. Richard},
  year = {1972},
  month = aug,
  journal = {The Astrophysical Journal},
  volume = {176},
  pages = {1},
  issn = {0004-637X},
  doi = {10.1086/151605},
  urldate = {2025-02-06}
}

@article{hoggOverdensitiesGalaxyEnvironments2003,
  title = {The {{Overdensities}} of {{Galaxy Environments}} as a {{Function}} of {{Luminosity}} and {{Color}}},
  author = {Hogg, David W. and Blanton, Michael R. and Eisenstein, Daniel J. and Gunn, James E. and Schlegel, David J. and Zehavi, Idit and Bahcall, Neta A. and Brinkmann, Jon and Csabai, Istvan and Schneider, Donald P. and Weinberg, David H. and York, Donald G.},
  year = {2003},
  month = jan,
  journal = {The Astrophysical Journal},
  volume = {585},
  number = {1},
  pages = {L5},
  issn = {0004-637X},
  doi = {10.1086/374238},
  urldate = {2025-02-01},
  langid = {english}
}

@article{houGroupDynamicsPlay2013,
  title = {Do Group Dynamics Play a Role in the Evolution of Member Galaxies?},
  author = {Hou, Annie and Parker, Laura C. and Balogh, Michael L. and McGee, Sean L. and Wilman, David J. and Connelly, Jennifer L. and Harris, William E. and Mok, Angus and Mulchaey, John S. and Bower, Richard G. and Finoguenov, Alexis},
  year = {2013},
  month = oct,
  journal = {Monthly Notices of the Royal Astronomical Society},
  volume = {435},
  pages = {1715--1726},
  issn = {0035-8711},
  doi = {10.1093/mnras/stt1410},
  urldate = {2024-12-06}
}

@article{kawinwanichakijEffectLocalEnvironment2017,
  title = {Effect of {{Local Environment}} and {{Stellar Mass}} on {{Galaxy Quenching}} and {{Morphology}} at 0.5 {$<$} z {$<$} 2.0},
  author = {Kawinwanichakij, Lalitwadee and Papovich, Casey and Quadri, Ryan F. and Glazebrook, Karl and Kacprzak, Glenn G. and Allen, Rebecca J. and Bell, Eric F. and Croton, Darren J. and Dekel, Avishai and Ferguson, Henry C. and Forrest, Ben and Grogin, Norman A. and Guo, Yicheng and Kocevski, Dale D. and Koekemoer, Anton M. and Labb{\'e}, Ivo and Lucas, Ray A. and Nanayakkara, Themiya and Spitler, Lee R. and Straatman, Caroline M. S. and Tran, Kim-Vy H. and Tomczak, Adam and {van Dokkum}, Pieter},
  year = {2017},
  month = oct,
  journal = {The Astrophysical Journal},
  volume = {847},
  pages = {134},
  issn = {0004-637X},
  doi = {10.3847/1538-4357/aa8b75},
  urldate = {2024-12-16}
}

@article{koekemoerCANDELSCosmicAssembly2011,
  title = {{{CANDELS}}: {{The Cosmic Assembly Near-infrared Deep Extragalactic Legacy Survey}}---{{The Hubble Space Telescope Observations}}, {{Imaging Data Products}}, and {{Mosaics}}},
  shorttitle = {{{CANDELS}}},
  author = {Koekemoer, Anton M. and Faber, S. M. and Ferguson, Henry C. and Grogin, Norman A. and Kocevski, Dale D. and Koo, David C. and Lai, Kamson and Lotz, Jennifer M. and Lucas, Ray A. and McGrath, Elizabeth J. and Ogaz, Sara and Rajan, Abhijith and Riess, Adam G. and Rodney, Steve A. and Strolger, Louis and Casertano, Stefano and Castellano, Marco and Dahlen, Tomas and Dickinson, Mark and Dolch, Timothy and Fontana, Adriano and Giavalisco, Mauro and Grazian, Andrea and Guo, Yicheng and Hathi, Nimish P. and Huang, Kuang-Han and {van der Wel}, Arjen and Yan, Hao-Jing and Acquaviva, Viviana and Alexander, David M. and Almaini, Omar and Ashby, Matthew L. N. and Barden, Marco and Bell, Eric F. and Bournaud, Fr{\'e}d{\'e}ric and Brown, Thomas M. and Caputi, Karina I. and Cassata, Paolo and Challis, Peter J. and Chary, Ranga-Ram and Cheung, Edmond and Cirasuolo, Michele and Conselice, Christopher J. and Roshan Cooray, Asantha and Croton, Darren J. and Daddi, Emanuele and Dav{\'e}, Romeel and {de Mello}, Duilia F. and {de Ravel}, Loic and Dekel, Avishai and Donley, Jennifer L. and Dunlop, James S. and Dutton, Aaron A. and Elbaz, David and Fazio, Giovanni G. and Filippenko, Alexei V. and Finkelstein, Steven L. and Frazer, Chris and Gardner, Jonathan P. and Garnavich, Peter M. and Gawiser, Eric and Gruetzbauch, Ruth and Hartley, Will G. and H{\"a}ussler, Boris and Herrington, Jessica and Hopkins, Philip F. and Huang, Jia-Sheng and Jha, Saurabh W. and Johnson, Andrew and Kartaltepe, Jeyhan S. and Khostovan, Ali A. and Kirshner, Robert P. and Lani, Caterina and Lee, Kyoung-Soo and Li, Weidong and Madau, Piero and McCarthy, Patrick J. and McIntosh, Daniel H. and McLure, Ross J. and McPartland, Conor and Mobasher, Bahram and Moreira, Heidi and Mortlock, Alice and Moustakas, Leonidas A. and Mozena, Mark and Nandra, Kirpal and Newman, Jeffrey A. and Nielsen, Jennifer L. and Niemi, Sami and Noeske, Kai G. and Papovich, Casey J. and Pentericci, Laura and Pope, Alexandra and Primack, Joel R. and Ravindranath, Swara and Reddy, Naveen A. and Renzini, Alvio and Rix, Hans-Walter and Robaina, Aday R. and Rosario, David J. and Rosati, Piero and Salimbeni, Sara and Scarlata, Claudia and Siana, Brian and Simard, Luc and Smidt, Joseph and Snyder, Diana and Somerville, Rachel S. and Spinrad, Hyron and Straughn, Amber N. and Telford, Olivia and Teplitz, Harry I. and Trump, Jonathan R. and Vargas, Carlos and Villforth, Carolin and Wagner, Cory R. and Wandro, Pat and Wechsler, Risa H. and Weiner, Benjamin J. and Wiklind, Tommy and Wild, Vivienne and Wilson, Grant and Wuyts, Stijn and Yun, Min S.},
  year = {2011},
  month = dec,
  journal = {The Astrophysical Journal Supplement Series},
  volume = {197},
  pages = {36},
  publisher = {IOP},
  issn = {0067-0049},
  doi = {10.1088/0067-0049/197/2/36},
  urldate = {2025-04-30}
}

@article{kraljicGalaxyEvolutionMetric2018,
  title = {Galaxy Evolution in the Metric of the Cosmic Web},
  author = {Kraljic, K. and Arnouts, S. and Pichon, C. and Laigle, C. and {de la Torre}, S. and Vibert, D. and Cadiou, C. and Dubois, Y. and Treyer, M. and Schimd, C. and Codis, S. and {de Lapparent}, V. and Devriendt, J. and Hwang, H. S. and Le Borgne, D. and Malavasi, N. and Milliard, B. and Musso, M. and Pogosyan, D. and Alpaslan, M. and {Bland-Hawthorn}, J. and Wright, A. H.},
  year = {2018},
  month = feb,
  journal = {Monthly Notices of the Royal Astronomical Society},
  volume = {474},
  pages = {547--571},
  issn = {0035-8711},
  doi = {10.1093/mnras/stx2638},
  urldate = {2024-12-16}
}

@misc{lazarAnalyticSurfaceDensity2024a,
  title = {An Analytic Surface Density Profile for \${{$\Lambda$}}\${{CDM}} Halos and Gravitational Lensing Studies},
  author = {Lazar, Alexandres and Bullock, James S. and Nierenberg, Anna and Moustakas, Leonidas and {Boylan-Kolchin}, Michael},
  year = {2024},
  month = jan,
  publisher = {arXiv},
  doi = {10.48550/arXiv.2304.11177},
  urldate = {2025-01-04},
  langid = {english}
}

@article{lemauxPersistenceColourdensityRelation2019,
  title = {Persistence of the Colour-Density Relation and Efficient Environmental Quenching to z {$\sim$} 1.4},
  author = {Lemaux, B. C. and Tomczak, A. R. and Lubin, L. M. and Gal, R. R. and Shen, L. and Pelliccia, D. and Wu, P. -F. and Hung, D. and Mei, S. and Le F{\`e}vre, O. and Rumbaugh, N. and Kocevski, D. D. and Squires, G. K.},
  year = {2019},
  month = nov,
  journal = {Monthly Notices of the Royal Astronomical Society},
  volume = {490},
  pages = {1231--1254},
  issn = {0035-8711},
  doi = {10.1093/mnras/stz2661},
  urldate = {2024-12-17}
}

@article{lesteMorphologicalAnalysisGalaxy2024,
  title = {A Morphological Analysis of the Galaxy Cluster {{XLSSC}} 122 at z = 1.98},
  author = {Leste, O. K. and Willis, J. P. and Canning, R. E. A. and Rennehan, D.},
  year = {2024},
  month = sep,
  journal = {Monthly Notices of the Royal Astronomical Society},
  volume = {533},
  pages = {2927--2947},
  issn = {0035-8711},
  doi = {10.1093/mnras/stae1967},
  urldate = {2024-12-06}
}

@article{liEVOLUTIONGROUPGALAXIES2012,
  title = {{{EVOLUTION OF GROUP GALAXIES FROM THE FIRST RED-SEQUENCE CLUSTER SURVEY}}},
  author = {Li, I. H. and Yee, H. K. C. and Hsieh, B. C. and Gladders, M.},
  year = {2012},
  month = apr,
  journal = {The Astrophysical Journal},
  volume = {749},
  number = {2},
  pages = {150},
  issn = {0004-637X, 1538-4357},
  doi = {10.1088/0004-637X/749/2/150},
  urldate = {2025-04-15},
  langid = {english}
}

@article{madauCosmicStarFormationHistory2014,
  title = {Cosmic {{Star-Formation History}}},
  author = {Madau, Piero and Dickinson, Mark},
  year = {2014},
  month = aug,
  journal = {Annual Review of Astronomy and Astrophysics},
  volume = {52},
  number = {1},
  pages = {415--486},
  issn = {0066-4146, 1545-4282},
  doi = {10.1146/annurev-astro-081811-125615},
  urldate = {2025-04-15},
  langid = {english}
}

@article{mantzXXLSurveyDetection2014,
  title = {The {{XXL Survey V}}: {{Detection}} of the {{Sunyaev-Zel}}'dovich Effect of the {{Redshift}} 1.9 {{Galaxy Cluster XLSSU J021744}}.1-034536 with {{CARMA}}},
  shorttitle = {The {{XXL Survey V}}},
  author = {Mantz, Adam B. and Abdulla, Zubair and Carlstrom, John E. and Greer, Christopher H. and Leitch, Erik M. and Marrone, Daniel P. and Muchovej, Stephen and Adami, Christophe and Birkinshaw, Mark and Bremer, Malcolm and Clerc, Nicolas and Giles, Paul and Horellou, Cathy and Maughan, Benjamin and Pacaud, Florian and Pierre, Marguerite and Willis, Jon},
  year = {2014},
  month = oct,
  journal = {The Astrophysical Journal},
  volume = {794},
  number = {2},
  pages = {157},
  issn = {1538-4357},
  doi = {10.1088/0004-637X/794/2/157},
  urldate = {2023-08-16}
}

@article{mantzXXLSurveyXVII2018,
  title = {The {{XXL Survey}}: {{XVII}}. {{X-ray}} and {{Sunyaev-Zel}}'dovich {{Properties}} of the {{Redshift}} 2.0 {{Galaxy Cluster XLSSC}} 122},
  shorttitle = {The {{XXL Survey}}},
  author = {Mantz, Adam B. and Abdulla, Zubair and Allen, Steven W. and Carlstrom, John E. and Logan, Crispin H. A. and Marrone, Daniel P. and Maughan, Benjamin J. and Willis, Jon and Pacaud, Florian and Pierre, Marguerite},
  year = {2018},
  month = dec,
  journal = {Astronomy \& Astrophysics},
  volume = {620},
  pages = {A2},
  issn = {0004-6361, 1432-0746},
  doi = {10.1051/0004-6361/201630096},
  urldate = {2023-08-16}
}

@article{mccarthyRamPressureStripping2008,
  title = {Ram Pressure Stripping the Hot Gaseous Haloes of Galaxies in Groups and Clusters},
  author = {McCarthy, I. G. and Frenk, C. S. and Font, A. S. and Lacey, C. G. and Bower, R. G. and Mitchell, N. L. and Balogh, M. L. and Theuns, T.},
  year = {2008},
  month = jan,
  journal = {Monthly Notices of the Royal Astronomical Society},
  volume = {383},
  pages = {593--605},
  publisher = {OUP},
  issn = {0035-8711},
  doi = {10.1111/j.1365-2966.2007.12577.x},
  urldate = {2025-03-27}
}

@article{mcgeeAccretionGalaxiesGroups2009,
  title = {The Accretion of Galaxies into Groups and Clusters},
  author = {McGee, Sean L. and Balogh, Michael L. and Bower, Richard G. and Font, Andreea S. and McCarthy, Ian G.},
  year = {2009},
  journal = {Monthly Notices of the Royal Astronomical Society},
  volume = {400},
  number = {2},
  pages = {937--950},
  issn = {1365-2966},
  doi = {10.1111/j.1365-2966.2009.15507.x},
  urldate = {2025-02-10},
  copyright = {{\copyright} 2009 The Authors. Journal compilation {\copyright} 2009 RAS},
  langid = {english}
}

@article{mcgeeOverconsumptionOutflowsQuenching2014,
  title = {Overconsumption, Outflows and the Quenching of Satellite Galaxies.},
  author = {McGee, S. L. and Bower, R. G. and Balogh, M. L.},
  year = {2014},
  month = jul,
  journal = {Monthly Notices of the Royal Astronomical Society},
  volume = {442},
  pages = {L105--L109},
  issn = {0035-8711},
  doi = {10.1093/mnrasl/slu066},
  urldate = {2025-02-06}
}

@article{mcnabGOGREENSurveyTransition2021,
  title = {The {{GOGREEN}} Survey: Transition Galaxies and the Evolution of Environmental Quenching},
  shorttitle = {The {{GOGREEN}} Survey},
  author = {McNab, Karen and Balogh, Michael L. and {van der Burg}, Remco F. J. and Forestell, Anya and Webb, Kristi and Vulcani, Benedetta and Rudnick, Gregory and Muzzin, Adam and Cooper, M. C. and McGee, Sean and Biviano, Andrea and Cerulo, Pierluigi and Chan, Jeffrey C. C. and De Lucia, Gabriella and Demarco, Ricardo and Finoguenov, Alexis and Forrest, Ben and Golledge, Caelan and Jablonka, Pascale and Lidman, Chris and Nantais, Julie and Old, Lyndsay and {Pintos-Castro}, Irene and Poggianti, Bianca and Reeves, Andrew M. M. and Wilson, Gillian and Yee, Howard K. C. and Zaritsky, Dennis},
  year = {2021},
  month = nov,
  journal = {Monthly Notices of the Royal Astronomical Society},
  volume = {508},
  pages = {157--174},
  issn = {0035-8711},
  doi = {10.1093/mnras/stab2558},
  urldate = {2024-11-27}
}

@book{moGalaxyFormationEvolution2010,
  title = {Galaxy {{Formation}} and {{Evolution}}},
  author = {Mo, Houjun and {van den Bosch}, Frank C. and White, Simon},
  year = {2010},
  month = may,
  journal = {Galaxy Formation and Evolution},
  doi = {10.1017/CBO9780511807244},
  urldate = {2025-06-06}
}

@article{momcheva3DHSTSurveyHubble2016,
  title = {The {{3D-HST Survey}}: {{Hubble Space Telescope WFC3}}/{{G141 Grism Spectra}}, {{Redshifts}}, and {{Emission Line Measurements}} for {\textasciitilde} 100,000 {{Galaxies}}},
  shorttitle = {The {{3D-HST Survey}}},
  author = {Momcheva, Ivelina G. and Brammer, Gabriel B. and {van Dokkum}, Pieter G. and Skelton, Rosalind E. and Whitaker, Katherine E. and Nelson, Erica J. and Fumagalli, Mattia and Maseda, Michael V. and Leja, Joel and Franx, Marijn and Rix, Hans-Walter and Bezanson, Rachel and Da Cunha, Elisabete and Dickey, Claire and F{\"o}rster Schreiber, Natascha M. and Illingworth, Garth and Kriek, Mariska and Labb{\'e}, Ivo and Ulf Lange, Johannes and Lundgren, Britt F. and Magee, Daniel and Marchesini, Danilo and Oesch, Pascal and Pacifici, Camilla and Patel, Shannon G. and Price, Sedona and Tal, Tomer and Wake, David A. and {van der Wel}, Arjen and Wuyts, Stijn},
  year = {2016},
  month = aug,
  journal = {The Astrophysical Journal Supplement Series},
  volume = {225},
  pages = {27},
  publisher = {IOP},
  issn = {0067-0049},
  doi = {10.3847/0067-0049/225/2/27},
  urldate = {2025-04-30}
}

@article{mooreGalaxyHarassmentEvolution1996,
  title = {Galaxy Harassment and the Evolution of Clusters of Galaxies},
  author = {Moore, Ben and Katz, Neal and Lake, George and Dressler, Alan and Oemler, Augustus},
  year = {1996},
  month = feb,
  journal = {Nature},
  volume = {379},
  pages = {613--616},
  issn = {0028-0836},
  doi = {10.1038/379613a0},
  urldate = {2025-02-06}
}

@article{nantaisEvidenceStrongEvolution2017,
  title = {Evidence for Strong Evolution in Galaxy Environmental Quenching Efficiency between z = 1.6 and z = 0.9},
  author = {Nantais, Julie B. and Muzzin, Adam and {van der Burg}, Remco F. J. and Wilson, Gillian and Lidman, Chris and Foltz, Ryan and DeGroot, Andrew and Noble, Allison and Cooper, Michael C. and Demarco, Ricardo},
  year = {2017},
  month = feb,
  journal = {Monthly Notices of the Royal Astronomical Society: Letters},
  volume = {465},
  number = {1},
  pages = {L104--L108},
  issn = {1745-3925},
  doi = {10.1093/mnrasl/slw224},
  urldate = {2025-02-10}
}

@article{navarroUniversalDensityProfile1997,
  title = {A {{Universal Density Profile}} from {{Hierarchical Clustering}}},
  author = {Navarro, Julio F. and Frenk, Carlos S. and White, Simon D. M.},
  year = {1997},
  month = dec,
  journal = {The Astrophysical Journal},
  volume = {490},
  pages = {493--508},
  publisher = {IOP},
  issn = {0004-637X},
  doi = {10.1086/304888},
  urldate = {2025-04-30}
}

@article{noordehQuiescentGalaxiesVirialized2021,
  title = {Quiescent Galaxies in a Virialized Cluster at Redshift 2: {{Evidence}} for Accelerated Size-Growth},
  shorttitle = {Quiescent Galaxies in a Virialized Cluster at Redshift 2},
  author = {Noordeh, E. and Canning, R. E. A. and Willis, J. P. and Allen, S. W. and Mantz, A. and Stanford, S. A. and Brammer, G.},
  year = {2021},
  month = sep,
  journal = {Monthly Notices of the Royal Astronomical Society},
  volume = {507},
  number = {4},
  pages = {5272--5280},
  issn = {0035-8711, 1365-2966},
  doi = {10.1093/mnras/stab2459},
  urldate = {2023-08-16}
}

@article{okabeHaloConcentrationGalaxy2019,
  title = {Halo Concentration, Galaxy Red Fraction, and Gas Properties of Optically Defined Merging Clusters{\dag}},
  author = {Okabe, Nobuhiro and Oguri, Masamune and Akamatsu, Hiroki and Hamabata, Akinari and Nishizawa, Atsushi J and Medezinski, Elinor and Koyama, Yusei and Hayashi, Masao and Okabe, Taizo and Ueda, Shutaro and Mitsuishi, Ikuyuki and Ota, Naomi},
  year = {2019},
  month = aug,
  journal = {Publications of the Astronomical Society of Japan},
  volume = {71},
  number = {4},
  pages = {79},
  issn = {0004-6264},
  doi = {10.1093/pasj/psz059},
  urldate = {2024-10-03}
}

@article{paccagnellaOmegaWINGSFirstComplete2017,
  title = {{{OmegaWINGS}}: {{The First Complete Census}} of {{Post-starburst Galaxies}} in {{Clusters}} in the {{Local Universe}}},
  shorttitle = {{{OmegaWINGS}}},
  author = {Paccagnella, A. and Vulcani, B. and Poggianti, B. M. and Fritz, J. and Fasano, G. and Moretti, A. and Jaff{\'e}, Yara L. and Biviano, A. and Gullieuszik, M. and Bettoni, D. and Cava, A. and Couch, W. and D'Onofrio, M.},
  year = {2017},
  month = apr,
  journal = {The Astrophysical Journal},
  volume = {838},
  number = {2},
  pages = {148},
  issn = {0004-637X, 1538-4357},
  doi = {10.3847/1538-4357/aa64d7},
  urldate = {2025-01-27},
  langid = {english}
}

@article{pengMassEnvironmentDrivers2010,
  title = {Mass and {{Environment}} as {{Drivers}} of {{Galaxy Evolution}} in {{SDSS}} and {{zCOSMOS}} and the {{Origin}} of the {{Schechter Function}}},
  author = {Peng, Ying-jie and Lilly, Simon J. and Kova{\v c}, Katarina and Bolzonella, Micol and Pozzetti, Lucia and Renzini, Alvio and Zamorani, Gianni and Ilbert, Olivier and Knobel, Christian and Iovino, Angela and Maier, Christian and Cucciati, Olga and Tasca, Lidia and Carollo, C. Marcella and Silverman, John and Kampczyk, Pawel and {de Ravel}, Loic and Sanders, David and Scoville, Nicholas and Contini, Thierry and Mainieri, Vincenzo and Scodeggio, Marco and Kneib, Jean-Paul and Le F{\`e}vre, Olivier and Bardelli, Sandro and Bongiorno, Angela and Caputi, Karina and Coppa, Graziano and {de la Torre}, Sylvain and Franzetti, Paolo and Garilli, Bianca and Lamareille, Fabrice and Le Borgne, Jean-Francois and Le Brun, Vincent and Mignoli, Marco and Perez Montero, Enrique and Pello, Roser and Ricciardelli, Elena and Tanaka, Masayuki and Tresse, Laurence and Vergani, Daniela and Welikala, Niraj and Zucca, Elena and Oesch, Pascal and Abbas, Ummi and Barnes, Luke and Bordoloi, Rongmon and Bottini, Dario and Cappi, Alberto and Cassata, Paolo and Cimatti, Andrea and Fumana, Marco and Hasinger, Gunther and Koekemoer, Anton and Leauthaud, Alexei and Maccagni, Dario and Marinoni, Christian and McCracken, Henry and Memeo, Pierdomenico and Meneux, Baptiste and Nair, Preethi and Porciani, Cristiano and Presotto, Valentina and Scaramella, Roberto},
  year = {2010},
  month = sep,
  journal = {The Astrophysical Journal},
  volume = {721},
  pages = {193--221},
  issn = {0004-637X},
  doi = {10.1088/0004-637X/721/1/193},
  urldate = {2024-12-20}
}

@article{pengMASSENVIRONMENTDRIVERS2012,
  title = {{{MASS AND ENVIRONMENT AS DRIVERS OF GALAXY EVOLUTION}}. {{II}}. {{THE QUENCHING OF SA}}TEL{{LITE GALAXIES AS THE ORIGIN OF ENVIRONMENTAL EFFECTS}}},
  author = {Peng, Ying-jie and Lilly, Simon J. and Renzini, Alvio and Carollo, Marcella},
  year = {2012},
  month = aug,
  journal = {The Astrophysical Journal},
  volume = {757},
  number = {1},
  pages = {4},
  issn = {0004-637X},
  doi = {10.1088/0004-637X/757/1/4},
  urldate = {2025-02-06},
  langid = {english}
}

@misc{poggiantiStarFormationHistories1999,
  title = {The {{Star Formation Histories}} of {{Galaxies}} in {{Distant Clusters}}},
  author = {Poggianti, Bianca and Smail, Ian and Dressler, Alan and Couch, Warrick and Barger, Amy and Butcher, Harvey and Ellis, Richard and Jr, Augustus Oemler},
  year = {1999},
  month = jan,
  publisher = {arXiv},
  doi = {10.48550/arXiv.astro-ph/9901264},
  urldate = {2024-10-18}
}

@article{raichoorStarFormationEnvironment2012,
  title = {Star Formation and Environment in Clusters up to z {\textasciitilde} 2.2},
  author = {Raichoor, A. and Andreon, S.},
  year = {2012},
  month = jan,
  journal = {Astronomy and Astrophysics},
  volume = {537},
  pages = {A88},
  issn = {0004-6361},
  doi = {10.1051/0004-6361/201117998},
  urldate = {2025-04-24}
}

@article{reevesGOGREENSurveyDependence2021,
  title = {The {{GOGREEN}} Survey: {{Dependence}} of Galaxy Properties on Halo Mass at z {$>$} 1 and Implications for Environmental Quenching},
  shorttitle = {The {{GOGREEN}} Survey},
  author = {Reeves, Andrew M. M. and Balogh, Michael L. and van der Burg, Remco F. J. and Finoguenov, Alexis and Kukstas, Egidijus and McCarthy, Ian G. and Webb, Kristi and Muzzin, Adam and McGee, Sean and Rudnick, Gregory and Biviano, Andrea and Cerulo, Pierluigi and Chan, Jeffrey C. C. and Cooper, M. C. and Demarco, Ricardo and Jablonka, Pascale and Lucia, Gabriella De and Vulcani, Benedetta and Wilson, Gillian and Yee, Howard K. C. and Zaritsky, Dennis},
  year = {2021},
  month = jul,
  journal = {Monthly Notices of the Royal Astronomical Society},
  volume = {506},
  number = {3},
  pages = {3364--3384},
  issn = {0035-8711, 1365-2966},
  doi = {10.1093/mnras/stab1955},
  urldate = {2024-12-13}
}

@article{roedigerNextGenerationVirgo2017a,
  title = {The {{Next Generation Virgo Cluster Survey}} ({{NGVS}}). {{XXIV}}. {{The Red Sequence}} to {\textasciitilde}106 {{L}} {$\odot$} and {{Comparisons}} with {{Galaxy Formation Models}}},
  author = {Roediger, Joel C. and Ferrarese, Laura and C{\^o}t{\'e}, Patrick and MacArthur, Lauren A. and {S{\'a}nchez-Janssen}, R{\'u}ben and Blakeslee, John P. and Peng, Eric W. and Liu, Chengze and Munoz, Roberto and Cuillandre, Jean-Charles and Gwyn, Stephen and Mei, Simona and Boissier, Samuel and Boselli, Alessandro and Cantiello, Michele and Courteau, St{\'e}phane and Duc, Pierre-Alain and Lan{\c c}on, Ariane and Mihos, J. Christopher and Puzia, Thomas H. and Taylor, James E. and Durrell, Patrick R. and Toloba, Elisa and Guhathakurta, Puragra and Zhang, Hongxin},
  year = {2017},
  month = feb,
  journal = {The Astrophysical Journal},
  volume = {836},
  pages = {120},
  publisher = {IOP},
  issn = {0004-637X},
  doi = {10.3847/1538-4357/836/1/120},
  urldate = {2025-06-20}
}

@article{rykoffRedMaPPerAlgorithmSDSS2014,
  title = {{{redMaPPer}}. {{I}}. {{Algorithm}} and {{SDSS DR8 Catalog}}},
  author = {Rykoff, E. S. and Rozo, E. and Busha, M. T. and Cunha, C. E. and Finoguenov, A. and Evrard, A. and Hao, J. and Koester, B. P. and Leauthaud, A. and Nord, B. and Pierre, M. and Reddick, R. and Sadibekova, T. and Sheldon, E. S. and Wechsler, R. H.},
  year = {2014},
  month = apr,
  journal = {The Astrophysical Journal},
  volume = {785},
  pages = {104},
  issn = {0004-637X},
  doi = {10.1088/0004-637X/785/2/104},
  urldate = {2025-02-18}
}

@article{sarazinXrayEmissionClusters1986,
  title = {X-Ray Emission from Clusters of Galaxies},
  author = {Sarazin, Craig L.},
  year = {1986},
  month = jan,
  journal = {Reviews of Modern Physics},
  volume = {58},
  number = {1},
  pages = {1--115},
  doi = {10.1103/RevModPhys.58.1},
  urldate = {2025-02-10}
}

@article{schechterAnalyticExpressionLuminosity1976,
  title = {An Analytic Expression for the Luminosity Function for Galaxies.},
  author = {Schechter, P.},
  year = {1976},
  month = jan,
  journal = {The Astrophysical Journal},
  volume = {203},
  pages = {297--306},
  publisher = {IOP},
  issn = {0004-637X},
  doi = {10.1086/154079},
  urldate = {2025-04-16}
}

@book{sersicAtlasGalaxiasAustrales1968,
  title = {Atlas de {{Galaxias Australes}}},
  author = {Sersic, Jose Luis},
  year = {1968},
  month = jan,
  journal = {Cordoba},
  urldate = {2025-05-02}
}

@article{skelton3DHSTWFC3selectedPhotometric2014,
  title = {{{3D-HST WFC3-selected Photometric Catalogs}} in the {{Five CANDELS}}/{{3D-HST Fields}}: {{Photometry}}, {{Photometric Redshifts}}, and {{Stellar Masses}}},
  shorttitle = {{{3D-HST WFC3-selected Photometric Catalogs}} in the {{Five CANDELS}}/{{3D-HST Fields}}},
  author = {Skelton, Rosalind E. and Whitaker, Katherine E. and Momcheva, Ivelina G. and Brammer, Gabriel B. and {van Dokkum}, Pieter G. and Labb{\'e}, Ivo and Franx, Marijn and {van der Wel}, Arjen and Bezanson, Rachel and Da Cunha, Elisabete and Fumagalli, Mattia and F{\"o}rster Schreiber, Natascha and Kriek, Mariska and Leja, Joel and Lundgren, Britt F. and Magee, Daniel and Marchesini, Danilo and Maseda, Michael V. and Nelson, Erica J. and Oesch, Pascal and Pacifici, Camilla and Patel, Shannon G. and Price, Sedona and Rix, Hans-Walter and Tal, Tomer and Wake, David A. and Wuyts, Stijn},
  year = {2014},
  month = oct,
  journal = {The Astrophysical Journal Supplement Series},
  volume = {214},
  pages = {24},
  publisher = {IOP},
  issn = {0067-0049},
  doi = {10.1088/0067-0049/214/2/24},
  urldate = {2025-04-30}
}

@article{strazzulloClusterGalaxiesXMMU2010,
  title = {Cluster Galaxies in {{XMMU J2235-2557}}: Galaxy Population Properties in Most Massive Environments at z {\textasciitilde} 1.4},
  shorttitle = {Cluster Galaxies in {{XMMU J2235-2557}}},
  author = {Strazzullo, V. and Rosati, P. and Pannella, M. and Gobat, R. and Santos, J. S. and Nonino, M. and Demarco, R. and Lidman, C. and Tanaka, M. and Mullis, C. R. and Nu{\~n}ez, C. and Rettura, A. and Jee, M. J. and B{\"o}hringer, H. and Bender, R. and Bouwens, R. J. and Dawson, K. and Fassbender, R. and Franx, M. and Perlmutter, S. and Postman, M.},
  year = {2010},
  month = dec,
  journal = {Astronomy and Astrophysics},
  volume = {524},
  pages = {A17},
  issn = {0004-6361},
  doi = {10.1051/0004-6361/201015251},
  urldate = {2025-02-14},
  langid = {english}
}

@article{strazzulloGalaxyPopulationsMost2019,
  title = {Galaxy Populations in the Most Distant {{SPT-SZ}} Clusters. {{I}}. {{Environmental}} Quenching in Massive Clusters at 1.4 {$\lessequivlnt$} z {$\lessequivlnt$} 1.7},
  author = {Strazzullo, V. and Pannella, M. and Mohr, J. J. and Saro, A. and Ashby, M. L. N. and Bayliss, M. B. and Bocquet, S. and Bulbul, E. and Khullar, G. and Mantz, A. B. and Stanford, S. A. and Benson, B. A. and Bleem, L. E. and Brodwin, M. and Canning, R. E. A. and Capasso, R. and Chiu, I. and Gonzalez, A. H. and Gupta, N. and {Hlavacek-Larrondo}, J. and Klein, M. and McDonald, M. and Noordeh, E. and Rapetti, D. and Reichardt, C. L. and Schrabback, T. and Sharon, K. and Stalder, B.},
  year = {2019},
  month = feb,
  journal = {Astronomy and Astrophysics},
  volume = {622},
  pages = {A117},
  issn = {0004-6361},
  doi = {10.1051/0004-6361/201833944},
  urldate = {2025-02-14}
}

@article{strazzulloPassiveGalaxiesTracers2015,
  title = {Passive Galaxies as Tracers of Cluster Environments at z {\textasciitilde} 2},
  author = {Strazzullo, V. and Daddi, E. and Gobat, R. and Garilli, B. and Mignoli, M. and Valentino, F. and Onodera, M. and Renzini, A. and Cimatti, A. and Finoguenov, A. and Arimoto, N. and Cappellari, M. and Carollo, C. M. and Feruglio, C. and Le Floc'h, E. and Lilly, S. J. and Maccagni, D. and McCracken, H. J. and Moresco, M. and Pozzetti, L. and Zamorani, G.},
  year = {2015},
  month = apr,
  journal = {Astronomy and Astrophysics},
  volume = {576},
  pages = {L6},
  issn = {0004-6361},
  doi = {10.1051/0004-6361/201425038},
  urldate = {2025-02-14}
}

@article{strazzulloRedSequenceBirth2016,
  title = {The {{Red Sequence}} at {{Birth}} in the {{Galaxy Cluster Cl J1449}}+0856 at z = 2},
  author = {Strazzullo, V. and Daddi, E. and Gobat, R. and Valentino, F. and Pannella, M. and Dickinson, M. and Renzini, A. and Brammer, G. and Onodera, M. and Finoguenov, A. and Cimatti, A. and Carollo, C. M. and Arimoto, N.},
  year = {2016},
  month = dec,
  journal = {The Astrophysical Journal},
  volume = {833},
  pages = {L20},
  issn = {0004-637X},
  doi = {10.3847/2041-8213/833/2/L20},
  urldate = {2025-02-14}
}

@article{trudeauMassiveDistantClusters2024,
  title = {The {{Massive}} and {{Distant Clusters}} of {{WISE Survey}} 2: {{A Stacking Analysis Investigating}} the {{Evolution}} of {{Star Formation Rates}} and {{Stellar Masses}} in {{Groups}} and {{Clusters}}},
  shorttitle = {The {{Massive}} and {{Distant Clusters}} of {{WISE Survey}} 2},
  author = {Trudeau, A. and Gonzalez, Anthony H. and Thongkham, K. and Lee, Kyoung-Soo and Alberts, Stacey and Brodwin, M. and Connor, Thomas and Eisenhardt, Peter R. M. and Moravec, Emily and Puvvada, Eshwar and Stanford, S. A.},
  year = {2024},
  month = aug,
  journal = {The Astrophysical Journal},
  volume = {972},
  number = {1},
  pages = {27},
  issn = {0004-637X},
  doi = {10.3847/1538-4357/ad5545},
  urldate = {2024-11-20},
  langid = {english}
}

@article{trudeauXXLSurveyXLIX2022,
  title = {The {{XXL}} Survey. {{XLIX}}. {{Linking}} the Members Star Formation Histories to the Cluster Mass Assembly in the Z=1.98 Galaxy Cluster {{XLSSC}} 122},
  author = {Trudeau, A. and Willis, J. P. and Rennehan, D. and Canning, R. E. A. and Carnall, A. C. and Poggianti, B. and Noordeh, E. and Pierre, M.},
  year = {2022},
  month = aug,
  journal = {Monthly Notices of the Royal Astronomical Society},
  volume = {515},
  number = {2},
  pages = {2529--2547},
  issn = {0035-8711, 1365-2966},
  doi = {10.1093/mnras/stac1760},
  urldate = {2023-08-16}
}

@article{urquhartEnvironmentalButcherOemlerEffect2010,
  title = {An Environmental {{Butcher-Oemler}} Effect in Intermediate-Redshift {{X-ray}} Clusters},
  author = {Urquhart, S. A. and Willis, J. P. and Hoekstra, H. and Pierre, M.},
  year = {2010},
  month = jul,
  journal = {Monthly Notices of the Royal Astronomical Society},
  volume = {406},
  pages = {368--381},
  issn = {0035-8711},
  doi = {10.1111/j.1365-2966.2010.16766.x},
  urldate = {2024-10-16}
}

@article{vanderburgGOGREENSurveyDeep2020,
  title = {The {{GOGREEN Survey}}: {{A}} Deep Stellar Mass Function of Cluster Galaxies at 1.0 {$<$} z {$<$} 1.4 and the Complex Nature of Satellite Quenching},
  shorttitle = {The {{GOGREEN Survey}}},
  author = {{van der Burg}, Remco F. J. and Rudnick, Gregory and Balogh, Michael L. and Muzzin, Adam and Lidman, Chris and Old, Lyndsay J. and Shipley, Heath and Gilbank, David and McGee, Sean and Biviano, Andrea and Cerulo, Pierluigi and Chan, Jeffrey C. C. and Cooper, Michael and De Lucia, Gabriella and Demarco, Ricardo and Forrest, Ben and Gwyn, Stephen and Jablonka, Pascale and Kukstas, Egidijus and Marchesini, Danilo and Nantais, Julie and Noble, Allison and {Pintos-Castro}, Irene and Poggianti, Bianca and Reeves, Andrew M. M. and Stefanon, Mauro and Vulcani, Benedetta and Webb, Kristi and Wilson, Gillian and Yee, Howard and Zaritsky, Dennis},
  year = {2020},
  month = jun,
  journal = {Astronomy and Astrophysics},
  volume = {638},
  pages = {A112},
  issn = {0004-6361},
  doi = {10.1051/0004-6361/202037754},
  urldate = {2024-10-16}
}

@article{vanderburgStellarMassFunction2018,
  title = {The Stellar Mass Function of Galaxies in {{Planck-selected}} Clusters at 0.5 {$<$} z {$<$} 0.7: New Constraints on the Timescale and Location of Satellite Quenching},
  shorttitle = {The Stellar Mass Function of Galaxies in {{Planck-selected}} Clusters at 0.5 {$<$} z {$<$} 0.7},
  author = {{van der Burg}, Remco F. J. and McGee, Sean and Aussel, Herv{\'e} and Dahle, H{\aa}kon and Arnaud, Monique and Pratt, Gabriel W. and Muzzin, Adam},
  year = {2018},
  month = oct,
  journal = {Astronomy and Astrophysics},
  volume = {618},
  pages = {A140},
  issn = {0004-6361},
  doi = {10.1051/0004-6361/201833572},
  urldate = {2025-04-16}
}

@article{vanderwel3DHST+CANDELSEvolutionGalaxy2014,
  title = {{{3D-HST}}+{{CANDELS}}: {{The Evolution}} of the {{Galaxy Size-Mass Distribution}} since z = 3},
  shorttitle = {{{3D-HST}}+{{CANDELS}}},
  author = {{van der Wel}, A. and Franx, M. and {van Dokkum}, P. G. and Skelton, R. E. and Momcheva, I. G. and Whitaker, K. E. and Brammer, G. B. and Bell, E. F. and Rix, H. -W. and Wuyts, S. and Ferguson, H. C. and Holden, B. P. and Barro, G. and Koekemoer, A. M. and Chang, Yu-Yen and McGrath, E. J. and H{\"a}ussler, B. and Dekel, A. and Behroozi, P. and Fumagalli, M. and Leja, J. and Lundgren, B. F. and Maseda, M. V. and Nelson, E. J. and Wake, D. A. and Patel, S. G. and Labb{\'e}, I. and Faber, S. M. and Grogin, N. A. and Kocevski, D. D.},
  year = {2014},
  month = jun,
  journal = {The Astrophysical Journal},
  volume = {788},
  pages = {28},
  publisher = {IOP},
  issn = {0004-637X},
  doi = {10.1088/0004-637X/788/1/28},
  urldate = {2025-05-02}
}

@article{vanderwelFundamentalPlaneField2004,
  title = {The {{Fundamental Plane}} of {{Field Early-Type Galaxies}} at z = 1},
  author = {{van der Wel}, A. and Franx, M. and {van Dokkum}, P. G. and Rix, H. -W.},
  year = {2004},
  month = jan,
  journal = {The Astrophysical Journal},
  volume = {601},
  pages = {L5--L8},
  issn = {0004-637X},
  doi = {10.1086/381887},
  urldate = {2025-02-06}
}

@article{wattsVERTICOEnvironmentallyDriven2023,
  title = {{{VERTICO V}}: {{The}} Environmentally Driven Evolution of the Inner Cold Gas Discs of {{Virgo}} Cluster Galaxies},
  shorttitle = {{{VERTICO V}}},
  author = {Watts, Adam B. and Cortese, Luca and Catinella, Barbara and Brown, Toby and Wilson, Christine D. and Zabel, Nikki and Roberts, Ian D. and Davis, Timothy A. and Thorp, Mallory and Chung, Aeree and Stevens, Adam R. H. and Ellison, Sara L. and Spekkens, Kristine and Parker, Laura C. and Bah{\'e}, Yannick M. and Villanueva, Vicente and {Jim{\'e}nez-Donaire}, Mar{\'i}a and Bisaria, Dhruv and Boselli, Alessandro and Bolatto, Alberto D. and Lee, Bumhyun},
  year = {2023},
  journal = {Publications of the Astronomical Society of Australia},
  volume = {40},
  eprint = {2303.07549},
  primaryclass = {astro-ph},
  pages = {e017},
  issn = {1323-3580, 1448-6083},
  doi = {10.1017/pasa.2023.14},
  urldate = {2025-04-25},
  archiveprefix = {arXiv}
}

@article{webbGOGREENSurveyPostinfall2020,
  title = {The {{GOGREEN}} Survey: Post-Infall Environmental Quenching Fails to Predict the Observed Age Difference between Quiescent Field and Cluster Galaxies at z {$>$} 1},
  shorttitle = {The {{GOGREEN}} Survey},
  author = {Webb, Kristi and Balogh, Michael L. and Leja, Joel and {van der Burg}, Remco F. J. and Rudnick, Gregory and Muzzin, Adam and Boak, Kevin and Cerulo, Pierluigi and Gilbank, David and Lidman, Chris and Old, Lyndsay J. and {Pintos-Castro}, Irene and McGee, Sean and Shipley, Heath and Biviano, Andrea and Chan, Jeffrey C. C. and Cooper, Michael and De Lucia, Gabriella and Demarco, Ricardo and Forrest, Ben and Jablonka, Pascale and Kukstas, Egidijus and McCarthy, Ian G. and McNab, Karen and Nantais, Julie and Noble, Allison and Poggianti, Bianca and Reeves, Andrew M. M. and Vulcani, Benedetta and Wilson, Gillian and Yee, Howard K. C. and Zaritsky, Dennis},
  year = {2020},
  month = nov,
  journal = {Monthly Notices of the Royal Astronomical Society},
  volume = {498},
  pages = {5317--5342},
  issn = {0035-8711},
  doi = {10.1093/mnras/staa2752},
  urldate = {2024-10-16}
}

@article{werlePoststarburstGalaxiesCenters2022,
  title = {Post-Starburst {{Galaxies}} in the {{Centers}} of {{Intermediate-redshift Clusters}}},
  author = {Werle, Ariel and Poggianti, Bianca and Moretti, Alessia and Bellhouse, Callum and Vulcani, Benedetta and Gullieuszik, Marco and Radovich, Mario and Fritz, Jacopo and Ignesti, Alessandro and Richard, Johan and Soucail, Genevi{\`e}ve and Bruzual, Gustavo and Charlot, Stephane and Mingozzi, Matilde and Bacchini, Cecilia and Tomicic, Neven and Smith, Rory and Kulier, Andrea and Peluso, Giorgia and Franchetto, Andrea},
  year = {2022},
  month = may,
  journal = {The Astrophysical Journal},
  volume = {930},
  pages = {43},
  issn = {0004-637X},
  doi = {10.3847/1538-4357/ac5f06},
  urldate = {2025-02-04}
}

@article{wernerSatelliteQuenchingWas2022,
  title = {Satellite Quenching Was Not Important for z {$\sim$} 1 Clusters: Most Quenching Occurred during Infall},
  shorttitle = {Satellite Quenching Was Not Important for z {$\sim$} 1 Clusters},
  author = {Werner, S V and Hatch, N A and Muzzin, A and {van der Burg}, R F J and Balogh, M L and Rudnick, G and Wilson, G},
  year = {2022},
  month = feb,
  journal = {Monthly Notices of the Royal Astronomical Society},
  volume = {510},
  number = {1},
  pages = {674--686},
  issn = {0035-8711},
  doi = {10.1093/mnras/stab3484},
  urldate = {2024-11-19}
}

@article{wetzelGalaxyEvolutionGroups2013,
  title = {Galaxy Evolution in Groups and Clusters: Satellite Star Formation Histories and Quenching Time-Scales in a Hierarchical {{Universe}}},
  shorttitle = {Galaxy Evolution in Groups and Clusters},
  author = {Wetzel, Andrew R. and Tinker, Jeremy L. and Conroy, Charlie and {van den Bosch}, Frank C.},
  year = {2013},
  month = jun,
  journal = {Monthly Notices of the Royal Astronomical Society},
  volume = {432},
  pages = {336--358},
  issn = {0035-8711},
  doi = {10.1093/mnras/stt469},
  urldate = {2025-02-06}
}

@article{willisDistantGalaxyClusters2013,
  title = {Distant Galaxy Clusters in the {{XMM Large Scale Structure}} Survey},
  author = {Willis, J. P. and Clerc, N. and Bremer, M. N. and Pierre, M. and Adami, C. and Ilbert, O. and Maughan, B. and Maurogordato, S. and Pacaud, F. and Valtchanov, I. and Chiappetti, L. and Thanjavur, K. and Gwyn, S. and Stanway, E. R. and Winkworth, C.},
  year = {2013},
  month = mar,
  journal = {Monthly Notices of the Royal Astronomical Society},
  volume = {430},
  number = {1},
  pages = {134--156},
  issn = {1365-2966, 0035-8711},
  doi = {10.1093/mnras/sts540},
  urldate = {2023-08-16}
}

@article{willisSpectroscopicConfirmationMature2020a,
  title = {Spectroscopic Confirmation of a Mature Galaxy Cluster at Redshift Two},
  author = {Willis, J. P. and Canning, R. E. A. and Noordeh, E. S. and Allen, S. W. and King, A. L. and Mantz, A. and Morris, R. G. and Stanford, S. A. and Brammer, G.},
  year = {2020},
  month = jan,
  journal = {Nature},
  volume = {577},
  number = {7788},
  pages = {39--41},
  issn = {0028-0836, 1476-4687},
  doi = {10.1038/s41586-019-1829-4},
  urldate = {2023-08-16}
}

@article{shibuyaMorphologies190000Galaxies2015,
  title = {Morphologies of {$\sim$}190,000 {{Galaxies}} at z = 0-10 {{Revealed}} with {{HST Legacy Data}}. {{I}}. {{Size Evolution}}},
  author = {Shibuya, Takatoshi and Ouchi, Masami and Harikane, Yuichi},
  year = {2015},
  month = aug,
  journal = {The Astrophysical Journal Supplement Series},
  volume = {219},
  pages = {15},
  publisher = {IOP},
  issn = {0067-0049},
  doi = {10.1088/0067-0049/219/2/15},
  urldate = {2025-06-27}
}

@article{okeAbsoluteSpectralEnergy1974,
  title = {Absolute {{Spectral Energy Distributions}} for {{White Dwarfs}}},
  author = {Oke, J. B.},
  year = {1974},
  month = feb,
  journal = {The Astrophysical Journal Supplement Series},
  volume = {27},
  pages = {21},
  publisher = {IOP},
  issn = {0067-0049},
  doi = {10.1086/190287},
  urldate = {2025-07-23}
}

%%%%%%%%%%%%%%%%% APPENDICES %%%%%%%%%%%%%%%%%%%%%

\appendix
\clearpage
\section{Image Alignment}
Alignment of the F140W and F105W mosaics required aligning drizzled images based on a pattern between regions. First, images were aligned to other images with the corresponding field name, then, these fields were aligned to the centre reference catalog. For F105W, this required a two step approach, first aligning to East and West portions of the central contiguous region, then aligning the drizzled products of east+west+centre to the reference catalog. The star galaxy separation (SGS) was used as an indicator of alignment, as well as a visual inspection of the final drizzled image, the number of objects used in alignment, and the weight images.

\section{Sample Spectra}
\begin{figure*}
	\includegraphics[width=\textwidth]{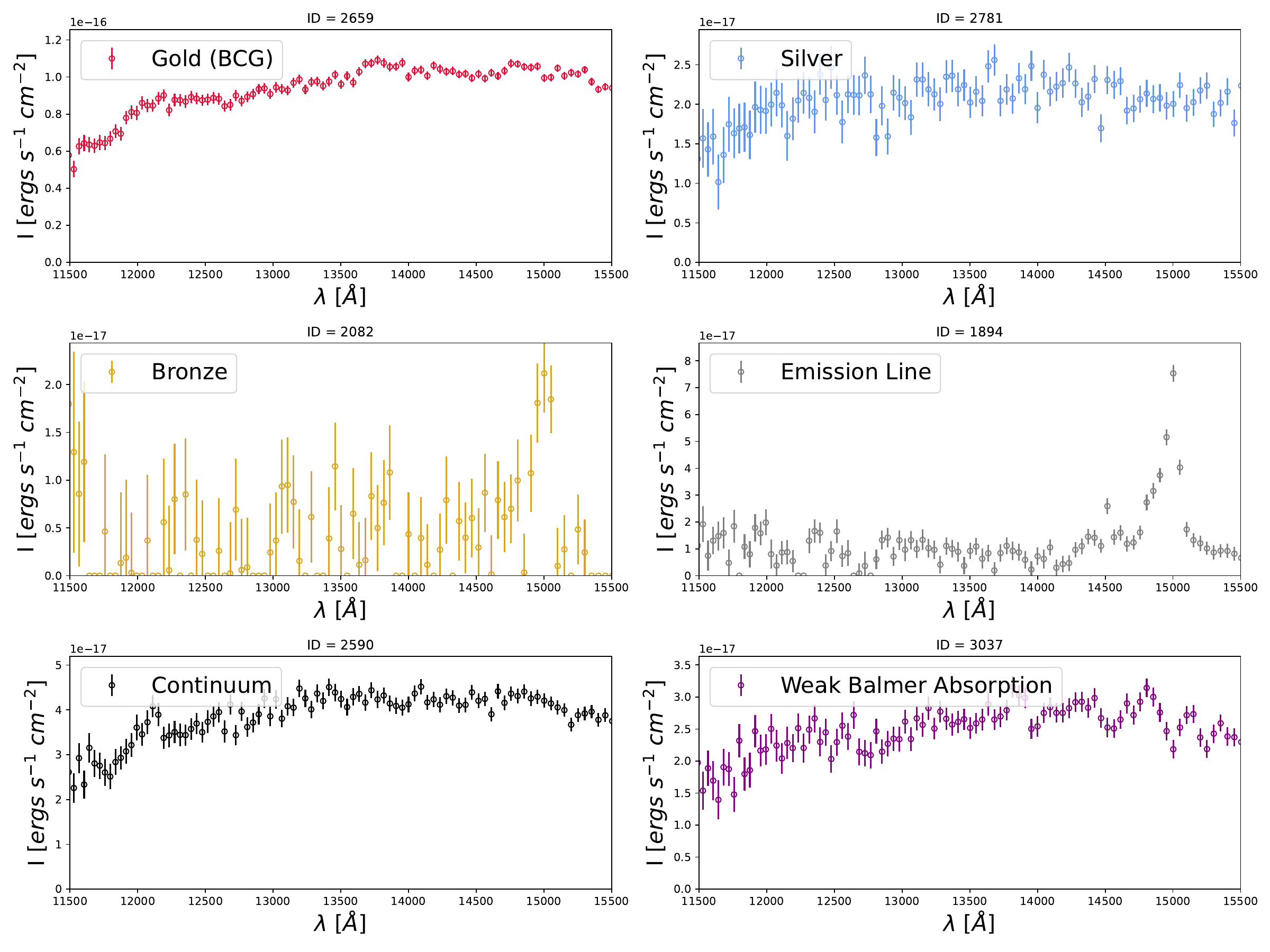}
    \caption{Sample optimally extracted, sensitivity corrected, spectra for 6 different types of galaxy in XLSSC 122. Red, blue, and yellow indicate Gold, Silver, and Bronze respectively. Grey, black, and purple indicate galaxies that are emission dominated, continuum dominated, or appear to have weak Balmer absorption respectively.}
    \label{fig:SampleSpectra}
\end{figure*}

In Fig. \ref{fig:SampleSpectra} I show the extracted spectra for a variety of galaxies. Gold members typically had high SNR, and bronze members tend to have low SNR. The continuum dominated galaxies show no sign of emission or absorption, whereas the galaxies with emission typically have [OIII]5007 and $H_b$ emission.

\section{\foreground \ Photometry and Cluster Membership}\label{sec: Foreground CMD}
\begin{figure*}
	\includegraphics[width=\textwidth]{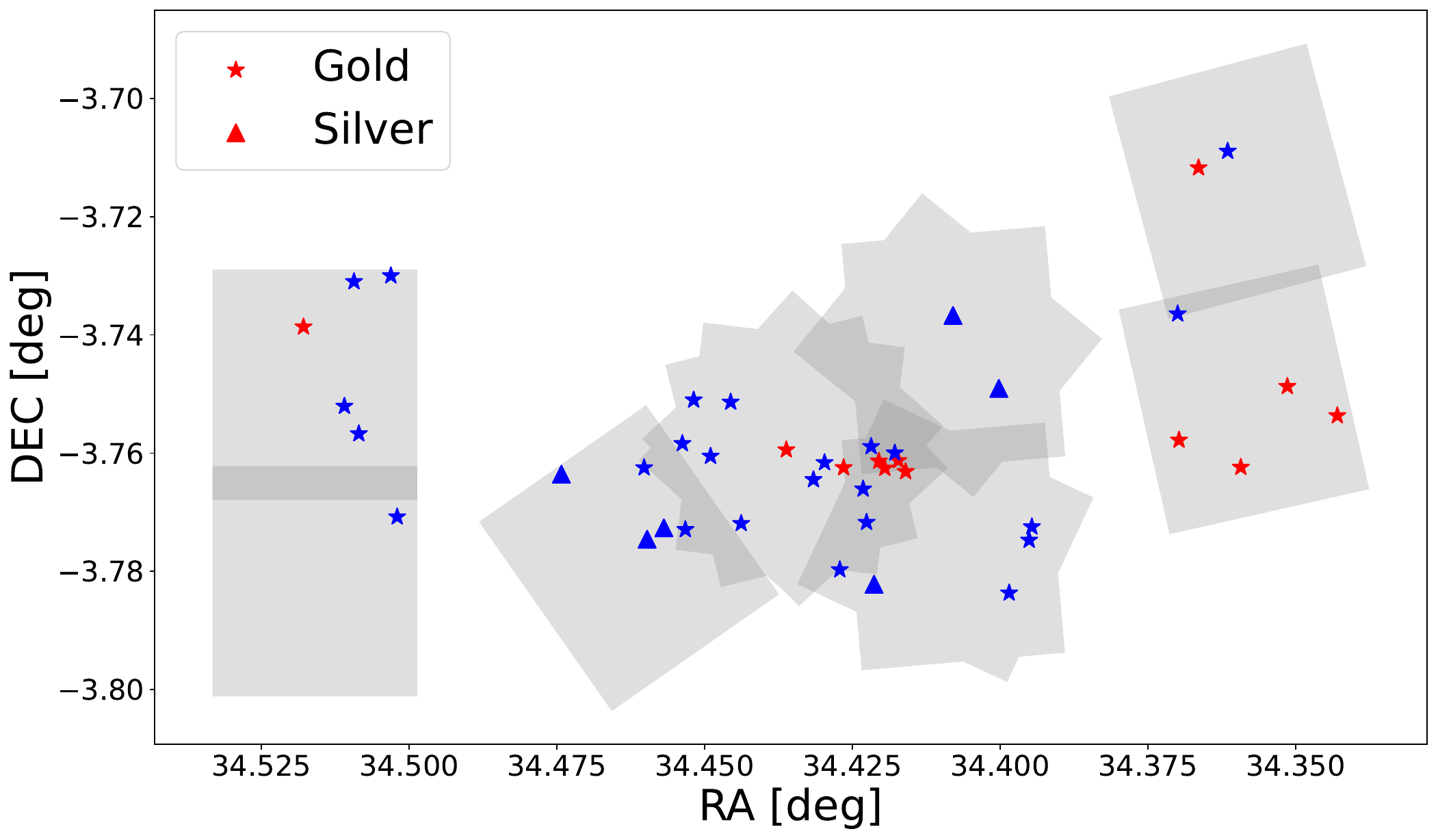}
    \caption{A footprint map with the F140W+G141 exposure in grey. Members of the Gold/Silver classification of z = 1.93 structure members is shown with stars/triangles. Cluster members are coloured red or blue based on the colour cuts in F105W-F140W = 1.15 in the central region, and F814W-F140W = 1.85 in the far regions.}
    \label{fig:F_cmemb_map}
\end{figure*}

When creating the updated catalogue of XLSSC 122 cluster members, a significant overdensity of galaxies around $z = 1.93$ was discovered close to XLSSC 122. The G141 grism's range allows for a similar spectroscopic redshift identification of these galaxies as for the primary cluster. Included below in Fig. \ref{fig:F_CMD_F105-F140} and Fig. \ref{fig:F_CMD_F814-F140} are the colour magnitude diagrams for the second cluster, \foreground. These are shown in both filter combinations used for the XLSSC122 figures, with the same colour cuts. The colour of the red and blue populations are similar to those seen in XLSSC 122, with a low scatter flat red coloration at an F105W-F140W colour of 1.4, and a blue cloud with high scatter around a colour of 0.5 from the same bands. There are 47 galaxies in the gold and silver categories of \foreground \ as seen in the cluster membership map in Fig. \ref{fig:F_cmemb_map}. Given the large number of these galaxies distributed around a central location, we label it a galaxy cluster. The central point of this cluster is listed as the barycentre of the 12 galaxies nearest the brightest classified galaxy. This cluster, though it may have complicated the classification of some members of XLSSC 122, is distinctly separated from that cluster, and is not dynamically interacting with the cluster given its co-moving distance of 76 megaparsecs \citep{willisSpectroscopicConfirmationMature2020a}. As such it has no additional bearing on the investigation of XLSSC 122. However, as a galaxy cluster it also shows a strong colour bimodality and the occasional post-starburst galaxy. \foreground \ does have a much larger proportion of blue galaxies, most galaxies in the cluster are blue, with a strong red core at the very centre. \foreground \ has a red fraction of $0.26\% \pm12\%$. These signs point to \foreground \ being a smaller, less mature cluster than XLSSC 122.

\begin{figure*}
	\includegraphics[width=\textwidth]{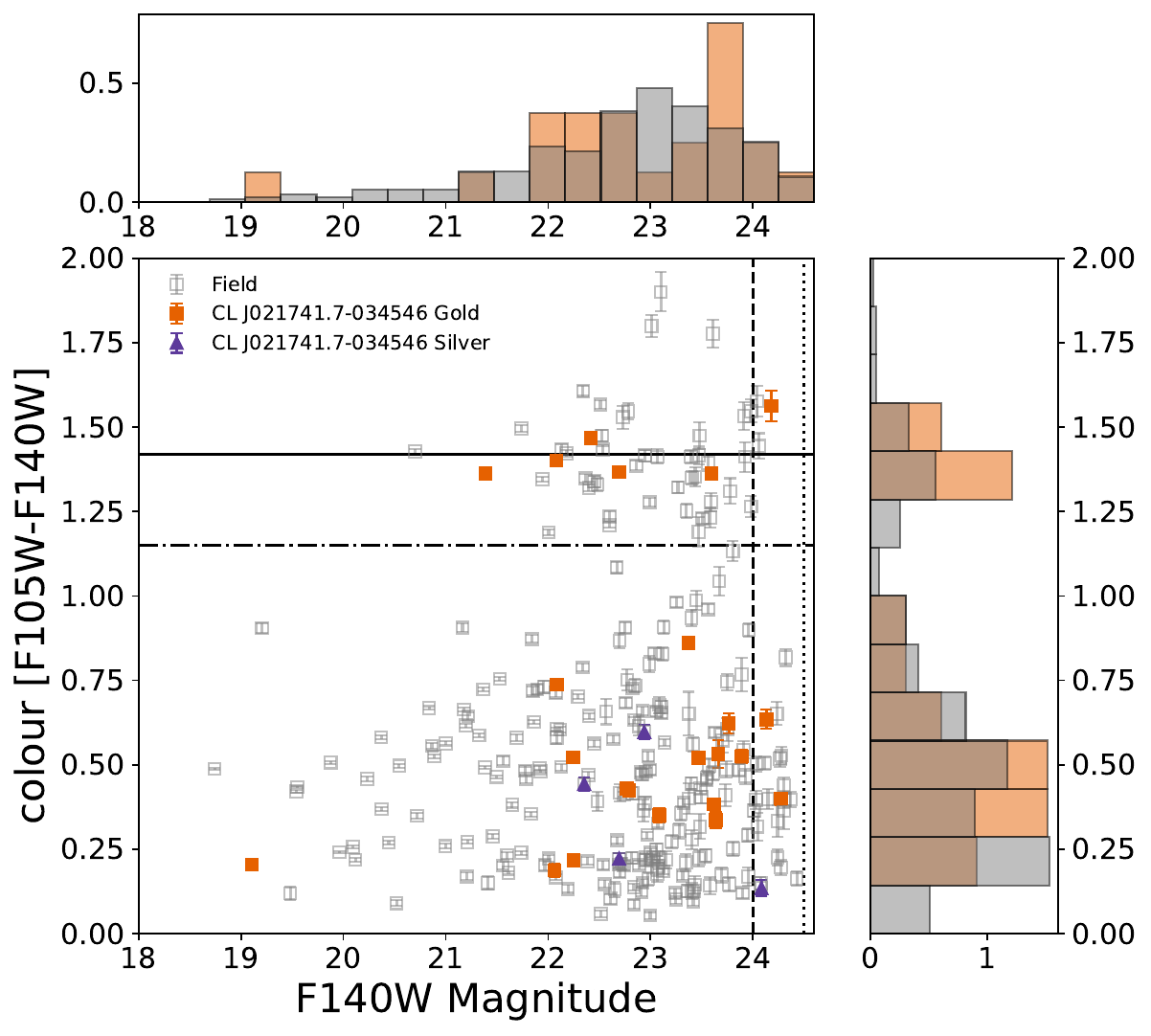}
    \caption{A Colour-Magnitude Diagram of the full \foreground \ photometric field with colour SNR > 5. Orange and purple points are gold and silver members respectively. Membership is determined by the MCMC probability within $z = (1.96,2.00)$. Grey points represent all photometrically extracted objects in the area. Vertical lines designate the 24 and 24.5 magnitude limits on cluster membership from prior core observations. The dash-dot horizontal line at 1.15 in colour indicates the red-sequence blue-cloud cut. The solid horizontal black line indicates the mean colour of the gold red sequence members. The histograms show the relative normalized number density of cluster members (orange) to non-cluster-members (grey) over magnitude.}
    \label{fig:F_CMD_F105-F140}
\end{figure*}
\begin{figure*}
	\includegraphics[width=\textwidth]{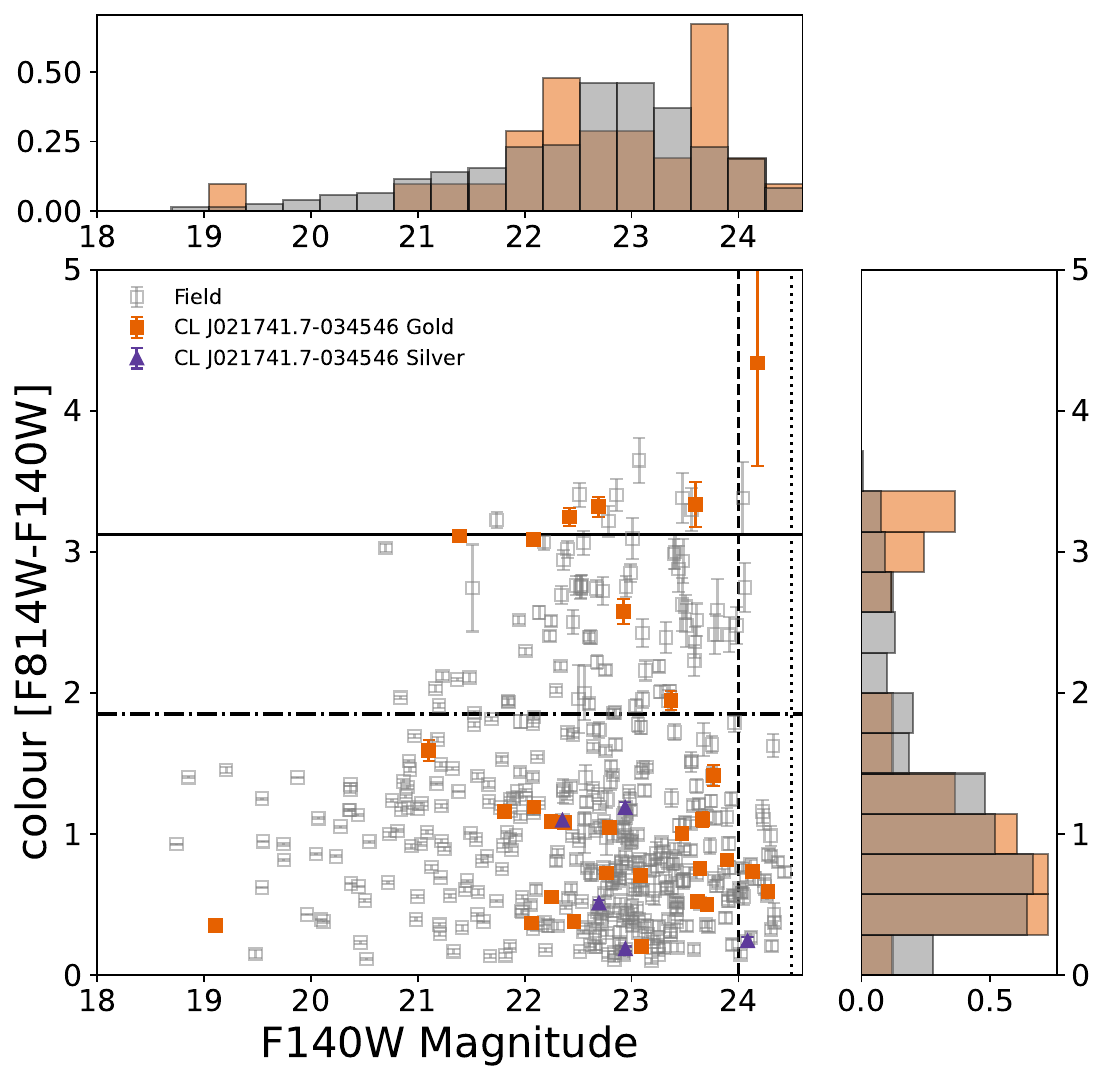}
    \caption{A Colour-Magnitude Diagram of the full \foreground \ photometric field with colour SNR > 5. Orange and purple points are gold and silver members respectively. Membership is determined by the MCMC probability within $z = (1.96,2.00)$. Grey points represent all photometrically extracted objects in the area not in either cluster. Vertical lines designate the 24 and 24.5 magnitude limits on cluster membership from prior core observations. The dash-dot horizontal line at 1.85 in colour indicates the red-sequence blue-cloud cut from \citep{noordehQuiescentGalaxiesVirialized2021} given as a quenched cut. The solid horizontal black line indicates the mean colour of the gold red sequence members. The histograms show the relative normalized number density of cluster members (orange) to non-cluster-members (grey) over magnitude.}
    \label{fig:F_CMD_F814-F140}
\end{figure*}

\clearpage

\section{Ram Pressure Stripping Model}\label{RamPressDerivation}
XLSSC 122 has a measured average temperature of $\mathrm{kT} = 5 \pm 0.7\,\mathrm{keV}$ \citep{mantzXXLSurveyXVII2018} within $R_{500} = 295\, \mathrm{kpc}$. The resolution of the grism spectra extracted for XLSSC 122 is insufficient to accurately measure radial velocity, so we will use an approximation instead. An individual ionized particle within the virial radius would follow the kinetic theory of gasses in thermal equilibrium, with a three dimensional velocity equal to $\sqrt{3}\sigma_v$, where $\sigma_v$, the velocity dispersion, is given in Eq. \ref{velocity dispersion of gas}
\begin{equation}\label{velocity dispersion of gas}
\scriptsize
    \sigma_v^2 = \frac{kT}{\mu m_p}
\end{equation}
derived from equating the kinetic and thermal energy of the gas shown in Eq. \ref{kinetic_to_thermal}.

\begin{equation}\label{kinetic_to_thermal}
\scriptsize
    \frac{1}{2}m<v^2> = \frac{3}{2}kT
\end{equation}

Here, $\mu$ is the mean molecular mass, and $m_p$ is the mass of a proton. For, XLSSC 122, we use $kT=5000\,\mathrm{eV}$, $\mu = 0.608$ (H:He = 3:1), and $m_p = 1.602\cdot10^{-27}\,\mathrm{kg}$  \citep{mantzXXLSurveyXVII2018}. With this, we find $v_{vir} = 1500\,\mathrm{km/s}$. We assume that at any radius $r<r_{vir} = 440\,\mathrm{kpc}$, $v = v_{vir}$, and for any $r \geq r_{vir} = 440\,\mathrm{kpc}$, $v=\frac{v_{vir}}{\sqrt{r/r_{vir}}}$. Since we do not have the average temperature within $R_{200}$, this is an overestimate of the virial temperature. We use the ram pressure stripping condition from \cite{mccarthyRamPressureStripping2008}, given in Eq. \ref{stripcond}.
\begin{equation}\label{stripcond}
\scriptsize
    \rho_{ICM}(r)\cdot v^2 > \frac{\pi}{2}\frac{GM_{gal}(R)\rho_{gas}(R)}{R}
\end{equation}
Where $\rho_{ICM}$ is the density of the ICM plasma at a distance r from the cluster centre, v is the three-dimensional velocity of the infalling galaxy and G is the gravitational constant. In this model, the infalling galaxy is assumed to be on a direct path to the centre of the cluster. We will now make assumptions about the member galaxies in XLSSC 122. For the ram pressure stripping equation $M_{gal}(R)$ and $\rho_{gas}(R)$ are the mass contained within and galaxy gas density at a distance R from the galaxy centre. 
We use the ICM density profile as found by \cite{mantzXXLSurveyXVII2018}, the velocities we calculate above and $\rho_{gas}$ from Eq. \ref{gasprofile} with scale radius $r_d = 3.5\,\mathrm{kpc}$.
The most prevalent component is that of atomic hydrogen in the interstellar medium, or ISM. The distribution of the atomic hydrogen in a late type galaxy galaxy can be approximated by a simple exponential falloff given in Eq. \ref{gasprofile} \citep{freemanDisksSpiralS01970}.

\begin{equation}\label{gasprofile}
\scriptsize
n(R) = n_0\exp\left(-\frac{R}{R_c}\right)
\end{equation}

Here, $n_0$ is a peak particle number density, n(R) is the particle density at a radius from galaxy centre R, with a given scale radius $R_c$. More complex models of the gas dynamics do exist, but this is a simple case \citep{binneyGalacticDynamics1987}.
From previous studies of the cluster core \citep{trudeauXXLSurveyXLIX2022} we know the stellar masses of the galaxies in XLSSC 122. Using this and a conversion from stellar mass to halo mass from \cite{behrooziLackEvolutionGalaxy2013,behrooziAverageStarFormation2013} for galaxies at $z = 2$, we can estimate the mass profiles of infalling galaxies. We use an NFW profile \citep{navarroUniversalDensityProfile1997} with a varying scale radius Rs from \cite{vanderwel3DHST+CANDELSEvolutionGalaxy2014} given in Eq. \ref{Redistribution}.

\begin{equation}\label{Redistribution}
\scriptsize
R_{eff,z=2} \approx \left(\frac{M_*}{5\times10^{10}M_\odot} \right)^\alpha
\end{equation}

Here $R_{eff}$ is the effective radius of the stellar mass according to a \cite{sersicAtlasGalaxiasAustrales1968} profile, $\alpha$ is a mass scaling parameter. At $z = 2$, we use $\alpha = 0.6$ for spherical galaxies, and $\alpha = 0.76$ for disk galaxies, from the fits to observations from \cite{vanderwel3DHST+CANDELSEvolutionGalaxy2014}. We use a Sersic index n = 1.5 for both galaxy types \citep{shibuyaMorphologies190000Galaxies2015}. For the NFW scale radius, $R_s = \frac{R_{vir}}{c}$. We use the evolution of the concentration parameter c from \cite{duttonColdDarkMatter2014}, which at $z = 2$ can be estimated by $c \approx 10^{0.65}\left(\frac{M_{h}}{10^{12}M_{\odot}} \right)^{-0.05}$. For the virial radius one can approximate a virial halo at $R_{vir} = R_{200}$, with resulting mass $M = 4\pi R_{vir}^{3}\rho_{crit}\Delta_c/3$ \citep{moGalaxyFormationEvolution2010} with $\Delta_c = 200$. Thus, $R_{vir} \propto M^{\frac{1}{3}}$. We will assume that the virial radius can then be estimated at $z = 2$ by Eq. \ref{Rvirdistribution} derived from \cite{moGalaxyFormationEvolution2010}.

\begin{equation}\label{Rvirdistribution}
   \scriptsize
    R_{vir,z=2} \approx \frac{200}{3}\left(\frac{M_{vir}}{10^{12}M_\odot} \right)^\frac{1}{3}
\end{equation}

Here, $M_{vir}$ is the mass contained within the virial radius. The constant of 200 is for $z = 0$, and the factor of $1+z = 3$ comes from the scaling of the Hubble parameter with redshift. We consider the gas profile for the galaxies we model using Eq. \ref{gasprofile}.

%%%%%%%%%%%%%%%%%%%%%%%%%%%%%%%%%%%%%%%%%%%%%%%%%%

% Don't change these lines
\bsp	% typesetting comment
\label{lastpage}
\end{document}